\documentstyle[prl,aps]{revtex}
\twocolumn

\begin{document}
\input{epsf}
\def\Fig#1{Figure \ref{#1}}
\def\Eq#1{Eq.~(\ref{#1})}
\draft
\twocolumn[\hsize\textwidth\columnwidth\hsize\csname@twocolumnfalse\endcsname

\title{Pattern formation in 2-frequency forced parametric waves}
\author{H. Arbell and J. Fineberg}
\address{The Racah Institute of Physics, The Hebrew
University of Jerusalem, Jerusalem 91904, Israel} \maketitle

\begin{abstract}
We present an experimental investigation of superlattice patterns
generated on the surface of a fluid via parametric forcing with 2
commensurate frequencies. The spatio-temporal behavior of 4
qualitatively different types of superlattice patterns is
described in detail. These states are generated via a number of
different 3--wave resonant interactions. They occur either as
symmetry--breaking bifurcations of hexagonal patterns composed of
a single unstable mode or via nonlinear interactions between the
two primary unstable modes generated by the two forcing
frequencies. A coherent picture of these states together with the
phase space in which they appear is presented. In addition, we
describe a number of new superlattice states generated by 4--wave
interactions that arise when symmetry constraints rule out 3--wave
resonances.
\end{abstract}
 \vskip1pc]
\narrowtext

\section{Background}\label{background}
Patterns are ubiquitous in the world around us.  The word
``pattern" describes an order, regularity, or a simple
mathematical description, which can be found either in a natural
or man--made system. Patterns often result from self--interactions
of driven nonlinear systems. Naturally, the first patterns to be
scientifically analyzed were the simplest ones, which can be
described by few mathematical variables. However, in recent years
we have learned to recognize and categorize patterns in systems
that were assumed to be formless, devoid of any order. Perhaps the
most obvious characteristics of these systems are the multiple
length and time scales that can be present simultaneously. One of
the most important mechanisms to explain such phenomena is the
nonlinear resonant interaction between the different modes that
are excited in these systems. In these interactions two or more
waves can interact to form ``new" waves. These waves have a
wavelength and frequency which is the sum or difference of the
basic waves. The system's energy can then be transferred between
these modes or dissipated at different scales. The purpose of the
work described in this paper was to explore this paradigm in the
experimental study of a simple controlled system: the parametric
excitation of waves on the surface of a fluid (the Faraday
system).

The general form of the external acceleration applied to the
system is given by:
\begin{equation}\label{2freqDriving1}
g(t)= A [ \cos(\chi)
\cos(m\omega_0t)+\sin(\chi)\cos(n\omega_0t+\phi)]
\end{equation}
This spatially uniform vertical excitation preserves the system's
spatial symmetries, while modifying its temporal ones.

 As first noted by Faraday, sinusoidal acceleration (in the
direction of gravity) of a fluid layer with angular frequency
$\omega$ induces a pattern, having a wavenumber $k(\omega)$, on
the fluid surface. Whereas waves excited by a single--frequency
have been studied extensively over the last 4 decades, the
response of the system for multi--frequency excitations has only
recently begun to be investigated. Single--frequency driving can
produce patterns of different symmetries. Patterns consisting of
rolls, squares, hexagons and 8,10,12-fold quasi--patterns have
been experimentally observed in
\cite{Douday,Binks,KudrolliPhysicaD,KumarBajaj}. The secondary
instabilities of these different patterns involve complicated
states that display either spatiotemporal chaos
\cite{Bosch,Ciliberto,Ezerskii,Gluckman,KudrolliPhysicaD,KudrolliChaos},
transverse amplitude modulations \cite{Daudet} or various defects
\cite{KudrolliPhysicaD,Daudet} that break the patterns' initial
global symmetry. The use of multiple--frequency driving enables us
to study the interactions of different excited modes in a
controlled way, as each excitation frequency can linearly excite a
well--defined wavenumber. In this way, we hope to be able to
slowly unfold the system's underlying behavior and, thereby
unravel the fundamental mechanisms that describe the waves'
interactions.

The purpose of this paper is to provide a coherent overview of the
wide variety of nonlinear states that result from 2-frequency
forcing. We will provide detailed descriptions of the spatial and
temporal behavior of these states. In doing so, we will provide a
characterization of these nonlinear states - depicting both the
resonant mechanisms and symmetry constraints giving rise to their
formation.

 This paper is organized in the following
fashion. In section \ref{background} we will briefly describe the
theoretical and experimental work that has, to date, been
performed in this system. The experimental apparatus and
measurement techniques used in our measurements will then be
described in section \ref{System}. We will then present, in
section \ref{Results1}, an overview of the phase diagram together
with a brief description of the different types of superlattice
states observed. Each type of superlattice, together with the
mechanisms that form it, will then be described in detail in the
subsequent sections.
 A codimension two point, at
$\chi_c$, exists in this system where both externally driven modes
simultaneously become linearly unstable. Subharmonic Superlattice
States (SSS), which bifurcate from an initial hexagonal state far
from $\chi_c$, will be described in section \ref{sectionSSS}. We
will then progress to the region of phase space in the vicinity of
$\chi_c$. Three different types of superlattice patterns will be
described in sections \ref{DHS}, \ref{Results2}, and
\ref{Results3}. All of the above superlattice patterns result from
different types of 3--wave resonant interactions. We will conclude
with section \ref{Results4} in which a number of superlattice
states generated by 4--wave resonant interactions are described.
We will show that these states can occur when 3--wave interactions
are forbidden.

\subsection{Notation}\label{notation}
The notation conventions used throughout this paper are as
follows. The driving function is specified in \Eq{2freqDriving1}.
To avoid confusion, we will always specify the driving frequency
ratio used in \Eq{2freqDriving1} by the ratio $m/n$ where $m$ and
$n$ are the two co--prime integers that describe the two
frequencies $\omega_1=m\omega_0$ and $\omega_2=n\omega_0$. We will
always assume that $n>m$. In our notation, $k_1$ and $k_2$ refer
to the wavenumbers excited, respectively, by the driving
frequencies $\omega_1$ and $\omega_2$. The angle $\chi$ in
\Eq{2freqDriving1} describes the relative mixing between the two
modes and the angle $\phi$ describes their phase difference, where
the relevant range is $0 < \phi <2\pi m/n$ . $\cos(\chi)$ and
$\sin(\chi)$ are sometimes \cite{Zhang2freq,Muller2freq} replaced
by the mixing coefficients $r$ and $1-r$.

In the following sections we will frequently characterize
eigenmodes by their temporal parity. Since the parity of $m$ and
$n$ is important we will use the notation odd/even (even/odd) to
described the classes of driving where $m$ ($n$) is odd and $n$
($m$) is even. Odd/odd describes driving where both $m$ and $n$
are odd. A state whose temporal response has a fundamental
frequency of $\omega_0/2$ will be denoted as a ``subharmonic"
state, whereas a ``harmonic" state is one with a fundamental
 frequency of $\omega_0$.

We shall use the following units. The total amplitude $A$,
appearing in \Eq{2freqDriving1}, is measured in units of
$g=981cm^2/sec$. The fluid's kinematic viscosity, $\nu$ is
measured in cS ($0.01 cm^2/s$), and the depth of the fluid layer
$h$, is measured in cm. In many cases, we will identify, for
simplicity, angular frequencies (e.g. $\omega$) with the
corresponding temporal ones (e.g. $\omega / (2\pi)$). Where
necessary, angular or temporal frequencies will be explicitly
denoted. Additional notation will be defined as needed.

\subsection{Linear analysis}\label{linear analysis}
The linear stability analysis of the problem was performed by
Besson et al. \cite{Besson} by numerically solving the linearized
Navier-Stokes equation via an extension of the technique developed
by Kumar and Tuckerman \cite{Kumar} for single frequency
excitations. As in the single frequency case, the
acceleration-wavenumber plane is characterized by alternating
tongues corresponding to the acceleration at which a given
wavenumber becomes linearly unstable. For mixing angles
$\chi=0^\circ$ and $\chi=90^\circ$ the tongue structure of single
frequency forcing with either $m\omega_0$ or $n\omega_0$ driving
is obtained. Increasing $\chi$ from 0 results in the appearance of
additional tongues whose dominant frequencies are spaced
$\omega_0/2$ apart, since the system's basic frequency is then
$\omega_0$. Each odd numbered tongue possesses a {\em subharmonic}
temporal dependence composed of only frequencies $\omega_0(p+1/2)$
whereas even-numbered tongues are temporally {\em harmonic} i.e.
composed of frequencies $p\omega_0$ (where $p$ is any whole
number). Although the time dependence of each tongue is given by
an infinite series, the {\em dominant} frequency of the $p^{th}$
odd (even) tongue corresponds to $(p+1/2)\omega_0$ ($p\omega_0$).
Generally, within the critical tongues the dominant frequencies
$n\omega_0/2$ or $m\omega_0/2$ have an order of magnitude greater
amplitude than the other components. Thus, the temporal response
of the critical modes for $m/n=$odd/odd driving ratios is always
subharmonic. For odd/even (even/odd)  the $k_1$ ``tongue" has a
subharmonic (harmonic) response while the $k_2$ has the opposite
parity.

 The system's critical wavenumber, $k_c$, corresponds to the lowest
acceleration, $a_c$, at which the flat, featureless state loses
stability. At a critical value of $\chi=\chi_c$, a codimension 2
point exists where two tongues having wavenumbers $k_1$ and $k_2$
corresponding, respectively, to $m\omega_0/2$ and $n\omega_0/2$,
simultaneously become unstable. For $\chi$ far from $\chi_c$ the
critical wavenumbers are close to the values of $k_c$ obtained for
single--frequency experiments. Near the codimension 2 point $k_c$
tends to differ from the corresponding single--frequency value by
up to 10\%. The calculated values of both $a_c$ and $k_c$ are in
excellent agreement with experiments \cite{Besson,Arbell1}. While
modes other than the critical ones are linearly damped, we will
see that they can play an important role in nonlinear wave
interactions.

\subsection{Experiments with two-frequency forcing}\label{2freq-exp}

In the case of single-frequency driving, the subharmonic time
dependence prohibits quadratic terms in the ``amplitude" equations
describing the nonlinear interactions between the amplitudes of
the excited modes. However, when using two-frequency driving with
odd/even or even/odd parity, both harmonic and subharmonic
temporal responses are possible. When one of the driving frequency
components is dominant, one can consider the smaller component as
a perturbation that breaks the system's temporal subharmonic
symmetry. The reflection invariance of the corresponding set of
coupled amplitude equations is then broken and, generically,
quadratic terms can appear. These quadratic terms are important
since they enable three--wave interactions between different
modes.

Edwards and Fauve were the first to study the two--frequency
driven Faraday instability \cite{Edwards92,Edwards93,Edwards94}.
They chose to focus most of their study on 4:5 driving although
they also explored other ratios (such as 6/7, 4/7, 8/9 and 3/5).
These experiments used a relatively viscous fluid and small fluid
layer height in order to minimize lateral boundary effects.
 As the viscosity of the fluid was rather high ($\nu=100$ cS) stripe
patterns occurred for single--frequency driving. The phase space
as presented in \cite{Edwards92} for (even/odd) 4:5 driving can be
divided into two parts: The harmonic (subharmonic) part where the
$k_1$ ($k_2$) wavevector is dominant and the leading temporal term
has the frequency of $4\omega_0/2$ ($5\omega_0/2$). In the
harmonic region, in place of the stripe patterns of wavenumber
$k_1$ appearing for pure $4\omega_0$ driving, a first order
transition to hexagons occurs for $\chi> 10^\circ$. In the
subharmonic region, striped patterns with wave number $k_2$ are
observed until the near vicinity of the codimension 2 point at
$\chi_c$. The first-order transition to the hexagonal state
results from the quadratic interactions mentioned above.

In the neighborhood of the codimension 2 point, a temporally
harmonic, twelvefold symmetric quasi-periodic pattern was
observed. These states appeared for only a small range of $\phi$
($\phi\sim 75^\circ \pm 5^\circ$). They evolved, via a first-order
bifurcation, from either the flat zero--amplitude state or the
subharmonic striped patterns. In the hysteretic region of these
states, Edwards and Fauve also observed solitary axial waves that
originated from the quasi--periodic pattern. Arbell and Fineberg
\cite{Arbell3} have shown that these highly locallized waves are
related to ``oscillons'' states observed \cite{OscillonsNature} in
driven granular systems.

Muller \cite{Muller2freq} later conducted two-frequency forcing
experiments using a 1/2 driving ratio. These experiments were
performed near $\chi_c$ in shallow fluid layers in various regions
of the $\chi-\phi$ phase space. Both temporally subharmonic
hexagon and triangle patterns were observed. Triangular patterns
are formed when the spatial phase associated with each of the
excited eigenmodes differs from $0^\circ$ or $180^\circ$. Muller
showed that amplitude equations with both cubic and quintic terms
(applying to temporally subharmonic waves), can form triangular
patterns. In contrast, amplitude equations with quadratic terms
have only stable hexagonal solutions. These experiments were later
modelled by Zhang and Vinals \cite{Zhang2freq} using a
quasipotential approach. Experimentally, Muller showed that the
addition of a third small--amplitude forcing frequency (which is
equivalent to perturbatively breaking the system's parity) could
stabilize either the hexagonal or triangular states.

More recent experimental studies in 2 frequency-forced systems
were performed by two groups, Kudrolli, Peir and Gollub
\cite{Kudrolli} and Arbell and Fineberg
\cite{Arbell1,Arbell3,Arbell2}. These studies were conducted both
in the near vicinity and far from $\chi_c$. They revealed a number
of qualitatively new, superlattice-type states in which new
scales, not directly introduced via the external forcing, were
evident.

In regions of phase space that can be relatively far from $\chi
_c$, superlattice states were observed as secondary bifurcations
from the harmonic ($m\omega_0 /2$) hexagonal states that occur for
odd/even or even/odd driving ratios. The primary hexagonal
symmetry with wavenumber $k_c$ is broken by additional modes with
wavenumbers $q < k_c$ whose temporal response possesses an
$m\omega_0/4$ component. These states include ``SL-II" states
observed for 4:5 driving by Kudrolli et al. \cite{Kudrolli} and
the Subharmonic Superlattice (``SSS") states observed for a large
number of driving ratios by Arbell et al. \cite{Arbell1}.

A second type of superlattice state is observed in the near
vicinity of the codimension 2 point, once again on the side
dominated by the {\em harmonic} driving component when even/odd or
odd/even forcing is used. Two variants of these states coined
``SL-I" \cite{Kudrolli} and double hexagonal states (``DHS")
\cite{Arbell3} have been observed. These states can be described
by the superposition of two hexagonal sets of wavevectors of
magnitude $k_c$. The two sets of six wavevectors are oriented at
an angle $\theta_r \simeq 22^\circ$ to each other. This specific
angle is not arbitrarily chosen. The sum and difference vectors
between the two wavevector sets produce a sublattice, spanned by
the smaller difference wavevectors. When the two sets of
wavevectors are oriented at specific angles of $\theta_r$, the
sublattice formed by the difference vectors becomes commensurate
with the two hexagonal lattices. This structure is one of the
generic possibilities that were anticipated on the basis of
symmetry arguments proposed by Silber and Proctor
\cite{SilberProctor}.

A qualitatively different type of superlattice occurs in the
vicinity of $\chi_c$ for all driving ratio parities. These states,
coined 2-mode superlattices (2MS) \cite{Arbell1}, are the most
general of the superlattice states observed. They are formed by an
interaction of the two linearly excited modes ($k_1$ and $k_2$)
with a third linearly--damped slaved mode that is nonlinearly
excited. The angle between $k_1$ and $k_2$ is chosen by the
following resonance mechanism: the vector difference
$\vec{k}_2-\vec{k}_1$ produces a third wavevector $\vec{k}_3$. The
magnitude of $k_3$ is determined by the dispersion relation,
$\omega(k)$, for the difference frequency
$\omega_3=\omega_2-\omega_1$.

An additional type of superlattice state has been observed for 2/3
and 4/5 driving in the vicinity of $\chi_c$ \cite{Arbell2}. This
state, which appears in place of the 2MS state, consists of a
rhomboid pattern that is formed by the simple nonlinear resonance:
$\vec{k}_2-\vec{k'}_2=\vec{k}_1$ where $|\vec{k}_2|=|\vec{k'}_2|$.
When the coupling angle $ \theta$ ($ \theta \equiv cos^{-1}[\vec
k_2 \cdot \vec k'_2/k_2^2]$) between $\vec{k}_2$ and $\vec{k'}_2$
is tuned to a value of $\theta \sim 2\pi / n$, 2--n fold
quasicrystalline patterns are naturally formed.

\subsection{Model equations and nonlinear analysis}\label{2freq-nonlinear}

Generally, two  methods have been used to study the Faraday
instability with two--frequency driving. The first uses simple
model systems that yield qualitative insights regarding the
behavior of the Faraday system. These use general assumptions
based mainly on symmetry considerations. The second method is to
start from the full nonlinear set of equations that describe the
system, employ carefully chosen approximations, and derive a set
of equations that describes the behavior of the system based on
the real physical parameters. Both methods have yielded valuable
insights.

\subsubsection{Model equations} \label{ModelEq}

The observation of quasi-crystalline patterns (``quasi--patterns")
generated using two--frequency driving by Edwards and Fauve
\cite{Edwards92} and via single frequency driving by Binks et al.
\cite{BinksHeight} provided a motivation to find model equations
that display similar behavior. Muller \cite{MullerModel} first
considered a system of N coupled Landau equations with cubic
nonlinear terms. These equations could be written as the gradient
of a Lyaponov functional. Muller showed, by minimization of this
functional, that regular N--fold patterns of different symmetries
can be stable. Pattern selection depended on the value of the
nonlinear coefficients coupling the linearly degenerate modes.
This mechanism may be related to both the appearance of
quasi--patterns in single frequency Faraday experiments and to the
quasi--patterns observed in the harmonic region of two--frequency
Faraday systems with even/odd driving.

A second mechanism that can create quasi--patterns is related to
quadratic interactions between degenerate nonlinear modes. Muller
proposed that a quadratic nonlinearity, generating the triad
interaction: $\vec k_2 -\vec k_2'=\vec k_1$ ($|\vec k_2|=|\vec
k_2'|$) could also lead to quasi--crystalline patterns. The angle
between $\vec{k_2}$ and $\vec{k'_2}$ is tunable by the ratio
$k_1/k_2$, with resonant angles $\theta=45^\circ$,
$\theta=36^\circ$, $\theta=30^\circ$ for 8--fold, 10--fold and
12--fold quasi--patterns respectively. These states were observed
in a system of two coupled Swift-Hohenberg equations, each with a
different unstable wavenumber. Frisch and Sonnino  \cite{Frisch}
also observed subcritical 10--fold symmetrical patterns in coupled
Swift-Hohenberg equations. This state was numerically shown to be
stable even when the dynamics are not derived from a free energy
functional. In addition to states with N-fold symmetry,
nonsymmetric, rhomboidal patterns were also seen to be stable for
some parameter values. Later, both the rhomboidal patterns and
resonant quasipatterns resulting from the above interactions were
observed experimentally by Arbell and Fineberg \cite{Arbell2}.

Lifshitz and Petrich \cite{Lifshitz} modeled the two--frequency
Faraday system with a single generalized Swift--Hohenberg type
equation for a {\em single} real field $u(x,y)$. This model is
simpler than the coupled equations used by Muller, Frisch and
Sonino, and Newell and Pomeau \cite{Newell}. The model equation
used was rotationally invariant with {\em two} built-in critical
wavenumbers. The equation contained a quadratic term that both
broke the system's up--down symmetry and allowed triad wave
interactions. Stable striped, hexagonal and 12-fold symmetric
patterns were observed for different values of the control
parameter. In addition to these N--fold symmetric states, a
compressed hexagon state, similar to the rhombic/stripe pattern
described by Muller, was observed.

The above model systems suggest that the existence of two unstable
wavenumbers together with the possibility of triad interactions
(provided by quadratic terms) is a sufficient condition for the
formation of quasi-periodic patterns. Another common feature of
these models is the existence of distinct regions of phase space
in which patterns that lack N--fold symmetry are stable.

\subsubsection{Nonlinear analysis}\label{Nonlinear-analysis}

In contrast to the simplified model systems described above, Zhang
and Vinals \cite{Zhang2freq} derived a description of the system's
dynamics from the governing equations for the 2--frequency Faraday
problem. To this end, they applied the quasi-potential approach
developed for single--frequency study\cite{Zhang1freq} to the
problem of two--frequency driving. This approach is strictly valid
in the limits of weak dissipation and infinite fluid depth.

To compare their results with Muller's experimental results, Zhang
et al. analyzed the special case of $1/2$ driving in depth. They
first used the linearized equation to study the location of the
codimension two point, $\chi_c$ as a function of the phase
difference $\phi$. The results were in qualitative agreement with
the experiments. The discrepancies were attributed to the high
damping used in Muller's experiment, which was outside the region
of validity of the theory. It is interesting to note that the
dependence of $\chi_c$ on $\phi$ is a special feature of $1/2$
driving and does not occur for other driving combinations.

Zhang and Vinals then, using a multiple scales approach, derived
standing wave amplitude equations. This weakly nonlinear analysis
assumed that the system was far from the codimension 2 point, so
that a single temporal mode dominated the dynamics.  For the case
of 1/2 driving, they first obtained a prediction for the relative
magnitudes of the different Fourier components of the weakly
nonlinear temporal response of the fluid surface. Then, assuming
$N$ degenerate modes, the coupled amplitude equations describing
these modes were derived. In contrast to \cite{MullerModel}, the
function $\beta(\theta_{ij})$ coupling the $i^{th}$ and $j^{th}$
modes, was computed from the physical parameters of the problem.
One interesting result of this calculation was that the phase
difference $\phi$ can have a strong effect on the coupling
function $\beta$ - and thereby a strong effect on the nonlinear
pattern selection. The relative stability of different N--fold
nonlinear states was then calculated by minimization of a Lyapunov
functional, as in \cite{MullerModel} and \cite{Lifshitz}.
Semi--quantitative agreement with the regions of stability
observed experimentally in \cite{Muller2freq} for different
patterns (squares, hexagons/triangles and quasi--patterns of
various orders) observed in the {\em subharmonic} region of the
$\phi-\chi$ phase space was obtained.

An important result of this work was that it suggested a new type
of physical mechanism that governs the selection process. These
calculations indicated that the value of the coupling coefficients
was strongly influenced by triad interactions between the linearly
excited modes (corresponding to the dominant excitation frequency)
and the {\em linearly stable} modes corresponding to the {\em
second} excitation frequency. Resonant coupling to these latter
modes served, in the region of phase space far from $\chi_c$, to
enhance the effective damping - as energy transferred to these
``slaved" modes is more efficiently dissipated. Far from $\chi_c$,
states that can {\em not} couple to the slaved modes are then
preferred by the system. As we shall later see, however, in the
{\em vicinity} of $\chi_c$ resonant triad coupling to the linearly
stable, slaved modes provides one of the main mechanisms for the
rich variety of nonlinear states observed.

Silber and Skeldon \cite{SilberSkeldon} were the first to
theoretically study the two-—frequency Faraday system in the
vicinity of the codimension 2 point. This study pointed out the
importance of accounting for the temporal symmetries of the
system. Silber and Skeldon focused on forcing ratios, $m/n$,
having either odd/even or even/odd parities, where interactions
between harmonic and subharmonic waves may occur.

As shown in \cite{Zhang2freq}, resonant mode interactions greatly
affect the mode coupling function $\beta(\theta)$. Using normal
form analysis, Silber and Skeldon showed that triad resonances
$\vec{k_1}\pm\vec{k'_1}=\vec{k_2}$ (where $k_1 = k'_1$) are only
possible when the temporal mode corresponding to $k_2$ is {\em
harmonic}. When $k_2$ has a subharmonic temporal dependence,
quadratic terms in normal form equations can be eliminated
\cite{Crawford91a,Crawford91b} - thereby decoupling the harmonic
modes from the subharmonic ones. This can be simply understood
since the product of two linear eigenfunctions (resulting from a
quadratic interaction term) results in the addition of their
temporal phases. The sum of {\em two} harmonic or subharmonic
temporal phases cannot produce a subharmonic one, therefore two
modes of like parity cannot couple quadratically to a subharmonic
state. Silber and Skeldon \cite{SilberSkeldon} went on to
demonstrate the above, by calculating the amplitude equations for
both  1/2 (odd/even) or 2/3 (even/odd) driving by means of the
quasi--potential approximation used in \cite{Zhang2freq}.

 The existence or suppression of three wave resonances can have a
significant effect on the qualitative features of the phase
diagram. When one is far from $\chi_c$ we have
seen\cite{Zhang2freq} that three--wave resonant coupling
influences pattern selection by enhancing dissipation via the
coupling to a heavily damped (slaved) mode. In this case, resonant
triads are strongly {\em suppressed}. On the other hand, when in
the near vicinity of $\chi_c$, resonant coupling to nearly
unstable linear modes can occur. Moreover, if these modes undergo
a first order bifurcation, their growth will be (to first order)
unchecked and resonant coupling to them may have a very
significant effect on the spatial-temporal behavior of the system.

Silber, Topaz and Skeldon \cite{SilberTopaz} have recently
demonstrated the importance of resonant coupling to slaved modes
for 6/7 forcing near $\chi_c$. Using the quasi-potential
approximation \cite{Zhang2freq} they showed how weakly damped
linear modes with wavenumbers $K < k_c$, quadratically couple to
the unstable modes to create the ``SL-I" states observed by
Kudrolli et al. \cite{Kudrolli}. In this case, the critical
wavevectors $|\vec k_i|=k_c$ could be constructed from a
commensurate hexagonal sublattice of wavevectors, $\vec K_i$, such
that $\vec{k_i}=q\vec{K_1}+p\vec{K_2}$. The ``SL-I" states are a
particular case where $(p,q)=(\pm 2,\pm 3)$ with $k_c/K=\sqrt{7}$.
This particular coupling was made possible by the existence of a
weakly damped linear ``tongue" with a wavenumber close to $K$. The
``SL-I" state was {\em not} observed for 2/3 forcing since, for
this forcing ratio, no linear tongues near this resonance exist
since there are no additional harmonic modes with $K<k_c$ ($K_i$
must be harmonic by \cite{SilberSkeldon} as they result from the
vector difference of two $\vec k_i$ modes).

Recent work by Tze, Rucklidge, Hoyle and Silber \cite{Tze} has
shown that the ``SL-II" states observed by Kudrolli et al.
\cite{Kudrolli} may be understood as resulting from a
symmetry--breaking bifurcation of an initial hexagonal symmetry.
Study of the possible invariant subgroups of the original $D_6$ +
$Z^2$ symmetry characterizing hexagonal standing waves revealed a
number of possible solution branches. One of these corresponds to
the spatial symmetry of temporally--averaged ``SL-II" type states.
Depending on the normal form coefficients, 5 additional possible
solution branches were predicted. It remains to be seen whether
these other branches are experimentally observed.

\section{Experimental System}\label{System}

Our experimental system consisted of a shallow fluid layer,
laterally bounded by a plastic sidewall and mounted on a
computer-controlled mechanical shaker.
 A 1 cm thick, black-anodized aluminum plate of 14.4 cm
diameter supported the fluid from below. This plate was machined
to a 10$\mu m$ flatness. The mechanical shaker used (either
Unholtz$-$Dickie model 5PM or VTS model 100) provided vertical
accelerations ranging from 0 to 15g.
 The cell acceleration, regulated to within 0.01g, was
monitored continuously by a calibrated accelerometer (Silicon
Designs, INC 1210L$-$010) attached directly to the armature of the
shaker. A feedback mechanism was used to control and stabilize the
amplitude $A$, mixing angle $\chi$, and phase $\phi$ to desired
values.

Most of our experiments were conducted with Dow Corning 200
silicone oil of different viscosities (DC200/10, DC200/20,
DC200/50 and DC200/100). Silicone oil has a typical density of
0.95 gr/cm$^3$ and surface tension of 21.5 dyne/cm. This fluid is
Newtonian for the viscosity range of 1--100 cS. Since the fluid
viscosity is highly temperature dependent, stabilizing the fluid
temperature was important. A stable fluid temperature of
$30\pm0.05 ^\circ C$ was used in all experiments. Resultant
viscosity variations were less than 0.04 cS.  A number of
experiments were also performed using TKO--77 vacuum pump fluid
with viscosities ranging between 221 cS at $33^\circ$ and 184 cS
At $30^\circ$. Both Dow Corning 200 and TKO-77 have very low vapor
pressures so there was no need to seal the cell against
evaporation. The results of our experiments showed no dependence
on the particular type of fluid used.

Our experiments were performed at frequencies between 20-150 Hz.
The selection of the frequency was influenced by the aspect ratio
of the patterns and the shaker's maximum acceleration and stroke.
Frequency selection was also influenced, to a lesser extent, by
limitations of the imaging and laser probe technique. Typically
the aspect ratio between the cell diameter, $L$, and the
wavelength $\lambda$, was between $5<L/\lambda<50$. The maximal
driving frequency of 150 Hz was governed by $a_c$, which increases
with increasing $\omega$.  The shaker's maximal stroke (2.5 cm
peak to peak) and boundary mode quantization at small aspect
ratios dictated the lower frequency limit.

\subsection{Boundary Conditions}

The lateral boundary conditions of the experimental cell can have
an important effect on the waves excited by the system. In our
experiments we attempted to minimize the role of the sidewalls. A
circular shape for the lateral boundary was chosen. This ensured
that no particular pattern was preferred. This is especially
significant when the system is only slightly dissipative (e.g low
viscosity fluids and/or large fluid depth). For a more highly
dissipative system (e.g high viscosity fluids and/or shallow
fluids), the boundary's shape does not influence the symmetry of
the excited pattern \cite{Edwards94}.
\begin{figure}[h]
\vspace{0cm} \hspace{0.0cm} \centerline{ \epsfxsize =7.5cm
\epsfysize =2.7cm \epsffile{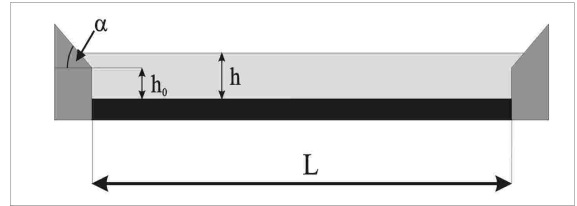}}
\caption{(top) A profile of the container used is shown: a Delrin
circular boundary (gray) is attached to the bottom plate (black).
The boundary consists of a vertical section of height $h_0$ and an
inclined section at an angle of $\alpha=20^\circ$ chosen to allow
the surface of the liquid used, Dow Corning 200, to be at zero
contact angle with the rim (liquid in light gray). Rings with
$h_0$ of 1 mm, 1.5 mm, 2.5 mm, 4 mm and 5 mm were used. $h$ could
be changed continuously by adding small amounts of fluid with a
calibrated pipette. } \label{Boundary}
\end{figure}
\begin{figure}[h]
\vspace{0cm} \hspace{0.0cm} \centerline{ \epsfxsize =7.5cm
\epsfysize =5.4cm \epsffile{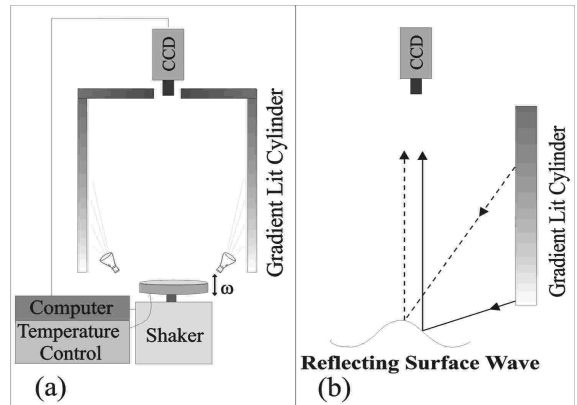}}
\caption{(a) A schematic diagram of the experimental system. The
imaging system consisted of a cylindrical screen, concentric with
the experimental cell, which was illuminated by halogen lamps
arranged in a circle. A CCD camera was mounted on the cylinder
axis above the fluid surface. (b) The cylinder's illumination
intensity was varied as a function of the height above the fluid.
At each point on the fluid surface the local slope reflects only a
single point from the cylinder surface into the CCD. Since the
lighting provides a unique intensity at each height along the
cylinder, the intensity reflected by each point is uniquely mapped
to the projection of the fluid surface's slope in the direction of
the cylinder axis. } \label{exp_imaging}
\end{figure}
As discussed by Douady (1990)\cite{Douday}, an additional effect
of sidewalls is the possible emission of waves (meniscus waves)
from the lateral boundaries. These waves are forced at the driving
frequency via forced height variations of the meniscus formed at
the contact line between the fluid and lateral boundaries.
Meniscus waves have no threshold and can therefore mask the
instability threshold of parametrically forced waves. To minimize
this effect, our system's lateral boundaries (shown schematically
in Fig \ref{Boundary}) were sloped at an angle conjugate to the
contact angle between the fluid and the material (Delrin) from
which the lateral rings were constructed. (A slope of
$\alpha=20^\circ$  was used for the Dow Corning 200 silicone oil.)
In this way, we ensured that the static fluid interface was nearly
flat.

An additional advantage of sloped lateral boundaries is the
elimination of reflected waves by impedance matching. Since the
instability threshold increases with decreasing $h$, a gradual
decrease (sloping sides) in fluid depth increases the effective
local threshold at the larger radii to far beyond $a_c$. Since the
typical height of the fluid layer in the sloped region was only
$0.1-0.8$mm, parametric waves could not be excited and any
meniscus waves emanating from the wall were strongly damped. In
practice, this boundary condition combined with the fluid
viscosities and depths used enabled us to obtain values of $a_c$
within 2\% of the calculated values
\cite{KumarTuckerman,Besson,Oleg2} for a system of infinite
lateral extent.

\subsection{Visualization}

\subsubsection{Imaging from above}\label{ImageAbove}
 To visualize the fluid surface, we employed a novel
imaging technique. The imaging system is schematically shown in
\Fig{exp_imaging}b. The experimental cell was illuminated by a
tall cylindrical screen whose axis was concentric with center of
the cell. The screen was illuminated from below by a ring of 12
small lamps. As a result, the light intensity along the screen
varied as a function of the height above the fluid. A CCD camera
was mounted on the cylinder axis, 1.4m above the cell. At each
point on the fluid surface (see \Fig{exp_imaging}b), the local
slope reflects only a single point from the screen onto the CCD.
Since the lighting provides a unique intensity at each height
along the cylinder, the intensity reflected by each point on the
fluid surface is uniquely mapped to the projection of its slope on
the cylinder axis. We used the CCD's high speed shuttering mode
($1/1000$ sec) to obtain instantaneous images of the fluid
surface.

Two methods of triggering were used to control the CCD camera. The
first method employed a trigger signal that was synchronized with
the driving. This signal both reset the camera and initiated
acquisition of the video frame at a desired phase relative to the
driving signal. To observe slow changes in the patterns over long
times, slow trigger rates that were commensurate with the driving
frequency were used. The short-term behavior of a state in its
different temporal phases was studied by the use of slightly
incommensurate trigger rates. This allowed nearly continuous
acquisition of the different temporal phases of a given state
without the need for very high-speed acquisition.

\begin{figure}[t]
\vspace{0cm} \hspace{0.0cm} \centerline{ \epsfxsize =7.5cm
\epsfysize =4.7cm \epsffile{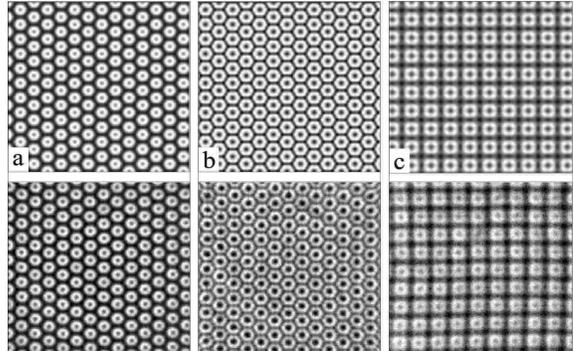}}
\caption{Using the method described in the text we calculated
model images (upper) of a simple hexagonal state in its peak state
(a) and crater state (b) and of a square state (c). The
corresponding experimental patterns are shown (bottom). To model
the surface waves we assumed an asymmetry between up hexagons and
down hexagons due to the the fact that the hexagon's amplitude was
large compared to the small layer height.
 } \label{HX-SQ}
\end{figure}

Our imaging technique, although providing quantitative
information, does not directly yield the surface wave height
function $h(x,y)$. The imaging yields a grayscale image, $I(x,y)$,
that is approximately the absolute value of the gradient of the
height function i.e.
\begin{equation}\label{imaging}
I(x,y)=\sqrt{(\partial_{x}h(x,y))^2+(\partial_{y}h(x,y))^2}
\end{equation}
One must then work backwards from $I(x,y)$ to determine the
function $h(x,y)$. This is done by inputting an assumed state into
Eq. \ref{imaging} and comparing the computed pattern to the state
observed. By iteration it is possible to arrive at fairly good
estimates of $h(x,y)$. Some examples are presented in \Fig{HX-SQ}.

For high-amplitude states, visualization from the side was
sometimes used. This was performed by illuminating from the side
and placing a video camera in the horizontal plane of the plate at
the height of the system's lateral boundary. This configuration
enabled direct quantitative measurements of the wave amplitudes
adjacent to the lateral boundaries of the cell. An additional
advantage of this imaging was that it allowed us to simultaneously
view both the lower plate's vertical movement together with the
wave's motion. In this way their relative phase could be directly
measured.

\subsection{High resolution temporal measurements}\label{System-PSD}
Most previous studies of the Faraday system have used imaging
techniques that mainly yield information about the symmetries of
the states observed. Time domain information was generally
extracted by the use of stroboscopic lighting at the frequency of
the shaker. This technique measures the lowest frequency in which
a state oscillates but yields no information about higher
harmonics. Two experimental studies used a laser beam probe to
extract information about the surface waves. Douady \cite{Douday}
used a laser beam deflected at an angle from the surface waves and
reflected on a screen to study the amplitude of the waves.  Gollub
and Simonelli \cite{SimonelliJFM} used a laser beam deflected from
the surface waves onto a photo detector to obtain temporal
information. This technique, however, could not give precise
information about the slope of the fluid surface since an average
intensity reflected from a single ``cell'' of the pattern was
measured.

We combined these two methods by imaging the reflection at the
fluid surface of a highly focused laser onto a position sensitive
detector (UDT SL20). This method yielded an accuracy of $1-5\%$ in
the surface slope and a temporal resolution of 0.02 mS. Good
signal to noise was obtained by amplitude-modulating the laser
signal and then deconvolving the resultant signal of the position
sensitive detector (PSD). The temporal response of the system was
only limited by the maximal sampling frequency of the PSD voltage.

\section{Overview of the phase diagram}
\label{Results1}

The phase space of the two--frequency Faraday system is very
large. Besides physical parameters such as fluid layer height and
viscosity, one also has to set the driving parameters. Introducing
two driving components with different frequencies, different
amplitudes and a nontrivial relative phase makes the task of
choosing a working regime and the relevant dimensionless
combinations of parameters a difficult one.

We have chosen to focus on the simplest commensurate driving
ratios as a first step. The $m/n$ ratio combinations used in our
experiments were numerous: 1/2, 2/3, 2/3, 2/5, 2/7, 3/4, 3/5, 3/7,
4/5, 4/7, 5/6, 5/7, 5/8, 6/7, 6/11, 41/60, 40/59, 21/50. Most of
our detailed experiments were performed with no phase difference
between the two frequency components ($\phi=0$ in
\Eq{2freqDriving1}). For system parameters that were seen to
excite special patterns, scans of $\phi$ were made. Changing
$\phi$ was found to be crucial for the existence of some of the
states and of no relevance to others. Phase diagrames were
constructed by fixing the mixing angle $\chi$ and increasing the
amplitude $A$ until a state of droplet ejection was reached.

\begin{figure}
\vspace{-.5cm} \hspace{0.0cm} \centerline{ \epsfxsize =7.5cm
\epsfysize =7.0cm \epsffile{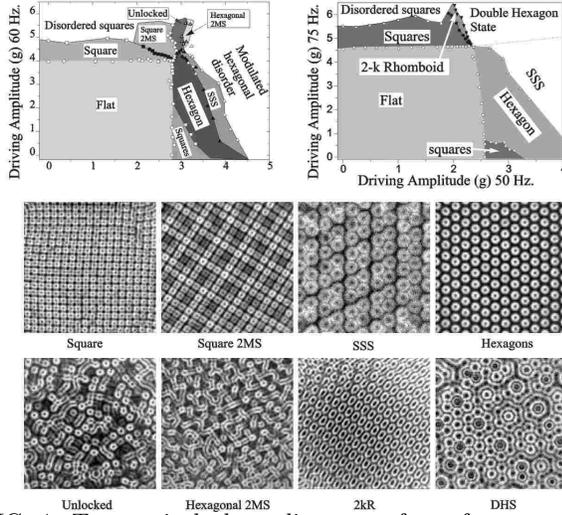}}
\caption{Two typical phase diagrams of two-frequency experiments
obtained for 2:3 driving with different system parameters:
$\omega_0=20$ Hz and $h=0.155$ cm (left) and  $\omega_0=25$ Hz and
$h=0.2$ cm (right). In both experiments $\nu=23$ cS and
$\phi=0^\circ$. Square regions exist in the near vicinity of
single frequency forcing. The square symmetry dominates in the
subharmonic regime to near $\chi_c$ while throughout most of the
harmonic region the hexagonal symmetry dominates. In the vicinity
of the co-dimension 2 point we observe three new states that exist
for many combinations of the driving ratio. These states are 2
mode superlattices (2MS), with underlying square or hexagonal
symmetries and spatially and temporally unlocked states
(``Unlocked" states). A resonant state that consists of a rhomboid
unit cell ($2kR$) was also observed (right). Unlike the 2MS and
unlocked states, which appear for many different driving ratios
(odd/odd, odd/even and even/odd), this state was observed for only
for 2/3 and 4/5 driving. In the harmonic region of phase space
where hexagons are initially dominant, a second bifurcation occurs
to either temporally subharmonic states (Subharmonic Superlattice
State, SSS) or high amplitude waves (as well as, at times,
localized ``oscillon" waves) that appear on a double hexagonal
superlattice (DHS). Symbols in the phase diagram describe measured
transitions for fixed $\chi$. (bottom) Typical photographs of
these states. } \label{PhaseSpaceBoth}
\end{figure}

Two typical two--frequency phase diagrams for even/odd driving are
presented in \Fig{PhaseSpaceBoth}. In single frequency experiments
rolls, squares, hexagons and quasipatterns of different symmetries
are known to exist depending on the viscosity, height, frequency
and amplitude above the threshold. In our parameter regime, the
dominant structure in regimes dominated by a single frequency is
squares for low viscosities ($\nu=8.7-23$ cS) and both squares and
rolls for higher viscosities ($\nu=47-87$ cS). As in other studies
of two--frequency driving \cite{Muller2freq,Edwards92,Edwards93},
two main regions, dominated by either $k_1$ or $k_2$, exist. Each
of these regions has not only a different wavelength but also
different temporal behavior. The $k_1$ and $k_2$ dominated regions
have a strong response at $\omega_1/2$ and $\omega_2/2$
respectively. This is true for all values of $m/n$. The temporal
response, however, consists of additional frequency components
that depend on the ratio $m/n$. When both $m$ and $n$ are odd
(odd/odd driving) the excited surface modes in both regions of
phase space have only subharmonic components (i.e.
$(p+1/2)\omega_0$ with $p$ an integer). In the case of odd/even
(even/odd) driving, the $k_1$ dominated region is temporally
subharmonic (harmonic) while the $k_2$ dominated regime is
temporally harmonic (subharmonic). At the critical value
$\chi=\chi_c$, a codimension 2 point exists where both wavenumbers
are simultaneously linearly unstable. The $k_1$ dominated region
occurs for $\chi<\chi_c$ while the $k_2$ dominated region occurs
for $\chi>\chi_c$.
 The interaction of $k_1$ and
$k_2$ leads to a variety of different nonlinear states in the
vicinity of $\chi_c$. Before describing these states, we will
first describe the effects of two--frequency driving in the two
main regions {\em far} from $\chi_c$.

In the regions dominated by harmonic states we have found a number
of non--trivial states that bifurcate from single--mode hexagonal
states while breaking both their spatial and temporal symmetries.
These symmetry--breaking bifurcations can even occur when $\chi$
is quite small. The subharmonic superlattice state (SSS)
(\Fig{PhaseSpaceBoth} (bottom)) is an example of such a symmetry
breaking state.  SSS states are formed in the $k_1$ dominated
regime with $\chi_c-5^\circ \geq \chi \geq 10^\circ$, when the
primary hexagonal state's symmetry with {\em harmonic} temporal
behavior is broken by an additional set of wavevectors of
magnitude $q<k_1$ with {\em subharmonic} temporal behavior
(frequency $\omega_1/4$) with respect to the primary $\omega_1/2$
frequency. We have observed two main types of SSS states, which
differ from each other in the orientation, magnitude and number of
$\vec{q}$ wavevectors. For example, the SSS state shown in
\Fig{PhaseSpaceBoth}, (SSS type I)  breaks the initial hexagonal
symmetry by the introduction of two wavevectors $\vec{q}$, which
are parallel to two of the three initial wavevectors $\vec{k_1}$.
The magnitude of $q$ in this case is $k_1/2$ which yields a simple
resonance condition $\vec{q}+\vec{q}=\vec{k_1}$. SSS have been
observed only for even/odd driving for {\em all} of the m/n
combinations listed above. The different SSS types and the
mechanisms that form them will be described in detail in Sec.
\ref{sectionSSS}

For even/odd driving, the effect of two frequency driving on the
pattern formation in the $k_2$ dominated region is quite different
than in the $k_1$ dominated region. In the $k_2$ region, square
symmetry dominates at threshold from $\chi=90^\circ$ to
$\chi\approx\chi_c$. Only in the vicinity of $\chi_c$ do we see
the effects of the two--frequency driving on the patterns formed.
It is interesting to note that although theory predicts that
hexagons are preferred for harmonic response and squares for
subharmonic response (see Sec. \ref{background}), we have observed
{\em square} symmetry in large parts of the harmonic region for
odd/even driving ($\chi > \chi_c$).

Let us now briefly describe the patterns formed for even/odd
driving in the vicinity of $\chi_c$. Starting with the $k_1$
dominated ($\chi<\chi_c$) region, two types of patterns are
observed near $\chi_c$. One pattern, which we call the ``Double
Hexagonal State" (DHS) is formed by two sets of hexagonally
arranged wavevectors (of length $k_1$) with a finite angle,
$\alpha$, between them. In the phase space shown in
\Fig{PhaseSpaceBoth},
 $\alpha \sim 22^\circ$. In
contrast to the SSS, this state does {\em not} break the temporal
symmetry of the harmonic hexagon state. Depending on various
system parameters, DHS are sometimes formed by a first--order
bifurcation. Perhaps their most outstanding characteristic is
their very high amplitude. The surface wave  maxima can reach
amplitudes much higher than the fluid layer's height. In
\cite{Arbell3} we have shown how the DHS states can form
oscillons, a highly localized large--amplitude nonlinear state
that has been observed \cite{Arbell3,OscillonsNature,mudoscillons}
in a number of periodically driven systems.

A special case of the DHS occurs for $\alpha=30^\circ$, whence one
obtains 12--fold quasipatterns such as first observed in
\cite{Edwards92}. We have seen the formation of such patterns for
4/5 driving in the same region where the DHS with
$\alpha=22^\circ$ appears for 2/3 driving.

Let us now move to the vicinity of $\chi_c$ both on the border of
the $k_1$ region and within the $k_2$ region. Here two linearly
unstable eigenvectors with different magnitudes can be
concurrently excited. There are numerous possible configurations
in which such a system can organize itself. Four different classes
of mixed mode states were found to exist. These will be described
in detail in Section \ref{Results2}.

\begin{itemize}
\item {\em Two mode superlattices (2MS)} These states formed by the
interaction of the dominant mode (e.g. $k_2$) with its original
symmetry and the weaker mode (e.g. $k_1$), which breaks the
symmetry of the original pattern. The symmetry of the dominant
mode can be either square or hexagonal depending on the proximity
of the nearest primary state in phase space. In
\Fig{PhaseSpaceBoth} we present two types of 2MS modes, a square
2MS obtained for $\chi>\chi_c$ and a hexagonal 2MS obtained for
$\chi<\chi_c$. The temporal behavior of the 2MS contains both the
$\omega_1/2$ and $\omega_2/2$ frequency responses and always
includes the subharmonic frequency of $\omega_0/2$. These states
appear for all types of driving (odd/even, even/odd and odd/odd)
although the precise structure of phase space depends on the
driving ratio used.
\item {\em Unlocked States} Between the square and hexagonal 2MS states an
intermediate region exists where both $k_1$ and $k_2$ appear but
no well--defined symmetry or spatial mode locking is observed.
Thus, no long range correlations in either space or time exist.
The basic time scale of the surface waves is $T=4\pi/\omega_0$ but
the pattern changes its local structure over time scales of order
$(10^2-10^3)T$.
\item {\em Rhomboidal states} Changing $h$, $\nu$ or $\omega_0$
 can lead to qualitative changes in the phase space.
\Fig{PhaseSpaceBoth} shows two different phase diagrams obtained
for 2/3 driving. The only difference between the two diagrams is
the fluid layer height and the value of basic frequency
$\omega_0$. Lioubashevski et al. \cite{Oleg1} describe how the
dimensionless number $\delta/h$, defined by the ratio between the
effective boundary layer depth, $\delta = \sqrt{\nu/\omega_0}$,
and the fluid height, $h$, affects single mode states selected by
the system. For a certain range of $\delta /h$ rhomboidal patterns
replace the 2MS and unlocked states (see \Fig{PhaseSpaceBoth}
bottom). These states couple two wavevectors of length $k_2$ with
one wavevector of length $k_1$. These wavevectors evolve {\em
spontaneously} from two circles of linearly degenerate states. We
will show that for special parameter values, n--fold quasipatterns
can naturally evolve from the rhomboidal structures.
\end{itemize}

In addition to the states described above, in Section
\ref{Results4} we will show examples of a number of other resonant
structures that are formed for different system parameters. The
richness of this system allows one to observe a wide variety of
different resonant patterns. We will show some common traits of
these resonant selection mechanisms that can lead to a more
comprehensive understanding of resonant interactions in pattern
forming systems.

\section{Subharmonic Superlattice States (SSS)}\label{sectionSSS}

 \begin{figure}
\vspace{0cm} \hspace{0.0cm} \centerline{ \epsfxsize =7.5cm
\epsfysize =4.2cm \epsffile{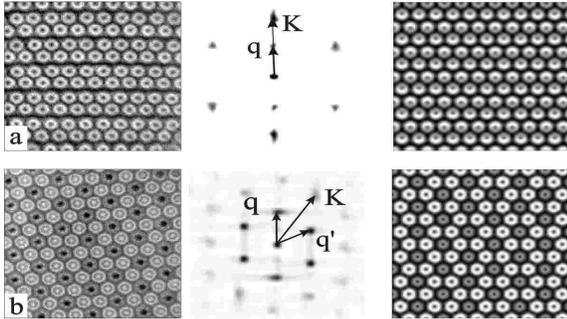}}
\caption{For 2/3 driving we observe two types of temporally
subharmonic superlattices, the SSS--I (a) SSS--II (b).
Experimental images (left) together with their spatial power
spectra (middle) and simulated images (right) are shown. In (a) we
see that a small vector of magnitude $q=K/2$ ($\vec q||\vec K$)
breaks the hexagonal symmetry formed by primary wavenumbers $\vec
K_i$, where $K_i=k_c$. In (b) the primary hexagonal symmetry is
broken by wavevectors $q=K/\sqrt{3}$ located at a $30^\circ$ angle
relative to the primary wavevectors. Both patterns can exist in
regions far from $\chi_c$, where the higher odd--frequency
component is weak (see \protect\Fig{PhaseSpaceBoth}). The patterns
obey the resonance conditions $2\vec{q}$=$\vec{k_c}$ (SSS--I) and
$\vec{q}+\vec{q'}=\vec{K}$ (SSS--II).} \label{SSS_intro}
\end{figure}
Hexagonal patterns can be formed in two--frequency experiments
when the basic subharmonic temporal invariance is broken and
quadratic nonlinear terms appear in the amplitude equations
describing the system. These terms can occur in regions of the
two--frequency phase space that have a harmonic response. By
increasing the amplitude of the driving, the primary hexagonal
symmetry can be broken and new {\em stable} structures appear even
in regions where the second externally driven mode cannot be
excited at all. In this section we will describe the
characteristics of these symmetry breaking patterns.

In the notation used in this section, $\vec{K}$ is the linearly
unstable wavevector, excited $\omega_1=m\omega_0$, that
characterizes the primary pattern. $\vec{q}$  is an additional
smaller wavevector that appears in states that bifurcate from the
primary pattern. Since in each pattern there is a degeneracy in
the direction of the wavevectors, we will use an index to number
the different wavevectors of the same magnitude e.g. $\vec{K_i}$
$i=1,2\ldots$.

Sub-harmonic superlattice states occur over a wide range of $\chi$
in the two-frequency phase diagram where the lower frequency,
$\omega_1$, is dominant. SSS were observed for most even/odd
frequency ratios tested, but were {\em not} seen for odd/even or
odd/odd ratios. Since the first SSS type state was found
\cite{Arbell1} additional variants have been observed for
different experimental parameters. Both types of SSS pattern,
SSS-I (\Fig{SSS_intro}a) and SSS-II (\Fig{SSS_intro}b) are shown
with their corresponding spatial power spectra. Both states share
the following characteristics:
\begin{itemize}
\item The basic temporal dependence of both SSS types is subharmonic with respect to the
the primary instability i.e. the system has a strong response at
$m\omega_0/4$.
\item These states are secondary bifurcations of temporally harmonic hexagonal states.
\item Both states occur in the same basic region of phase space (for $0< \chi < \chi_c$).
\item In both states $k_2$ wavevectors are {\em not} observed.
\end{itemize}

The different states, in general, occur for different fluid
parameters such as fluid viscosity, fluid depth and
$\omega_1/\omega_2$ ratios. The most obvious differences can be
seen in \Fig{SSS_intro}. Examining their spatial power spectra, we
see that while both SSS-I and SSS-II's spectra are constructed by
three evenly spaced wavevector pairs, $\vec{K_i}$. The spectra
include additional sets of wavevectors, $\vec{q_i}$ (where
$|q_i|<K$), of smaller magnitude. The orientations and magnitudes
of $\vec{q_i}$ {\em differ} in SSS-I and SSS-II states.

In SSS-I $\vec{q_i}$
 are oriented solely along (some or all of the) axes
defined by the $\vec{K_i}$. The lengths of the $q_i$ in the SSS-I
are usually $K/2$, but in some cases the $q_i \neq K/2$.

SSS-II states are formed by a set of $\vec q_i$ which {\em always}
consists of three evenly spaced wavevector pairs of magnitude
$q=K/\sqrt{3}$ arranged at an angle of $30^\circ$ relative to the
direction of the $\vec K_i$ triad. The magnitude and orientation
of $q_i$ yield the simple resonance condition
$\vec{q_1}+\vec{q_2}=\vec{K_i}$. While the SSS-I have been
observed for a wide range of driving ratios (2/3, 4/7, 2/5, 2/7,
4/5), the SSS-II have only been seen for 2/3 and 4/5 driving. Both
types of SSS states have recently been identified as
representations of different invariant subgroups when hexagonal
symmetry is broken \cite{SilberRucklidge,RSF}.

In the next subsections we will present a detailed description of
each of the SSS types together with a mechanism that can explain
their formation.

\subsection{Subharmonic Superlattice type I (SSS--I)}\label{SSS-I}

The spatio--temporal behavior of the SSS-I state can be modelled
by a simple equation for the surface height function:
\begin{eqnarray}\label{eq:SSS-I}
\nonumber h(r,t)=
\cos(\frac{1}{2}m\omega_0t)\sum_{i=1}^{3}A_i\cos(\vec{K_i}\cdot\vec{r}+\alpha_i)+
\\
\cos(\frac{1}{4}m\omega_0t+\gamma)\sum_{i=1}^{M}B_i\cos(\vec{q_i}\cdot\vec{r}+\beta_i)
\end{eqnarray}
with
\begin{equation}
\vec{q_i}\|\vec{K_i},\text{  } \beta_i=0,90^\circ \nonumber
\end{equation}
where $M$ is the number of axes with broken symmetry, $\gamma$ is
the {\em temporal} phase difference between the two sets of modes
and $\alpha_i$ and $\beta_i$ are the respective {\em spatial}
phase of the $\vec{K_i}$ and $\vec{q_i}$ components. Because the
pattern has hexagonal symmetry we assume $\alpha_i=0^\circ$ and
not $\alpha_i=2\pi/3$ as in patterns having triangular symmetry
\cite{Muller2freq}. Eq. \ref{eq:SSS-I} summarizes the most
important features of the SSS-I in a compact way. We will now
present experimental evidence for the validity of this equation
and describe the relevance of each term ($A_i$, $B_i$, $\gamma$,
$\beta_i$ and $M$) in the experimentally observed states.
\begin{figure}
\vspace{0cm} \hspace{0.0cm} \centerline{ \epsfxsize =7.5cm
\epsfysize =3.0cm \epsffile{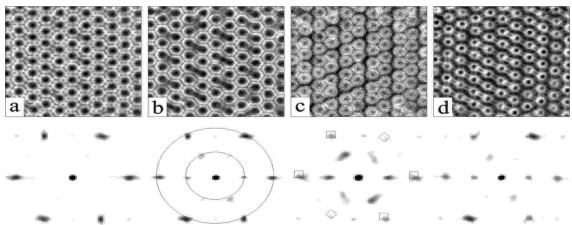}}
\caption{Typical temporal sequences of an SSS--I state (a) taken
at constant values of the driving parameters for the frequency
ratio 40/60 Hz. The spatial Fourier spectra are composed solely of
wavenumbers $K$, corresponding to $\omega_1/2$, and $q=K/2$. The
{\em locations} of the peaks are fixed in the different temporal
phases. For the system parameters $\omega_0/(2\pi)=20$ Hz,
$\nu=23$ cS, $\phi=0^\circ$, and $h=0.155$ cm the hexagonal
symmetry is broken in {\em two} directions by the $K/2$ vectors.
Circles of radii $K$ and $K/2$ are drawn in (b). The relative
intensities of the different wavevectors can be seen; in (a)
 $\vec q_i$ are nearly absent while in (c) their intensities are almost
equal to the $\vec K_i$. The symmetry breaking is also revealed in
the intensities of the primary hexagonal vectors, as can be seen
in (c), where the two strong intensity wavevectors are enclosed in
a square and the weak one in diamonds.
 } \label{SSS_1_time}
\end{figure}

\begin{figure}
\vspace{-.5cm} \hspace{0.0cm} \centerline{ \epsfxsize =7.5cm
\epsfysize =5.5cm \epsffile{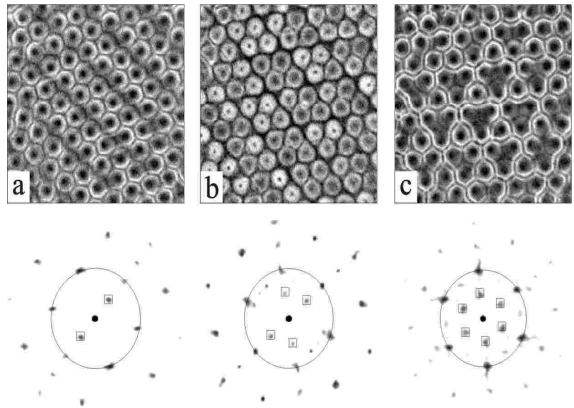}}
\caption{Images (top) and power spectra (bottom) of an SSS--I type
state with broken symmetry in one, (a) two, (b) and three (c)
directions. The circles indicate the primary hexagon wavevector's
magnitude, $K$. The symmetry breaking vectors of magnitude $q=K/2$
are enclosed in squares. All images were obtained for the same
system parameters of $\nu=23$ cS ($40/70$ Hz and $h=0.155$ cm,
$\chi=43^\circ$ and $\phi=0^\circ$). This pattern is not stable
and drifts slowly (order of 10--60 seconds) between these three
states.
} \label{1d2d3d_SSS}
\end{figure}

In \Fig{SSS_1_time} we present a sequence of SSS--I states taken
at different times for constant values of the driving parameters.
Although the state's appearance changes with time, their spatial
Fourier spectra reveal that the state results from the interaction
of two specific spatial scales; the primary wavenumber
$|\vec{K}|=|\vec{k_1}|$ that is excited by the $\omega_1$
frequency component, and its spatial sub-harmonic, $\vec q=\vec
K/2$. The SSS spectra show that while the $\vec K_i$ have 6--fold
symmetry, two $\vec q_i$ with relatively large amplitude and a
third smaller amplitude $\vec{q}_i$ have broken this symmetry. The
amplitudes of the $q_i$ vary with time. Within the temporal phases
shown in \Fig{SSS_1_time}a the $q_i$ amplitudes have little power
while in \Fig{SSS_1_time}c the $\vec{q}_i$ are stronger than the
$\vec{K}_i$ components. This behavior is reflected by the temporal
phase, $\gamma$, in Eq. \ref{eq:SSS-I}. The symmetry breaking is
reflected by the relative strengths of both the $\vec{q}_i$ and
$\vec{K}_i$. It is clear that the two $K$ wavevectors enclosed in
squares, have different strength than the wavevector enclosed in
diamonds. This symmetry--breaking is also seen in the relative
power of the $q_i$ wavevectors.

As demonstrated in \Fig{1d2d3d_SSS}, SSS states can have broken
symmetry in one, two or three directions ($M=1,2,3$ in Eq.
\ref{eq:SSS-I}). In most cases, a specific number of symmetry
breaking directions was selected. However, for the experiment
described in \Fig{1d2d3d_SSS}, the pattern drifted slowly (on the
scale of seconds) between the three possible symmetry--broken
states.

\begin{figure}
\vspace{0.cm} \hspace{0.0cm} \centerline{ \epsfxsize =7.5cm
\epsfysize =7cm \epsffile{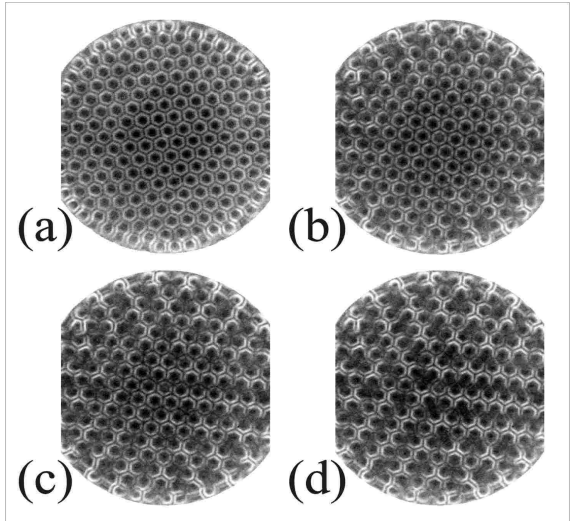}}
\caption{The transition between Hexagonal and the SSS state is
displayed for a typical experiment (40/60). This is a gradual
process in which the basic symmetry of a perfect hexagonal pattern
(a) is broken first at the circular boundary of the cell. As the
amplitude of the external forcing increases, the area of broken
symmetry grows inwards (b,c) until  the SSS state fills the entire
cell (d). This process can also occur in the reverse direction.}
\label{Trans-HX-SSS1}
\end{figure}

The transition between the primary hexagonal pattern and the SSS
states is (within $\approx 1\%$) {\em non}-hysteretic and occurs
via a circular front that propagates slowly inward from the
plate's lateral boundaries. The process is continuous and
reversible (see \Fig{Trans-HX-SSS1}). Before continuing the study
of the spatial characteristics of the SSS-I pattern let us digress
briefly and examine the temporal behavior of the system using the
laser probe method (see Sec. \ref{System-PSD}).

\begin{figure}
\vspace{-.0cm} \hspace{0.0cm} \centerline{ \epsfxsize =6.5cm
\epsfysize =11.5cm \epsffile{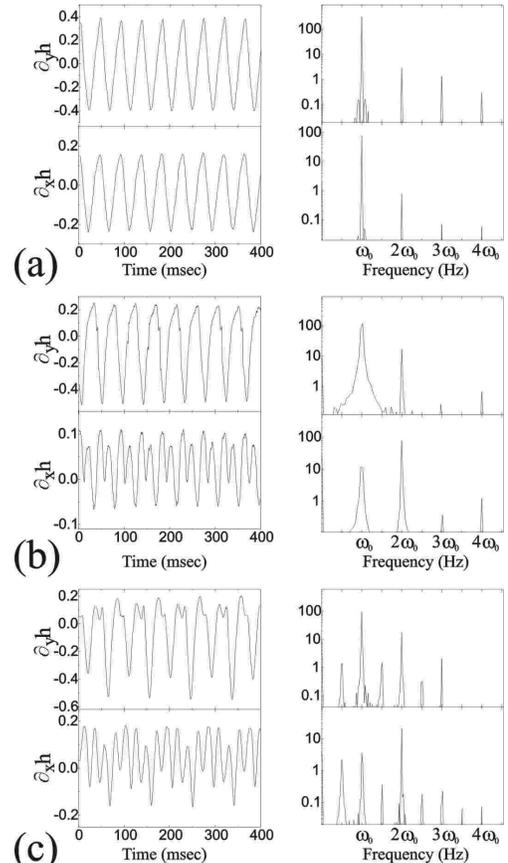}}
\caption{The time dependence of the SSS--I state as studied by the
reflection of a laser by the surface waves. In this experiment,
for system parameters $\omega_0/(2\pi)=22$ Hz, $\nu=23$ cS,
$h=0.2$ cm, $\phi=0^\circ$ and $\chi=36.3^\circ$, increasing the
driving amplitude from 2.5 g (a) to 2.9 g (b) and 3.7 g (c)
results in (a) Low amplitude hexagons (b) developed hexagons and
(c) SSS-I states. In (a) both $\partial_{x}h$ and $\partial_{y}h$
have the same peaks in their power spectrum. Increased driving
amplitude yields a different temporal response in the two
directions $x$ and $y$. This phenomenon also occurs in single
frequency experiments where squares are dominant and may be a
general feature of the Faraday instability in viscous fluids. In
the SSS--I state (c) a subharmonic temporal response at
$\omega_0/2$ occurs.
} \label{laserSSS}
\end{figure}

Typical time series of the $x$ and $y$ components of the fluid
surface gradient at a single point are presented in \Fig{laserSSS}
for three different accelerations. These describe the temporal
behavior of hexagons at threshold (a), developed hexagons (b) and
SSS-I (c). At threshold, the response is harmonic with respect to
the total period of $2 \pi / \omega_0$. The response is strongest
at the frequency of $\omega_1/2$. Increasing the driving amplitude
results in a bifurcation to a state with a strong {\em
superharmonic} response at the frequency of $\omega_1$. This
phenomenon also occurs for single frequency experiments.

A further increase of the driving amplitude yields a second
bifurcation. In this bifurcation the temporal response becomes
{\em subharmonic} with respect to the period of $2\pi /\omega_0$.
As can be seen in \Fig{laserSSS} the superharmonic component does
not disappear and can be quite strong. It is important to note
that $2/n$ driving has some special relations between the various
frequencies that are not present for higher order driving (such as
4/5,4/7...6/7...). For $2/n$ driving $\omega_1/2=\omega_0$. The
temporal response in the harmonic region has the same frequency as
the common frequency $\omega_0$. In other driving ratios, such as
4/5 driving, the subharmonic is $4\omega_0/2=2\omega_0$ where the
common frequency is $\omega_0$. It is possible that such a
relation can enhance certain resonant mechanisms and help
stabilize certain patterns such as the $2kR$ and oscillon states
\cite{Arbell3,Arbell2}.

Wagner et al. \cite{WagnerNew} studied a two mode system generated
by single--frequency excitation at a bicritical point where both
harmonic and subharmonic tongues become unstable. They describe
two different superlattice states that exist in the transition
region between  subharmonic squares to harmonic hexagonal states.
In the first superlattice state, the square symmetry is broken by
a small wavevector $\vec k_D$ that is equal to
$\vec{k_H}-\vec{k_S}$, where $\vec{k_H}$, $\vec{k_S}$ are,
respectively, the primary hexagonal and square wavevectors. The
relative phase of the symmetry breaking $k_D$ mode compared to the
primary square $k_S$ mode can either be $0^\circ$ or $90^\circ$
according to the sign of the nonlinear coefficient in the
amplitude equation for the symmetry breaking mode. The
experimental observation shows that the phase selected for the
first transition state is $90^\circ$. A second superlattice state
\cite{WagnerNew} was observed that is similar to the SSS--II
states found in our experiments. The spatial phase difference,
$\beta$, between the harmonic hexagonal mode and the subharmonic
symmetry--breaking mode was found to be $0^\circ$ although from
amplitude equation considerations $90^\circ$ is also a possible
solution.

\begin{figure}
\vspace{0.0cm} \hspace{0.0cm} \centerline{ \epsfxsize =7.5cm
\epsfysize =3.5cm \epsffile{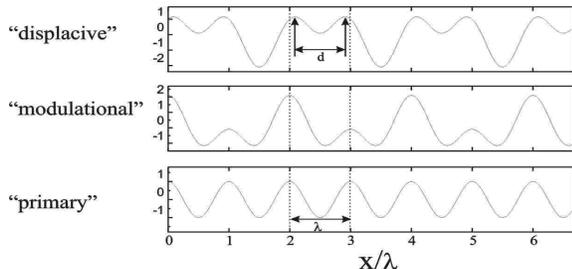}}
\caption{A simple one--dimensional superposition of a harmonic
function (bottom) $\cos(2\pi x/\lambda)$ with its subharmonic can
have two basic combinations. The ``displacive'' mode shown results
from the superposition of $\cos(2\pi x/\lambda)+b\cos(\pi
x/\lambda+\pi/2)$ (top) ($b=2.2$ is arbitrarily chosen), whereas a
``modulational'' mode results from the superposition of $\cos(2\pi
x/\lambda)+b\cos(\pi x/\lambda)$ (middle). It can be seen that in
the ``displacive'' mode the distance between local maxima is
either $d$ or $2\lambda-d$ with $\lambda-d \ll \lambda$, while in
the ``modulational'' mode the distance between the local maxima
remains the same as in the primary mode, $\lambda$. }
\label{Muller}
\end{figure}

\begin{figure}
\vspace{0cm} \hspace{0.0cm} \centerline{ \epsfxsize =7.5cm
\epsfysize =7.0cm \epsffile{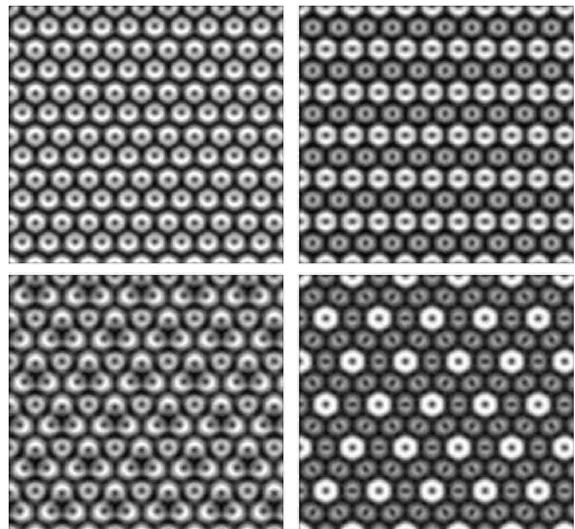}}
\caption{``Displacive'' (left) and ``modulational'' patterns
(right) are shown for both symmetry-breaking in 1 (top) and 3
directions (bottom). The pattern shown is simulated using our
imaging model applied to \protect\Eq{eq:SSS-I} with M=1 (M=3) for
1 direction (3 directions) and with $\beta_i=\pi/2$ ( $\beta_i=0$)
for the ``displacive'' (``modulational'') patterns. 2:3 forcing
was used with amplitudes of all modes taken to be equal
($A_i=B_i$). All SSS-I patterns were found to be displacive in
character (compare to \protect\Fig{1d2d3d_SSS}).}
\label{Muller_images}
\end{figure}

We will now describe in detail the effects of this phase
difference on SSS-I states with broken symmetry. Let us first
consider the effect of adding a $K/2$ modulation in one dimension.
In \Fig{Muller} we can see that adding the $K/2$ mode with
$\beta_i=0^\circ$ results in a modulation of the original cosine
form. The waves' local maxima remain at the same spatial location
but their amplitudes are now modulated with a $4\pi/K$ periodicity
resulting in one large peak followed by a smaller peak. In
contrast to this, adding the $K/2$ mode with $\beta_i=90^\circ$
results in a different effect. While all of the maxima have the
{\em same} amplitude, their spatial {\em locations} are modulated
with a $4\pi/K$ periodicity. This effect \cite{WagnerNew} is
called ``displacive''. The spatial effect in images can be quite
pronounced. This is demonstrated in \Fig{Muller_images} where we
present a comparison between the ``displacive'' (left) and
``modulational'' (right) effects with spatially subharmonic
patterns and symmetry breaking along one direction (top) or three
directions (bottom). The effects of symmetry breaking in three
directions are more complicated but the qualitative effect
remains. We find that SSS-I states are {\em always} of the
displacive nature ($\beta=90^\circ$). This can be seen by
comparing the experimental images shown in \Fig{1d2d3d_SSS} with
the simulated displacive patterns shown in \Fig{Muller_images}.

 It is interesting to note that the ``modulational'' pattern shown in
\Fig{Muller_images} (bottom right) was seen in an experimental
study of the single--frequency Faraday system in a viscoelastic
fluid \cite{WagnerVisco}. When changing the driving frequency,
Wagner et al. found both a harmonic region for low driving
frequencies and a subharmonic response for higher ones. In the
vicinity of the transition frequency, Wagner observed a hexagonal
superlattice composed of both the subharmonic and harmonic
wavevectors. Since the subharmonic wavevector is exactly half of
the harmonic one, the simple resonance of
$\vec{k_S}+\vec{k_S}=\vec{k_H}$ is retained (where the S index
stands for subharmonic and the H for harmonic). Although the power
spectra of this superhexagon described by Wagner are similar to
the SSS-I power spectra (in the case of three direction symmetry
breaking), the real space patterns are different, indicating a
$0^\circ$ spatial phase difference between $k_S$ and $k_H$, in
contrast to the $\beta_i=90^\circ$ seen for SSS-I.

\begin{figure}
\vspace{0.0cm} \hspace{0.0cm} \centerline{ \epsfxsize =7.5cm
\epsfysize =7.5cm \epsffile{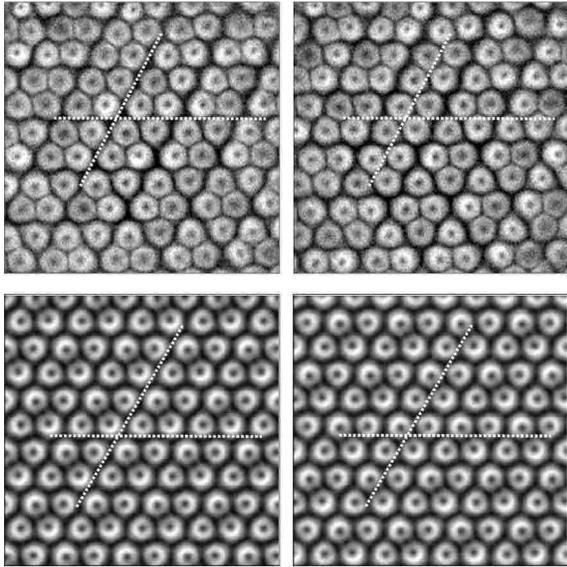}}
\caption{The ``jittering'' effect in 2d. Experimental (top) and
model (bottom) images of the SSS--I states at two temporal phases
separated by a $2\pi/\omega_0$ interval. The white crossed lines
are fixed in both frames. The two pictures demonstrate that the
time displacement is equivalent to a spatial displacement of
$2\pi/K$ in the symmetry-breaking directions.} \label{SSSjitter2d}
\end{figure}

The combination of the spatial displacive mechanism and the
subharmonic temporal dependence of the $K/2$ mode results in an
interesting ``jittering'' effect in time. If we consider Eq.
\ref{eq:SSS-I} we see that when $t \longrightarrow t+2\pi/\omega_0
$ the first part of l.h.s of Eq. \ref{eq:SSS-I} is invariant
whereas the second part changes sign due to the different time
dependence. In our model calculation the temporal displacement of
$2\pi/\omega_0$ is equivalent to a spatial displacement of
$2\pi/K$ in each of the symmetry-breaking directions. Images
photographed at time intervals of $2\pi/\omega_0$ appear to jitter
between these two displaced images, as demonstrated in
\Fig{SSSjitter2d}.

\subsection{Subharmonic Superlattice type II (SSS--II)}\label{SSS-II}

As mentioned above, a qualitatively different type of SSS type
pattern, SSS--II, has been observed. Increasing both viscosity and
height but keeping the dimensionless parameter $\delta/h$ constant
(by changing $\omega_0$), results in a different symmetry breaking
scenario. Though the SSS--II appears in the same region of phase
space as the SSS-I and shares its subharmonic temporal behavior,
it is qualitatively different from the SSS-I. The excited vectors
$\vec{q_i}$ are now aligned at an angle of $\pi/6$ relative to the
vectors $\vec{K_i}$ (in contrast to $\vec{q_i}||\vec{K_i}$ in type
SSS--I). The symmetry breaking wavevectors are of magnitude $|q_i|
= |K_i|/ \sqrt{3}$ and these states exhibit {\em no} spatial
symmetry breaking. For each $\vec{K_i}$ there is a corresponding
$\vec{q_i}$.

Our experiments suggest that SSS--II can be described by:

\begin{eqnarray}\label{eq:SSS-II}
\nonumber
h(r,t)=\cos(\frac{1}{2}m\omega_0t)\sum_{i=1}^{3}A_i\cos(\vec
K_i\cdot\vec{r}+\alpha_i)+\\
\cos(\frac{1}{4}m\omega_0t)\sum_{i=1}^{3}B_i\cos(\vec{q_i}\cdot\vec{r}+\beta_i)
\end{eqnarray}
$$\vec{K_1}=K(0,-1);\vec{K_2}=K(\frac{\sqrt{3}}{2},\frac{-1}{2});\vec{K_3}=K(\frac{-\sqrt{3}}{2},\frac{-1}{2});$$
 and
$$\vec{q_1}=q(-1,0);\vec{q_2}=q(\frac{1}{2},\frac{\sqrt{3}}{2});\vec{q_3}=q(\frac{1}{2},\frac{-\sqrt{3}}{2});q=\frac{K}{\sqrt{3}}$$

\begin{figure}
\vspace{0cm} \hspace{0.0cm} \centerline{ \epsfxsize =8.5cm
\epsfysize =2.0cm \epsffile{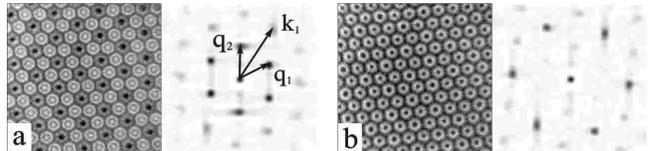}}
\caption{Images (a,b left) with corresponding power spectra (a,b
right) of different temporal phases of an SSS--II state observed
for 2/3 driving in the harmonic region of phase space at $\nu=47$
cS ($\omega_0/(2\pi)=25$ Hz and $h=0.2$ cm). In contrast to the
SSS--I pattern, where the primary hexagonal symmetry is broken by
wave vectors parallel to $\vec K_i$, SSS--II feature a different
symmetry breaking. A second hexagonal lattice of smaller magnitude
wave vectors forms, obeying the resonance condition
$\vec{q_1}+\vec{q_2}=\vec{K}$. At certain temporal phases the
effect can be easily seen (a) while at others only the hexagonal
symmetry is dominant (b).
} \label{SubSupHx_1_f2-3}
\end{figure}

Looking at this state at different temporal phases, the effect of
the $q_i$ can be easily seen at some phases
(\Fig{SubSupHx_1_f2-3}a) while, at others, only the hexagonal
symmetry is apparent (\Fig{SubSupHx_1_f2-3}c). This symmetry is a
representation of one possible invariant subgroup when hexagonal
symmetries are broken \cite{SilberRucklidge,RSF,Logvin}.

\begin{figure}
\vspace{0cm} \hspace{0.0cm} \centerline{ \epsfxsize =8.5cm
\epsfysize =1.8cm \epsffile{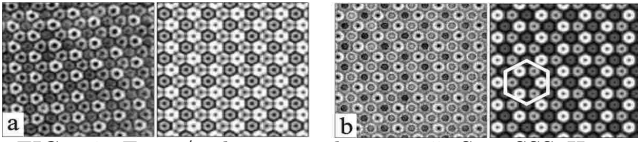}}
\caption{For 2/3 driving  and $\nu=0.47$cS, a SSS--II type pattern
with a different spatial dependence occurs. Two phases of this
state are shown in (a,b left) and a simulation of these pattern
are shown in (a,b right). These patterns are formed when the
spatial phases of the smaller wavevectors, ($\vec q_i$), are
shifted by $\pi/2$ relative to $\vec K_i$. These states were
observed for higher frequencies ($\omega_0>35$ Hz) than those for
which SSS--II having a zero phase shift are observed.}
\label{SubSupHx_sin}
\end{figure}

In the 1/2 driving experiments performed by Muller
\cite{Muller2freq} a transition between hexagonal and triangle
patterns was observed that corresponds to a change of
$\sum_i\alpha_i$ in \Eq{eq:SSS-II} from the value of $0$ to
$3\pi/2$ (where the second term of the R.H.S. of  \Eq{eq:SSS-II}
is zero).  A similar phenomenon can take place with SSS--II
states. As shown in \Fig{SubSupHx_sin}, a variant of the SSS-II
states occurs with a rotational symmetry of $2\pi/3$ in contrast
to the $2\pi/6$ that is typical for hexagon patterns. These SSS-II
variants are observed for 2/3 driving when relatively higher
frequencies ($\omega_0>40Hz$) are used. Using our imaging model,
we find that these patterns are formed by the same resonance as
the SSS--II but when spatial phase angles of the $\vec q_i$ modes
are $\beta_i=\pi/2$ for $i=1,2,3$ or $\sum_i\beta_i=3\pi/2$. These
SSS--II states have symmetries that are similar to those of the
time--averaged symmetries and wavenumbers of the ``SL2" states
observed in \cite{Kudrolli}. The instantaneous images of SL2
states, however, are more reminiscent of SSS--I states (as shown
in \Fig{SSS_1_time}). Patterns similar to this SSS-II variant have
recently been observed in a forced ferrofluid system \cite{Lee} in
the vicinity of a bicritical point where harmonic and subharmonic
solutions collide. Muller shows that when considering a harmonic
region amplitude equation with quadratic terms for a single
wavenumber model, only solutions where $\sum_i\beta_i=0$ are
possible. It appears that the second set of equations for the $q$
wavenumber modes, which are temporally {\em subharmonic}, does
allow the existence of modes with $\sum_i\beta_i=3\pi/2$
solutions.  The mechanism that selects the spatial phase in
multi--mode systems still must be clarified.

SSS--II type patterns have also been observed in the experimental
study of optical pattern formation in sodium vapor
\cite{Ackemann}. In the next section we will address the question
of a selection criterion between the SSS--I and SSS--II states.

\subsection{Selection between SSS--I and SSS--II}\label{Selection}

Both SSS-I and SSS-II break the temporal symmetry of the initial
hexagon state by temporal period-doubling to a basic frequency of
$m\omega_0/4$. The two states, however, differ in their spatial
behavior. For the case of SSS-I states, the excited wavevectors,
$\vec q_i$,  are  both parallel to the linearly unstable
wavevectors $\vec K_i$, and, in many cases $q=K/2$. In contrast,
the excited wavevectors in SSS-II states are rotated by $\pi/6$
relative to $\vec K_i$. What mechanism governs the selection of
both the two different states and the values of $q$ that are
excited?

 As Silber, Topaz and Skeldon have suggested \cite{SilberTopaz},
the symmetry-breaking wavevectors of harmonic patterns may
correspond to minima of linearly stable tongues that can be
excited via nonlinear coupling to the $\vec K_i$ modes. Since the
subharmonic frequency, $m\omega_0/4$, is excited by all of the SSS
states, the linearly stable tongue with a dominant $m\omega_0/4$
frequency would be a likely candidate to be selected. The
wavenumber, $q$, corresponding to these waves, can be
well-approximated by the linear dispersion relation
$q=k(m\omega_0/4)$. Note that $q$ is not constant for a given
value of $\omega_0$, but can be strongly dependent on the
parameters $\nu$ and $h$.

\begin{figure}[t]
\vspace{0.0cm} \hspace{0.0cm} \centerline{ \epsfxsize =8cm
\epsfysize =5.0cm \epsffile{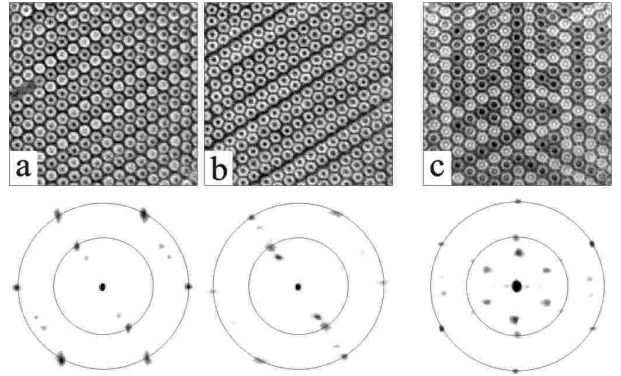}}
\caption{Images (top) and corresponding power spectra (bottom) of
SSS-I superlattices observed for 2/3 driving in which $q \neq
K/2$. Circles of a radii $K$ (outer) and $q$, corresponding to the
$m\omega_0/4$ frequency (inner) are drawn. The primary hexagonal
symmetry is broken either in one direction (a,b) or in three
directions (c). (a,b) show the same state at different temporal
phases. The relative intensities of $\vec{q}$ and
$\vec{K}-\vec{q}$ vary in the different phases. All experiments
were with $\nu=23$ cS. (a,b) for $h=0.2$ cm and (c) for $h=0.25$
cm. The basic frequency was $\omega_0/(2\pi)=35$ Hz for (a,b) and
$\omega_0/(2\pi)=30$ Hz for (c).  The value of $q/K$ is $\approx
0.6$ for all experiments.
} \label{SSS_Strange}
\end{figure}

Let us now examine the following premise. The system will
generically prefer to undergo spatial period doubling to $q=K/2$.
If, however, $\vec{q}(m\omega_0/4)$ is close to a wavevector
$\vec{q}$ with a magnitude that is substantially different than
$K/2$, one possible solution of the system is to lock to either
SSS--I or SSS--II patterns with $q\neq K/2$. If $q$ is near
$K/\sqrt{3}$, the system will lock to this value, thereby
fulfilling the spatial resonance condition:
$\vec{q_i}+\vec{q_j}=\vec{K_i}$.  SSS-II patterns will then occur.
Values of $q$ sufficiently far from either spatial resonance will
result in an SSS-I state with $q\neq K/2$. Such a state is
presented in \Fig{SSS_Strange} where symmetry breaking occurs in
either one direction (\Fig{SSS_Strange}a,b ) or three
(\Fig{SSS_Strange}c) but the symmetry-breaking wavevector's
magnitude was {\em not} $K/2$. Instead, vectors parallel to $\vec
K_i$, with magnitudes $q \neq K/2$ and $ K- q$ are observed, with,
empirically, $q\sim0.6k_c$.

\begin{figure}
\vspace{0.0cm} \hspace{0.0cm} \centerline{ \epsfxsize =7.5cm
\epsfysize =4.8cm \epsffile{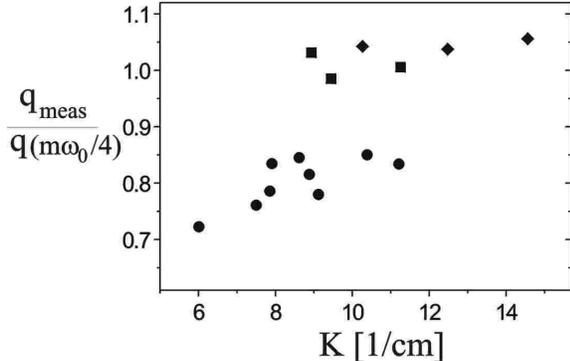}}
\caption{The measured wavenumber, $q_{meas}$, of the symmetry
breaking wavevectors in SSS--type patterns, normalized by the
linear wavenumber calculated for the frequency $m\omega_0/4$ as a
function of the measured critical wavenumber of the primary
hexagon pattern, $K$. The symbols correspond to different types of
SSS patterns; SSS-I patterns with $q=K/2$ (circles), SSS-I with $q
\neq K/2$ (diamonds) and SSS--II patterns with $q=K/\sqrt{3}$
(squares). The data indicate that the first type of pattern
($q=K/2$) is generally preferred by the system unless the value of
$q(m\omega_0/4)$ is either close to $q=K/\sqrt{3}$ or sufficiently
far from either of these preferred modes. } \label{PlotK-SSS}
\end{figure}

This premise is checked in \Fig{PlotK-SSS}, where we plot the
value of the ratio between the experimentally measured values of
$q$ and the wavenumber computed for single frequency driving using
$m\omega_0/4$ with $K$ for parameters where different SSS--type
patterns were observed. The plot shows that for both SSS--II and
the SSS-I with $q \neq K/2$ the approximation $q=k(m\omega_0/4)$
is correct to within $4\%$. In SSS--I states where $q=K/2$ is the
symmetry breaking mode, the ratio $q/k(m\omega_0/4)$ varies {\em
systematically} between $77\%-85\%$. This suggests that the
$\vec{q}=\vec{K}/2$ resonance is strong enough to induce this
``locking'' or detuning of $q$.

\begin{figure}[t]
\vspace{0cm} \hspace{0.0cm} \centerline{ \epsfxsize =7.5cm
\epsfysize =7.80cm \epsffile{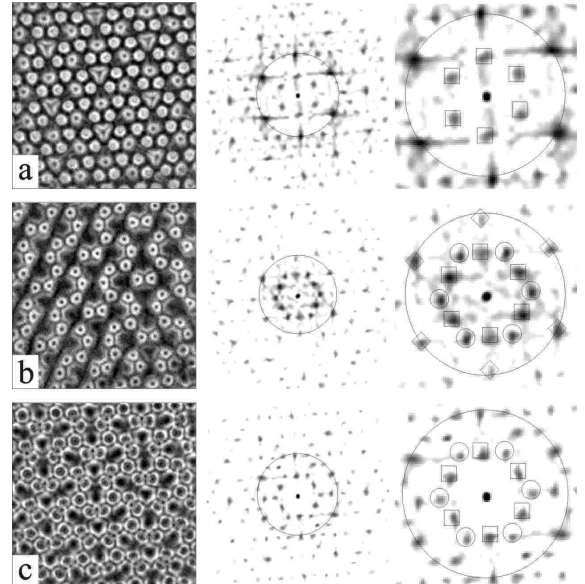}}
\caption{Reducing $h$ from $h=0.33$ cm where only SSS--II was
observed to $h=0.31$ cm results in a state where the two types of
SSS patterns SSS--I and SSS--II can coexist. In (a-c) the combined
state images (left), full power spectra (middle), and expanded
power specta right) are presented at different temporal phases.
The primary hexagonal symmetry (of magnitude $\vec{k}_\diamond$)
is broken by two different sets of wave vectors arranged in two
6--fold sublattices. The set corresponding to the SSS-I is
enclosed in squares. The set corresponding to the SSS-II is
enclosed in circles. This state was observed for 2/3 driving and
system parameters of $\nu=47$ cS, $\omega_0/(2\pi)=25$ Hz and
$h=0.31$ cm, $\chi=62.5^\circ$ and $\phi=0^\circ$.
} \label{SubSub_k}
\end{figure}

For 2/3 driving we observed the appearance of SSS--I states at
lower liquid layer depth and the SSS--II at higher depths. As $h$
is reduced, we found that instead of an abrupt transition between
the SSS--I and SSS--II states, both types of symmetry breaking can
occur {\em simultaneously}. As shown in \Fig{SubSub_k}, the
primary hexagonal symmetry of this state is broken by two 6--fold
sets of wavevectors.

One set corresponds to SSS--I with magnitude $q_0=K/2$ while the
other set corresponds to SSS--II, with $q_1=K/\sqrt{3}$. In real
space the characteristic pattern of SSS--II (\Fig{SubSub_k}b) is
broken by superimposed stripes in one direction. This state has
both modulative and displacive effects implying that the two
symmetry--breaking modes retain their respective spatial phase
characteristics.

\begin{figure}
\vspace{-.5cm} \hspace{0.0cm} \centerline{ \epsfxsize =7.5cm
\epsfysize =11cm \epsffile{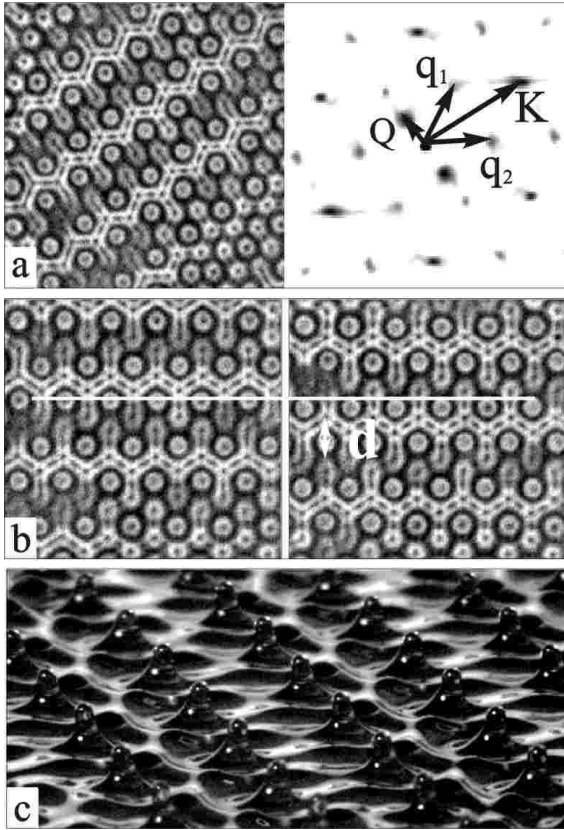}}
\caption{(a) The image (left) and power spectrum (right) of
``cascaded" SSS states appearing for 4/5 driving near $\chi_c$.
The two mechanisms appearing in SSS-I and SSS--II coexist at
different scales. A hexagonal sublattice of $q_i=K/\sqrt{3}$
bifurcates from the original hexagonal pattern. In addition, a
{\em third} wavevector $\vec{Q}\neq\vec{q}$ breaks the symmetry of
this sublattice in a single direction. This state has an overall
temporal periodicity of $2\pi/\omega_0$. This is demonstrated in
(b) where two images (b left, right) were taken at an interval of
$\pi/\omega_0$. The images are displaced by a $d=4\pi/K$ length
scale. The horizontal white line indicates the location of the
peaks in the first phase (b left). (c) A the side view of this
state. The peaks are of high amplitude and have the characteristic
shape of the oscillons described by Arbell and Fineberg
\protect\cite{Arbell3} with a small asymmetry in the direction of
$\vec{Q}$. These images were observed for system parameters of
$75/60$ Hz, $\nu=23$ cS, $h=0.2$ cm, $\chi=56^\circ$ and
$\phi=0^\circ$. } \label{SSS_60_75first}
\end{figure}

As demonstrated in \Fig{PlotK-SSS}, the mechanism that forms both
the SSS--I and the SSS-II patterns depends on a ``slaved" linear
eigenmode. In the 2/3 driving described above the only existent
linearly stable subharmonic tongue occurs for
$\omega_0/2=m\omega_0/4$.  In contrast to 2/3 driving, even/odd
driving ratios with $m/n$ values such as 4/5, 6/7 etc. possess
linearly stable tongues whose dominant frequencies differ from
$m\omega_0/4$. For example, in 60/75 Hz driving, besides the 15 Hz
($=m\omega_0/4$) response observed for the SSS state, a response
at 7.5 Hz ($=\omega_0/2$) is also possible. Recent theoretical
work by Silber et al. \cite{SilberTopaz} has suggested that these
additional slaved modes can influence the character of the
selected nonlinear state. We find that these additional slaved
modes can indeed appear. In 4/5 driving typical SSS--I states with
a frequency response of $m\omega_0/4$ are observed far from
$\chi_c$. Near $\chi_c$, both the 12--fold temporally harmonic
quasicrystalline states first described by Edwards and Fauve
\cite{Edwards94}, as well as a state composed of a {\em cascade}
of symmetry-breaking bifurcations occur. This second type of
pattern, which has a subharmonic time dependence, is shown in
\Fig{SSS_60_75first}a.

This ``cascaded" state appears as either a secondary bifurcation
of an initial hexagon pattern or as a bifurcation from a 12--fold
quasipattern state. In its spatial power spectrum the two
mechanisms that appeared in both SSS--I and SSS--II type patterns
are cascaded and appear at different scales. An SSS-II type
resonance occurs where wavevectors $\vec{q_i}$ with magnitudes
$q_i=K/\sqrt{3}$ appear. A {\em third} vector $\vec {Q}$, however,
is also present. $\vec{Q}$ is half the magnitude of $\vec{q_i}$
and, echoing the mechanism forming SSS--I states, breaks the
6--fold symmetry of the $\vec{q_i}$ by aligning itself parallel to
a single vector, $\vec{q}$.

\Fig{SSS_60_75first}b demonstrates that this state has an overall
temporal periodicity of $2\pi/\omega_0$. The two images (b left,
right) were taken at an interval of $\pi/\omega_0$. The images
look exactly the same but with a transverse displacement of the
$d=4\pi/K$ length scale. The symmetry breaking by the $\vec{Q}$ is
similar to the SSS-I mechanism where a mode with a phase of
$\pi/2$ produces a displacive effect, as shown in
\Fig{SSS_60_75first}b. The peaks are of high amplitude and have
the characteristic shape of the oscillons described by Arbell and
Fineberg \cite{Arbell3} modified by the asymmetry that is produced
by the displacive effect of the $\vec{Q}$ mode.

\begin{figure}[t]
\vspace{-.5cm} \hspace{0.0cm} \centerline{ \epsfxsize =7.5cm
\epsfysize =5.0cm \epsffile{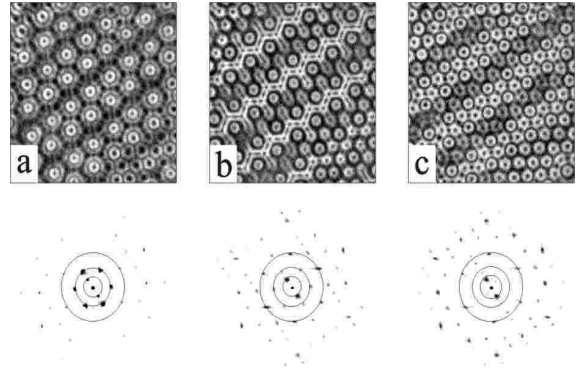}}
\caption{Images (a-c top)and the corresponding power spectrum (a-c
bottom) of different temporal phases of the SSS type pattern shown
in \protect\Fig{SSS_60_75first} that appears in 4/5 driving
experiment.  The original hexagonal wave vector is drawn as the
outer circle in the power spectrum shown (a,b bottom). It can be
seen that the different wave vectors have different relative
amplitudes at different temporal phases. In (a) the two smaller
wave vectors $\vec{Q}$ and $\vec{q}$ have higher intensities than
the wavevector $\vec{K}$ along the outer circle, where in (c) the
center circle of radius $q$ has a very small intensity. }
\label{SSS_60_75}
\end{figure}

As demonstrated by \Fig{SSS_60_75}, different wavevectors are
dominant at different temporal phases. These states were not
observed in experiments with higher $m/n$ values such as 6/7 and
8/9 driving ratio.

\section{Double hexagon superlattices (DHS)}\label{DHS}

\begin{figure}
\vspace{-.5cm} \hspace{0.0cm} \centerline{ \epsfxsize =7.5cm
\epsfysize =4.5cm \epsffile{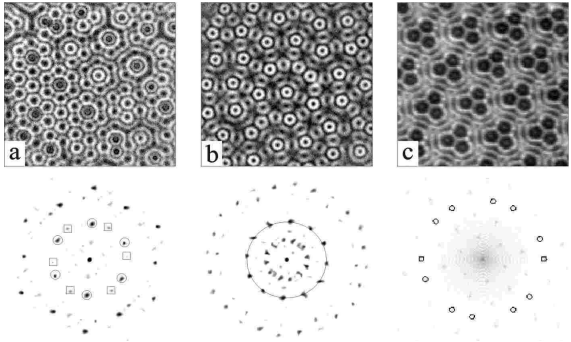}}
\caption{ When using even/odd driving, a class of temporally
harmonic states exists that is composed of two distinct hexagonal
sublattices (delineated by squares and circles) of size $K$, the
critical wavenumber of the harmonic driving component. (a) Double
Hexagon State (DHS) formed by both the critical hexagonal lattice
and an additional hexagonal sublattice of smaller amplitude which
is rotated by an angle of $\sim 22^\circ$. (b) 12--fold
quasicrystalline state (c) A similar (``SL1") state observed by
Kudrolli et al. \protect\cite{Kudrolli} for 6/7 driving. This
state is formed by a resonance similar to (a) but with spatial
phase of $2\pi/3$ in each of the sublattice components. (a) and
(b) were obtained for 4/5 ($60/75$ Hz) forcing. (c) was reproduced
with permission from \protect\cite{Kudrolli}.}
\label{Hx2experiment}
\end{figure}

Historically, the first two--frequency experiment focused on
quasicrystalline patterns with 12--fold symmetry. These were
observed in the vicinity of $\chi_c$ in systems driven with an
even/odd driving ratio and were found to bifurcate either from the
flat state or as a second bifurcation from the harmonic hexagon
region. In this section we will describe both the 12--fold
quasipatterns mentioned above and two other superlattice states
(see \Fig{Hx2experiment}). All of these states share a harmonic
temporal response for odd/even driving and appear in the same
general area of phase space. An important factor in pattern
selection is the nonlinear coupling coefficient that depends
strongly on resonant locking (see Sec. \ref{Nonlinear-analysis}).
In contrast to the SSS states, we will see below that no {\em new}
wavenumbers are needed to construct these states. This class of
superlattice patterns are differentiated by both the relative
angular orientation and spatial phase relations of the critical
wavevectors that form them.

The 12--fold quasipattern (see \Fig{Hx2experiment}a) was observed
for 4/5 driving in the vicinity of $\chi_c$. This state's temporal
behavior is harmonic with respect to $\omega_0$ This state can be
formed near the linear threshold for small values of $\epsilon
\sim 0.01$. Increasing the amplitude $A$ causes a bifurcation to
the cascaded--type superlattice described in \ref{Selection} (see
\Fig{SSS_60_75first}). The 12--fold quasipattern may also be
understood as being formed by two sets of hexagonal wavevector at
$30^\circ$ to each other. As in all quasicrystals, this state does
{\em not} have long range order or a well defined sub--unit cell.
It is interesting to note that unlike the quasipattern described
in \cite{Edwards93} that appeared only for $\phi\sim75^\circ$ the
quasipattern we describe exists for $\phi=0^\circ$.

\begin{figure}
\vspace{0cm} \hspace{0.0cm} \centerline{ \epsfxsize =7.5cm
\epsfysize =6.5cm \epsffile{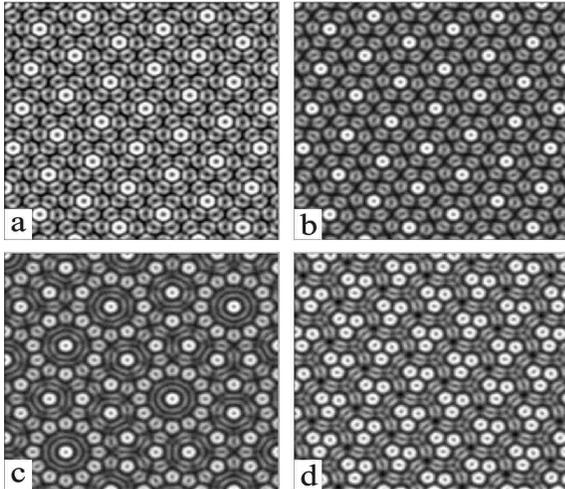}}
\caption{Using our numerical modelling technique, we calculated
the images of a perfect double hexagon state (DHS) formed by two
hexagonal lattices rotated by $22^\circ$. (a) $\alpha_i=\beta
_i=0$ (see \protect\Eq{eq:DHS}) shows a negative crater state and
(b) a positive peak state. (c) shows the calculated image of a
12--fold quasi--pattern. (d) A calculated DHS state with spatial
phases of $\alpha_i=0$ and $\beta_i=2\pi/3$ as in
\protect\Fig{Hx2experiment}(c). } \label{Hx2theory}
\end{figure}

Another superlattice pattern (\Fig{Hx2experiment}c) which occurs
in this regime of phase space was first studied by Kudrolli et al.
with 6/7 driving ratio with $\omega_0/(2\pi)=16.44Hz$,
$\nu=20CSt$, and $h=0.3cm$. They found a hexagonal superlattice
pattern composed of two hexagonal sets of critical wavevectors
with harmonic temporal response. The equation that describes the
surface height function of this pattern can be generally written
down as:
\begin{equation}\label{eq:DHS}
h(r)= \sum_{i=1}^{3}A_i\cos(\vec{K_i}\cdot\vec{r}+\alpha_i)+
\sum_{i=1}^{3}B_i\cos(\vec{K'_i}\cdot\vec{r}+\beta_i),
\end{equation}
where  $|K|=|K'|=k_{c}$ and
$$\vec{K_1}=K(1,0);\vec{K_2}=K(-\frac{1}{2},\frac{\sqrt{3}}{2});
\vec{K_3}=K(-\frac{1}{2},-\frac{\sqrt{3}}{2});$$

and $\vec{K'_i}$ can be obtained by rotating $\vec{K_i}$ by an
angle of $\theta=22^\circ$. In \Fig{Hx2theory}a,b,d we show
simulated images of this equation for different values of $\beta$
and $A_i/B_i$.

The pattern (\Fig{Hx2experiment}c) described by Kudrolli et al.
was found to have an angle $\theta$ with the value
$\theta=2\sin^{-1}(1/2\sqrt{7})\approx 22^\circ$, equal amplitude
coefficients $|A_i|=|B_i|$ for $i=1,2,3$ and spatial phase angles
$\alpha_i=0^\circ$ and $\beta_i=120^\circ$. For this special value
of $\theta$, resonance conditions such as
$2\vec{K'_1}-\vec{K'_3}=2\vec{K_1}-\vec{K_3}$ were shown
\cite{Kudrolli} to exist.

Double hexagon states formed by the superposition of two hexagon
sets of critical wavevectors oriented at a relative angle
$\theta=2\sin^{-1}(1/2\sqrt{7})$ are
 one example \cite{SilberProctor,SilberRucklidge} of a wider class of superlattices. This
superlattice class is composed of a periodic lattice formed by a
wavenumber smaller than the critical wavenumber of the excited
surface waves. This smaller wavenumber corresponds to the six
wavevectors formed by the difference between adjacent wavevectors
of the two hexagonal sets i.e. by
$\vec{Q_i}$=$\vec{K_i}$-$\vec{K'_i}$
 for $i=1,2...6$. Any two of these wavevectors, say
$\vec{Q_1}$ and $\vec{Q_2}$, are primitive vectors of the lattice
and the vectors $\vec{K_i}$ and $\vec{K'_i}$ are points on the
$Q$--lattice given by $\vec{K_i}=n_1\vec{Q_1}+n_2\vec{Q_2}$.
Silber and Proctor \cite{SilberProctor} show that only a discrete
countable set of $\vec{Q_i}$  where
$K/Q=\sqrt{n_1^2+n_2^2-n_1n_2}$ can satisfy this condition.  The
angle $\theta$ between the two lattices is given by
$\theta=\cos^{-1}({n_1^2+2n_1n_2-2n_2^2\over
2(n_1^2-n_1n_2+n_2^2)})$. In this formulation the experimentally
observed patterns are obtained for $n_1=3$, $n_2=2$ and
$K/Q=\sqrt{7}$, giving $\theta \approx 22^\circ$. Only the
simplest DHS have been observed to date.

One characteristic aspect of the DHS states is the relative
spatial phase of the different wavevectors.  The pattern described
by Kudrolli et al. \cite{Kudrolli} (coined ``SL-1'', see
\Fig{Hx2experiment}c) consisted of triangular unit cells that were
produced by a DHS with $\alpha_i=0^\circ$ and $\beta_i=120^\circ$.
Silber and Proctor describe, in simulations of thermal convection,
a DHS state formed with both $\alpha_i=0^\circ$ and
$\beta_i=0^\circ$ \cite{SilberProctor}. The DHS superlattice in
our experiments with a 2/3 driving ratio (see
\Fig{Hx2experiment}a) differs from these superlattices in that the
two hexagonal sublattices that form them possess {\em different}
amplitudes. States similar to these been observed in nonlinear
optics \cite{Herrero}. In addition, in our case, there is no
spatial phase difference between the two sets of hexagonal
wavevectors ($\alpha_i=\beta_j$ for all $i,j$).

\begin{figure}
\vspace{0cm} \hspace{0.0cm} \centerline{ \epsfxsize =8.5cm
\epsfysize =2.6cm \epsffile{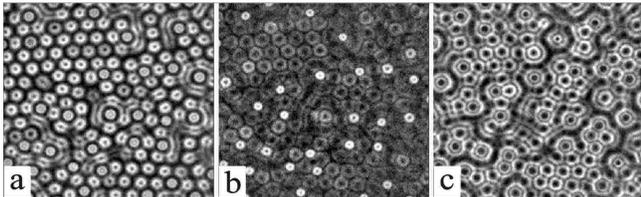}}
\caption{Typical images of the double hexagon state (DHS) taken at
different temporal phases. (a) shows a crater phase (b) an
intermediate phase  and (c) a peak phase for 2/3 driving with
$\omega_0=30$ Hz, $h=1.55$ cm, $\nu=23$ cS and $\chi=60.4^\circ$.
The time interval between (a) and (b) is $0.25T$ and between (a)
to (c) is ($0.4T$)
} \label{Hx2exp_time}
\end{figure}

Although this state is stable close to the threshold, increasing
the driving amplitude results in the appearance of many defects
and eventual temporal disorder. At high driving amplitudes the
spatial symmetry is hard to discern due to the many defects and
domains within the fluid cell. In \Fig{Hx2exp_time} we show a
typical DHS state at different temporal phases. The state
oscillates between a crater--like phase (a) and a positive
oscillon--like phase with a frequency of $\omega_0$. High
amplitude oscillons appear in the center of the hexagonal
sub--unit cell of the DHS state. Similar oscillons were also
observed for the 12--fold quasipattern state shown in
\Fig{Hx2experiment}b \cite{Arbell3}.

Let us summarize the common characteristics of DHS states. All of
the patterns reveal a harmonic time dependence and the lack of any
{\em fundamental} wavevectors other than the critical wavevectors
excited by the harmonic frequency. All of these patterns were
observed to be in the vicinity of $\chi_c$ on the harmonic side.
In all cases, the mixing angle was greater than that needed for
obtaining SSS states and within the range $\chi_c^\circ >
\chi>(\chi_c-12^\circ)$. Another common characteristic of these
states is that all exist in the vicinity of a first--order
transition of the hexagonal patterns from the featureless fluid
state. This implies that quadratic interactions can play an
important part in describing these states. Finally, two of these
patterns also generate oscillons as described in detail in
\cite{Arbell3}.

\section{Two mode superlattices (2MS)} \label{Results2}

In the vicinity of the critical mixing angle, $\chi_c$, two modes
with different wavelengths can be excited concurrently. These
modes can interact in different ways to produce a variety of
different patterns. In the next two sections we will describe two
distinct types of two--mode states that are formed near $\chi_c$.
The first of these states are two--mode superlattices (2MS). These
states are formed by the interaction of the two linearly excited
modes with a third ``slaved" mode that is selected via a temporal
resonance.
\begin{figure}[t]
\vspace{0cm} \hspace{0.0cm} \centerline{ \epsfxsize =7.5cm
\epsfysize =7.3cm \epsffile{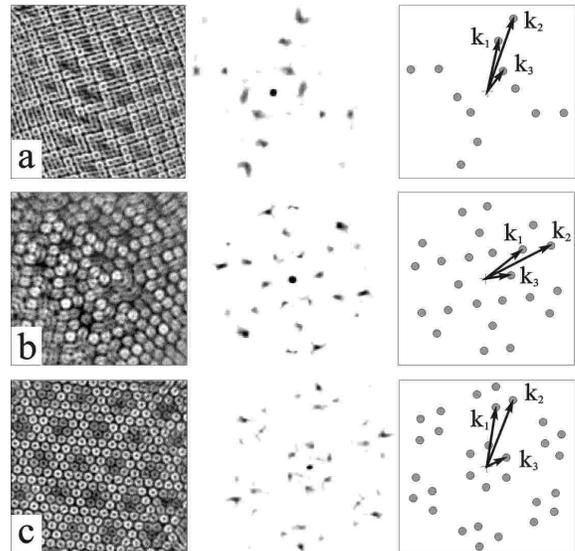}}
\caption{2MS patterns consist of a nonlinear superposition of {\em
both} eigenmodes, $k_1$ and $k_2$, excited by the driving
frequencies $\omega_1$ and $\omega_2$. Although $\omega_1$ and
$\omega_2$ determine the lengths of $k_1$ and $k_2$, their
relative orientations are determined by the condition:
$\vec{k_3}=\vec{k_2}-\vec{k_1}$ where the wavenumber, $k_3$, is
determined by the temporal resonance condition:
$\omega_3=\omega_2-\omega_1$. (a) 2MS patterns for 50/80 Hz
driving where $k_2$ with square symmetry is dominant. (b)
$k_2$--dominant 2MS with hexagonal symmetry for 45/60 Hz driving
and (c) $k_1$--dominant states having hexagonal symmetry for 40/60
Hz. Shown are the spatial spectra (center) with the resonant
triads (right) $\vec{k_3}=\vec{k_2}-\vec{k_1}$, highlighted.}
\label{2MS_intro}
\end{figure}

\Fig{2MS_intro} shows the three main types of 2MS states; $k_2$
dominated (with either square or hexagonal symmetries) and $k_1$
dominated (with hexagonal symmetry). 2MS states exist in both
harmonic and subharmonic regions of phase space in the vicinity of
$\chi_c$ (see \Fig{PhaseSpaceBoth}). In the region of phase space
between the square 2MS and the hexagonal 2MS a spatially
disordered ``unlocked'' state exists. Unlocked states, which are
formed by the same wavenumbers that form the 2MS, have no
well--defined spatial symmetries. As in the case of SSS states,
the transition to 2MS from either square or flat states is
non--hysteretic and occurs via propagating fronts.

The 2MS are qualitatively different than SSS states. They result
from spatial phase locking of {\em both} $k_1$ and $k_2$ whereas
the SSS states result from a resonance condition that is {\em
independent} of the $k_2$ mode. 2MS states are the most general of
the superlattice states described here. They are observed for all
types of driving parities (odd/even, even/odd and odd/odd) and, as
shown in \Fig{2MS_intro}, appear with either square or hexagonal
symmetry.
\subsection{2MS Resonance condition}\label{ResonanceCondition}

2MS spatial spectra, as shown in \Fig{2MS_intro}, are composed of
peaks of length $k_1$ and $k_2$ and their linear combinations. The
strongest secondary peaks, indicated in the figure, are given by
$\vec{k_3}= \vec{k_2}-\vec{k_1}$ where the magnitude of $k_3$ is
consistent with the linear value of $k$ calculated for a single
frequency excitation  at the difference frequency $\omega_3 =
\omega_2 - \omega_1$. Our calculated value of $k_3$ was obtained
using the {\em linear} single frequency code of \cite{Kumar} at
threshold. The difference between the calculated and measured
values of $k_3$ varies between $5-20 \%$. This shift between the
measured and calculated values is constant for a given value of
the difference frequency, $\omega _3$, and systematically
decreases as $\omega_3$ increases. We believe the shift to be the
result either finite size effects in the cell or the fact that the
$\omega(k)$ used is the linear dispersion relation for a {\em
featureless} state (not one with pre--existing waves).

\begin{figure}
\vspace{0cm} \hspace{0.0cm} \centerline{ \epsfxsize =7.5cm
\epsfysize =5.0cm \epsffile{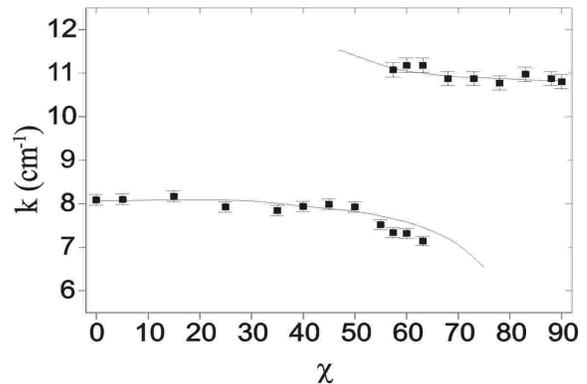}}
\caption{A comparison of $k$ (in $cm^{-1}$), calculated using the
two--frequency linear calculation \protect\cite{Besson} (solid
line) ,with measured values of $k$ as a function of $\chi$. Note
that both values of $k$ are nearly constant away from
$\chi_c=58^\circ$. Near $\chi_c$, $k_1$ (corresponding to
$\omega_1 =40$ Hz) decreases by nearly $10\%$ while $k_2$
(corresponding to $\omega_2 = 60$ Hz) increases slightly. }
\label{kscan}
\end{figure}

A subtle point in the interpretation of the power spectra of 2MS
states, is the evaluation of the magnitude of the two wavevectors
that appear. As can be seen in \Fig{kscan} the values of $k_1$ and
$k_2$ measured in the vicinity of $\chi_c$ are significantly
different than the values of $k$ excited by single frequency
excitation. The two--frequency linear stability code of Tuckerman
\cite{Besson,KumarTuckerman}, reproduces this effect and agrees to
within $1-2\%$ with the experimentally measured values (see
\Fig{kscan}). This enables us to accurately calculate values of
both $k_1$ and $k_2$ in the vicinity of $\chi_c$.

The resonant conditions stated above suggest that the orientation
of the wavevectors building the 2MS is selected by nonlinear
interactions that are resonant both in space and time. Thus, the
temporal resonance condition dictates the spatial orientation of
the vectors $\vec k_1$ and $\vec k_2$. Such 3--wave resonant
interactions have been predicted to occur in non--linear
interactions of surface waves \cite{Hammack} and are well known in
the physics of plasmas. Similar states were observed as a result
of nonlinear mixing of a multiple--mode optical beam
\cite{Pampaloni}. The selection of $k_3$ via the temporal
resonance condition yielding $\omega_3$ is non--trivial and can
not be accounted for by experimental artifacts such as possible
nonlinearities in the imaging. This 3--wave resonance condition
occurred for {\em all} frequency ratios tested.

As in SSS, the dominant 2MS wavevector retains its initial
symmetry, while the relative orientation of the other linearly
excited wavevectors are determined by the above resonance
condition. For odd/even driving hexagonal (square) symmetry
dominates for $\chi<\chi_c$ ($\chi>\chi_c$). Thus, square 2MS
(\Fig{2MS_intro}a) bifurcate from the $k_2$ square pattern that
dominates the $\chi>\chi_c$ region. Similarly, hexagonal 2MS
(\Fig{2MS_intro}b) bifurcate from the $k_1$ hexagonal pattern that
dominates the $\chi<\chi_c$ region. In 3/4 driving, hexagonal 2MS
states are excited whose dominant scale is that of the larger
wavenumber, $k_2$ (\Fig{2MS_intro}c). It is known that in
single--frequency driving experiments different symmetries can
arise for different system parameters even when the temporal
behavior is solely subharmonic. In two--frequency driving
experiments the parity of the dominant frequency does not
automatically dictate the symmetry selected. For example, in 5/8
driving (50/80 Hz) 2MS states are observed with square symmetry in
the temporally {\em harmonic} region.

\subsection{Temporal behavior of 2MS}

\begin{figure}
\vspace{0cm} \hspace{0.0cm} \centerline{ \epsfxsize =8.5cm
\epsfysize =3.3cm \epsffile{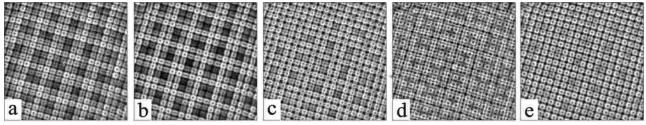}}
\caption{A typical temporal sequence of square 2MS states taken at
constant parameter values for 50/70 Hz driving. $k_1$ (middle
circle), $k_2$ (outer circle) and $k_3$ (inner circle) are
indicated in (a). In (d) the $k_3$ vectors are absent but enclosed
in squares are vectors that are the vector sum
$\vec{k'_3}+\vec{k_3}$. System parameters are $\omega_0/(2\pi)=20$
Hz, $\nu=23$ cS,$\phi=0^\circ$, $\chi=65.4^\circ$, and $h=0.155$
cm. } \label{2MS_50_70_time}
\end{figure}

\begin{figure}
\vspace{0cm} \hspace{0.0cm} \centerline{ \epsfxsize =7.5cm
\epsfysize =4.5cm \epsffile{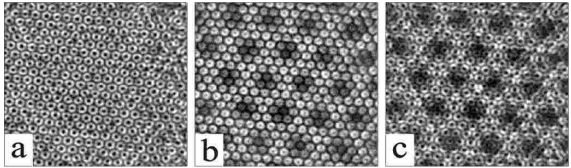}}
\caption{A typical sequence of {\em hexagonal} 2MS (a-e) at
different temporal phases taken at constant parameter values for
60/80 Hz driving. Circles of radii $k_1$ (middle), $k_2$ (outer)
and $k_3$ (inner) are drawn in (c). In (b) the $k_3$ and $k_1$
vectors interact to produce the vectors enclosed squares. $k_1$
and $k_2$ form two sets of hexagonally arranged vectors.
 } \label{2MS_HX_60_80_time}
\end{figure}

Using our strobed imaging technique, we studied the
spatio--temporal behavior of the patterns by inspecting them at
different temporal phases. The 2MS inherit the basic temporal
periodicity of {\em both} primary eigenfunctions, which are
periodic in time with a basic frequency of either $\omega_0$
(harmonic) or $\omega_0/2$ (subharmonic). The superposition of the
two modes always displays a subharmonic response ($\omega_0/2$).
Like the SSS, the state's appearance changes qualitatively with
time, with the spatial wavenumbers having fixed {\em locations}.
2MS time--dependence stems from the temporal dependence of the
{\em amplitudes} of these modes. In \Fig{2MS_50_70_time} and
\Fig{2MS_HX_60_80_time} we see typical sequences of both square
and hexagonal 2MS states taken at constant values of the driving
parameters. The relative intensities of the different wavevectors
change in the different phases; in \Fig{2MS_50_70_time}e the $k_1$
and $k_3$ wavevectors are almost absent, resulting in a pattern
whose dominant square symmetry has a $2\pi/k_2$ scale, while in
the remainder of the phases all vectors appear. In
\Fig{2MS_HX_60_80_time} the resonance condition allows $\vec{k_1}$
and $\vec{k_2}$ to be nearly co--linear and two sets of 6--fold
wavevectors appear in the power spectrum.  As in the square 2MS,
one can see phases with significantly stronger $k_1$ peaks
(\Fig{2MS_HX_60_80_time}(a)) as well as those where either the
$k_3$ (\Fig{2MS_HX_60_80_time}(b)) or $k_2$
(\Fig{2MS_HX_60_80_time}(c)) are stronger.

\begin{figure}
\vspace{0cm} \hspace{0.0cm} \centerline{ \epsfxsize =7.5cm
\epsfysize =5.0cm \epsffile{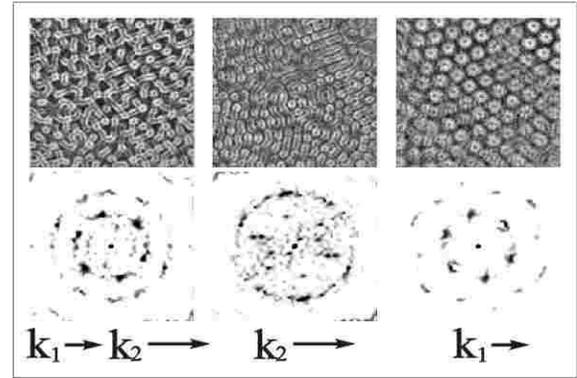}}
\caption{A typical sequence of $k_1$--dominant hexagonal 2MS
states taken at constant parameter values for 40/70 Hz driving.
These states generally appear in domains. (Wavenumbers are noted
by arrows). } \label{HX-T-2MS-time}
\end{figure}

The relative stability of 2MS hexagonal states (e.g.
\Fig{2MS_intro}b,c) is dependent on whether $k_1$ or $k_2$ is
dominant. Let us first consider hexagonal 2MS states where $k_1$
dominates. This state is found for
 even/odd driving for $\chi<\chi_c$ (see
\Fig{PhaseSpaceBoth}) and a typical time sequence is presented in
\Fig{HX-T-2MS-time}. Again, different wavenumbers are dominant in
different temporal phases.

\noindent The region of phase space where this state appears is
not as large as the $k_2$ dominant 2MS, but these states appear
for all even/odd driving ratios used. As in the case of square 2MS
and the $2kR$ states discussed in Section \ref{Results3}, these
states are not observed for all combinations of $\phi$ and
$\delta/h$ used. It is difficult to obtain a hexagonal 2MS state
of this kind that extends over the entire system and
$k_1$--dominant hexagonal 2MS states generally occur within
domains. Thus the spatial spectra (as seen in \Fig{HX-T-2MS-time})
appears sometimes smeared, as the various domains have different
angular orientations. Although the scenario described in
\Fig{HX-T-2MS-time} is typical, we have observed stable global
$k_1$--dominated 2MS state for particular values of $\phi$ and
$\delta/h$ (see e.g. \Fig{Transfer_Ph120_first}).

A typical time sequence and power spectra for the $k_2$ dominant
hexagonal 2MS state (shown in \Fig{2MS_intro}c) is presented in
\Fig{2MS_HX_60_80_time}. Here, in contrast to $k_1$ dominant
hexagonal 2MS, the pattern is global with a well--defined symmetry
in all of its temporal phases. Hexagonal symmetry was {\em not}
seen to be preferred for all odd/even driving and was only
observed for the simple ratios of 1/2 and 3/4. This might be a
feature of simple driving ratios that can sometimes have unique
properties which are related to temporal locking
(\cite{Zhang2freq,SilberTopaz}).

2MS states are not always as highly ordered as those in
\Fig{2MS_50_70_time} and \Fig{2MS_HX_60_80_time}. When highly
ordered 2MS states occur, the resonance condition for these
parameters allows $\vec{k_1}$ $\vec{k_2}$ and $\vec{k_3}$ to be
nearly co--linear. When the angles between the resonant
wavevectors are not small the pattern is usually not global and a
number of domains coexist. In highly ordered states we find a
locking of the wavevectors' magnitude to small natural number
ratios. The locking ratios of $k_1:k_2:k_3= 1:3:4$ and 1:4:5 were,
respectively, obtained in \Fig{2MS_50_70_time} and
\Fig{2MS_HX_60_80_time}. This ``spatial locking'' seems to
stabilize global patterns in a way similar to the global SSS
states presented in Section \ref{Results1}.

\begin{figure}
\vspace{0cm} \hspace{0.0cm} \centerline{ \epsfxsize =7.5cm
\epsfysize =7.5cm \epsffile{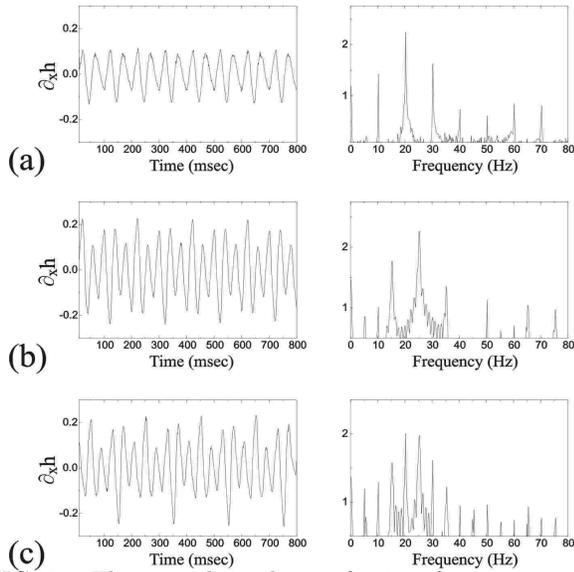} }
 \caption{The time
dependence of a two--frequency experiment with even/odd driving
was studied using the reflection of a laser by the surface waves.
The temporal response is subharmonic (harmonic) with respect to
$\omega_0=10$ Hz for $\chi<\chi_c$ ($\chi>\chi_c$).  Time series
(left) and power spectra (right) for three different regimes: (a)
Typical temporal response for hexagonal pattern found for
$\chi<\chi_c$. Although the $4\omega_0/2=20$ Hz component is the
strongest, an $\omega_0=10$ Hz harmonic component with all of its
higher harmonics is present. (b) Typical time dependence for
square patterns (found for $\chi>\chi_c$). The strongest response
is for $4\omega_0/2=25$ Hz as expected. A weaker response at the
basic subharmonic $\omega_0/2=5$ Hz is observed together with
stronger responses at other harmonics. The ordering of the
harmonics agrees qualitatively with \protect\cite{Zhang2freq}. (c)
Typical temporal response of a square 2MS state at $\chi_c$. The
power spectrum shows a strong response at both $\omega_1$ (20 Hz)
and $\omega_2$ (25 Hz). All other harmonics are present with a
different power distribution than in (a) and (b). System
parameters here are: $50/40Hz$, $\nu=23$ cS, $h=0.155$ cm,
$\phi=0^\circ$ and $\chi=40^\circ$ (a), $\chi=60^\circ$ (b) and
$\chi_c=53.5^\circ$ (c).
} \label{Laser40_50}
\end{figure}

 The local laser probe technique
provides more detailed, quantitative, information of the temporal
behavior of 2MS states. In \Fig{Laser40_50} we show one component
of the slope of surface waves obtained for 4/5 driving. In
\Fig{Laser40_50}(a) we show the typical waveform of a hexagonal
pattern for $\chi<\chi_c$. Besides the strong component at
$\omega_1/2=4\omega_0/2$ and its harmonics (40,60,80 Hz ...),
peaks appear at values of $j\omega_0, j=1,2...$. Those peaks are
expected from the linear theory. In \Fig{Laser40_50}(b) we see a
typical waveform of a square pattern for $\chi>\chi_c$. Although
the strongest frequency response is at $\omega_2/2$, the basic
frequency, $\omega_0/2$, together with its higher harmonics also
appear. Linear theory predicts a different distribution of energy
in the peaks for the unstable mode at threshold. Zhang and Vinals
\cite{Zhang2freq} non--linear theory accounts for the peak's
strength in a semi--quantitative way, as it correctly predicts the
ordering of the strongest peaks.
In \Fig{Laser40_50}(c) the temporal response of the 2MS state is
presented. This state has a temporal response that includes both
the frequencies of the harmonic mode (a) for $\chi<\chi_c$ and
those of the subharmonic one (b) for $\chi>\chi_c$. Although the
power spectrum shows that both main peaks at $\omega_1/2$ and
$\omega_2/2$ are of the same strength, their relative strengths
can vary with the location of the laser probe.

We have seen that the 2MS state contains the two linearly excited
eigenmodes both in space and in time. Does each mode keep its
distinct space--time behavior, or is there a complete mixing of
the spatial and temporal components? To clarify this question, we
write down a simple model for the surface height of a square 2MS
state. For simplicity, we will assume a square 2MS state where
both $\vec{k_1}$ and $\vec{k_2}$ are collinear with the same
spatial phase:
\begin{eqnarray}\label{SpaceTime}
h(r,t)=F_1(t)(\cos(k_1 x)+\cos(k_1 y))
 \\
 \nonumber
+F_2(t)(\cos(k_2 x)+\cos(k_2 y))
\end{eqnarray}

\begin{eqnarray}\nonumber
\frac{\partial h}{\partial
x}=k_1F_1(t)\cos(k_1x)+k_2F_2(t)\cos(k_2x)
\\ \nonumber
\frac{\partial h}{\partial
y}=k_1F_1(t)\cos(k_1y)+k_2F_2(t)\cos(k_2y)
\end{eqnarray}
where $F_1(t) \sim \cos(\omega_1t/2)+...$ and
$F_2(t)\sim(\cos(\omega_2t/2)+...$. It is easily seen that if one
selects a point $(x,y)=(\pi/2k_2,\pi/2k_1)$ then each of the
components of the partial derivative $\partial h/\partial x$,
$\partial h/\partial y$ has a temporal dependence of $F_1(t)$ and
$F_2(t)$ respectively. Both $F_1(t)$ and $F_2(t)$ include time
dependent terms retaining the parity of the linear modes: odd
multiples of $(p+1/2)\omega_0$ for a subharmonic response and even
multiples of $p\omega_0$ for harmonic response.

\begin{figure}
\vspace{0cm} \hspace{0.0cm} \centerline{ \epsfxsize =7.5cm
\epsfysize =7.5cm \epsffile{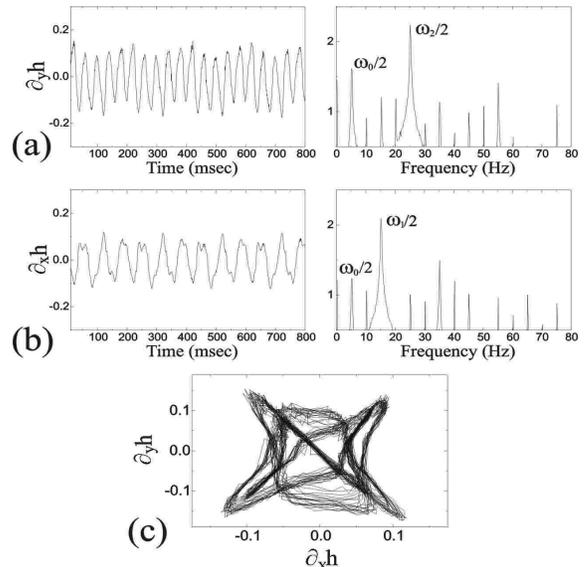} }
\caption{The time dependence of a square $2MS$ state measured via
the reflection of a laser at the fluid surface. Time series (left)
of surface gradients, $\partial_{y}h$ (a) and $\partial_{x}h$ (b)
are shown with their power spectra (right). (c) $\partial_{y}h$ as
a function of $\partial_{x}h$. The separation of the temporal
response (a,b) was obtained by choosing x and y along the symmetry
directions.  $\partial_{x}h$ is dominated by a $3\omega_0/2=15$ Hz
response whereas $\partial_{y}h$ undergoes a $5\omega_0/2=25$ Hz
response. System parameters used were $50/30$ Hz, $\nu=23$ cS,
$h=0.155$ cm, $\phi=0^\circ$ and $\chi_c=56.3^\circ$.
} \label{Laser30_50_second}
\end{figure}

As suggested by Eq. \ref{SpaceTime}, experiments (see
\Fig{Laser30_50_second}) confirm the separation of the time
dependence of the two modes within the 2MS state.  The
$\partial_{x}h$ component has a strong $\omega_1/2$ response
whereas the $\partial_{y}h$ component's strongest peak is at
$\omega_2/2$. This strong separation of spatial time dependencies
can only be observed for a few points in the (x,y) plan and, in
general, the two frequencies are mixed.

\subsection{The ``unlocked'' state and transition
regions}\label{Transitions}

\begin{figure}
\vspace{0cm} \hspace{0.0cm} \centerline{ \epsfxsize =7.5cm
\epsfysize =4.8cm \epsffile{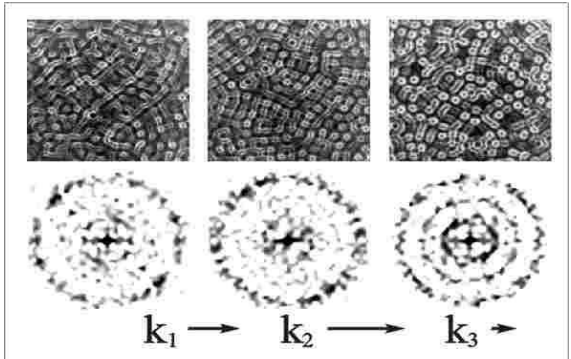} }
 \caption{In the
``unlocked'' state no orientational order is apparent.
 As in 2MS states, $k_1$ , $k_2$ and $k_3$ exist simultaneously
 but ``unlocked'' state spectra are diffuse and show little
angular correlation. (top) Typical views of an unlocked state with
corresponding spatial spectra (bottom) observed for 40/60 Hz
driving at different times and fixed driving parameters. The
different values of $k$ appear with different magnitudes and
orientations in each phase. Their lengths are indicated by the
arrows.
 }
\label{Unlocked_time}
\end{figure}

\begin{figure}
\vspace{0cm} \hspace{0.0cm} \centerline{ \epsfxsize =7.5cm
\epsfysize =7.1cm \epsffile{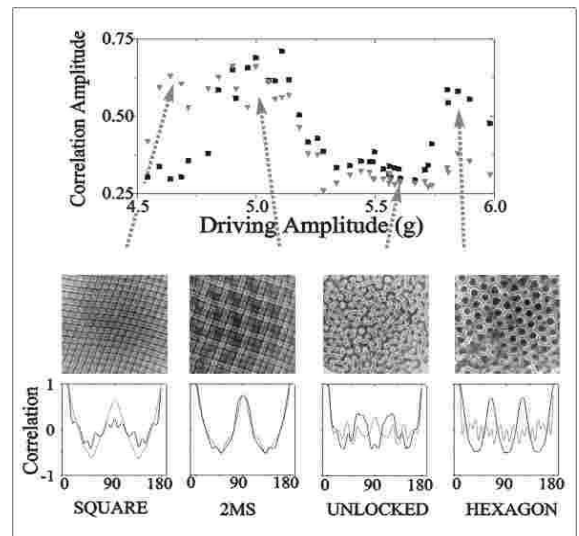} }
 \caption{Within the
``unlocked'' states, the correlation amplitude, $Q_k$ drops
significantly and the angular correlation function, $C_k(\theta)$,
indicates only a small amount of residual order. (upper) $Q_{k_1}$
(triangles) and $Q_{k_2}$ (squares), averaged over a single period
along the line $\chi=58^\circ$ for 40/60 Hz driving as a function
of the driving amplitude, $A$. This line traverses the square,
square--2MS, ``unlocked'', and hexagonal--2MS phases. Typical
patterns in each phase are displayed (center). The symmetry of the
different phases is highlighted by $C_k(\theta)$ for $k_1$ (grey
line) and $k_2$ (black line) computed for typical states ($\theta$
is in degrees).  The power (log scale) of $k_2$ relative to $k_1$
in each $C_k(\theta)$ is 8.3 (square), 1.1 (2MS), 0.9
(``unlocked'') and 0.2 (hexagon).
 }
\label{Correl}
\end{figure}

Let us now consider the ``unlocked'' state that appears in the
near vicinity of $\chi_c$. In \Fig{Unlocked_time} we present a
typical time series of the ``unlocked'' state and its
corresponding spatial spectra. The spatial behavior of the state
varies rapidly, over time scales of order ($2\pi/\omega_0$). In
contrast to the SSS and 2MS states, in the ``unlocked'' state no
orientational order is apparent.
\noindent Both $k_1$ and $k_2$ exist simultaneously in their
spatial spectra but spatial mode locking does not occur as in the
2MS states. This is evident in their power spectra, where,
generally, entire circles of radii $k_1$ and $k_2$ appear. As
\Fig{Unlocked_time}d indicates, additional peaks of wavenumber
$k_3$ corresponding to $\omega_3=\omega_2-\omega_1$ are sometimes
observed. The ``unlocked state" is a well-defined state that
exists in a relatively wide region of phase space. This can be
seen by defining (as in \cite{Binks,Arbell1}) the following
``orientational correlation function", $C_k(\theta)$ for each
value of $k$:
\begin{equation}
C_k(\theta)\equiv  \Sigma _\alpha [ f_k(\alpha )\cdot f_k(\alpha +
\theta) ]/ \Sigma _\alpha [ f_k(\alpha )\cdot f_k(\alpha )]
\label{2}
\end{equation}
where $ f_k(\alpha )$ is the Fourier transform of the wavenumber
$k$ at the polar angle $\alpha$. The correlation amplitude,
$Q_k=1/2[max{C_k(\theta)}-min{C_k(\theta)}]$, varies between 0 and
1 for, respectively, minimal and maximal orientational order. As
\Fig{Correl} shows, both the 2MS and hexagonal states have clear
orientational order while very little residual order is apparent
in the unlocked state.

\begin{figure}
\vspace{0cm} \hspace{0.0cm} \centerline{ \epsfxsize =8.5cm
\epsfysize =1.4cm \epsffile{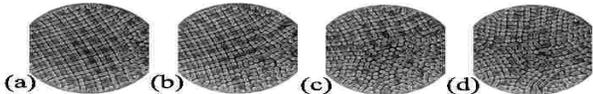} }
 \caption{The
transition between the square 2MS state and the ``unlocked'' state
occurs through a gradual process in which both states co--exist in
different domains. At the transition, increasing the amplitude for
constant $\chi$ constant results in (a) the global 2MS state is
first disturbed by small defects at the cell's rim. (b,c) The
disturbance spreads to the cell's center until, finally, the
entire pattern is in ``unlocked'' state(d). This process can also
occur in the reverse direction.
 }
\label{2MS-unlock}
\end{figure}

\begin{figure}
\vspace{0cm} \hspace{0.0cm} \centerline{ \epsfxsize =8.5cm
\epsfysize =1.5cm \epsffile{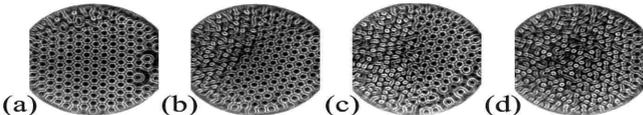} }
 \caption{A time
sequence of the transition from the regular hexagonal state (a) to
Hexagonal 2MS states is shown. The $k_2$ (60 Hz) component
propagates via a front (b,c )  until it dominates the entire plate
and co$-$exists with the large length scale (d). The driving
parameters are constant for this sequence. } \label{HX-T-HX}
\end{figure}

As is apparent from \Fig{Correl}, the orientational amplitude
drops sharply as the boundary between the 2MS and ``unlocked''
phases is crossed. This is demonstrated in \Fig{2MS-unlock}a--d
which corresponds to the range of 5.1--5.25g in \Fig{Correl}.
Between the pure hexagonal state and the ``unlocked'' state the
hexagonal 2MS state exists. This state displays strong $k_1$
dominant hexagonal symmetry at most temporal phases but at other
temporal phases some of the power is in the $k_2$ mode. Since
$Q_k$ is an average of equally time separated temporal phases the
hexagonal 2MS state has a $Q_k$ which is slightly smaller than
that of the pure hexagon state.
The parameter, $Q_k$,  does not completely characterize the
different transitions, since spatial FFT power spectra cannot
differentiate between global and local ordering. A closer look at
real--space images of the transitions presented in \Fig{Correl}
that correspond to square--2MS, ``unlocked" and Hexagonal--2MS are
shown respectively in \Fig{2MS-unlock} and \Fig{HX-T-HX}. Both
transitions involve an advancing front that separates two
well--defined domains. The clear separation between unlocked and
2MS domains provides further evidence that the unlocked state is
indeed a distinct nonlinear state and not, simply, a transition
region.

The transitions between unlocked states and the 2MS states with
different symmetries differ in two ways. The sensitivity to any
change of the driving parameters is much higher for the unlocked
to 2MS--Hexagon transition. Whereas the transition between square
2MS and ``unlocked'' states occurs for a relative change of
amplitude of $<5\%$, the transition between hexagonal 2MS and
hexagons can occur via a change smaller than $0.25\%$. The time
scales of the induced transitions are also different. The first
transition takes place in a nearly quasi--static reversible way,
whereas the second transition (as shown in \Fig{HX-T-HX}) can
occur over typical time scales of 50--1000 oscillation periods
with a hysteresis of less than 0.1\%. This rather sharp transition
is, perhaps, due to the effects of the quadratic interactions
inherent in the harmonic states. The precise duration of this
transition depends on the initial and final driving parameters. As
is typical of front propagation processes, the deeper one is
within the hexagonal regime the faster the transition time
\cite{JayVictor}.

\section{Rhomboidal states ($2kR$) and quasi--patterns}\label{Results3}

In the previous section we studied the three wave resonance
mechanism that can form 2MS states from the interaction:
$\vec{k_3}=\vec{k_2}-\vec{k_1}$ and $\omega_3=\omega_2-\omega_1$.
Is the only resonant mechanisms found in the vicinity of $\chi_c$?

In this section we will describe the spontaneous formation of
nonlinear rhomboidal patterns, formed as a result of solely {\em
spatially} resonant 3--wave coupling between wavevectors with
distinctly different wavenumbers. This state has only previously
been observed in a nonlinear optical system where the orientations
of the interacting wavevectors \cite{Saffman} were externally
imposed. Rhomboidal patterns have also been recently observed in
parametrically driven ferrofluids \cite{Lee}. The rhomboidal
patterns described below spontaneously couple the two circles of
linearly degenerate wavenumbers. These states qualitatively differ
from 2MS states in that they are composed solely of the linearly
excited wavevectors, $k_1$ and $k_2$, in contrast to the
additional slaved mode ($k_3$) necessary for 2MS formation.

The rhomboidal states observed in this system result from the
non-linear interaction of $k_1$ and $k_2$, which are waves with
significantly different wavelengths. Such states have been
observed numerically in a Swift--Hohenberg type models
\cite{MullerModel,Lifshitz} as discussed in section
\ref{background}. They have also been anticipated in anisotropic
models where two degenerate wavevectors are resonant with an
externally imposed wavenumber \cite{Zimmermann}, in nonlinear
optical systems \cite{Scroggie,Hoyuelos}, and in the analysis of
the Faraday instability excited with two frequencies
\cite{SilberTopaz,SilberSkeldon}. Both rhomboidal states and
superlattice patterns have also been recently predicted to occur
as a result of two bistable modes coupling to a zero--mode
\cite{Dewel}.

The rhomboidal states observed in this system differ distinctly
from rhomboids resulting from slightly ``distorted" hexagonal
states \cite{Kuznetsov,Gunaratne}. ``Distorted hexagons",
predicted to be stable in models with derivative-coupled quadratic
terms, may arise due to either initial or boundary conditions
\cite{Matthews}. These states have been observed in
reaction-diffusion systems \cite{Gunaratne}, convection in an
imposed shear flow \cite{Hall}, and flux line lattices in
superconductors \cite{Yethiraj}.

\subsection{Experimental conditions}\label{2kR-experiments}

We have observed rhomboid states using both Dow-Corning 200 oils
with kinematic viscosities, $\nu$, of 8.7, 23, 47, 87  cS (at
$30^\circ$C) and TKO--77 vacuum pump fluid with $\nu=184$ cS (at
$33^\circ$C) in fluid layers whose depth varied between
$0.1<h<0.55$ cm. The 2 wavenumber rhombic ($2kR$) states described
in this section were all generated with $m/n=2/3$ and $12<
\omega_0/2\pi < 45$ Hz or $m/n=4/5$ with $10< \omega_0/2\pi < 20$
Hz. Frequency combinations of 25/50, 30/60, 40/80, 40/56, 45/63,
48/68, 50/70, 55/75, 55/77, 30/50, 40/70, 45/60, 50/65, 50/80,
52/68, 60/84, 60/100 did {\em not} generate $2kR$ states.

\begin{figure}
\vspace{-1.5cm} \hspace{0.0cm} \centerline{ \epsfxsize =7.5cm
\epsfysize =3.5cm \epsffile{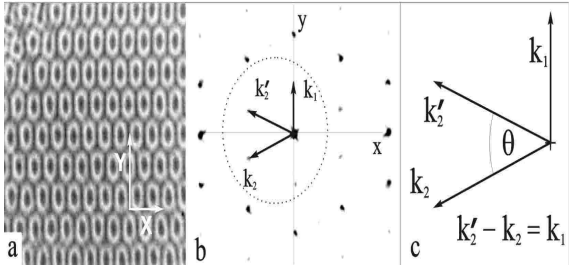} }
 \caption{(a) A
typical $2k$ Rhomboid ($2kR$) state observed for 2/3 driving and
$\omega_0/(2\pi)=25$ Hz, $\nu=23$ cS, and $h=0.2$ cm. The axis $y$
is defined by the direction of  $k_1$. (b) The power spectrum (c)
illustrates the simple resonance condition $\vec k_2'-\vec
k_2=\vec k_1$ that generates these states. $\vec k_2'$ and $\vec
k_2$ correspond to $\omega_2=3\omega_0/2$ while $\vec k_1$
corresponds to $\omega_1=2\omega_0/2$. $\theta$ here is $41^\circ
$.} \label{2kR_intro}
\end{figure}

In \Fig{2kR_intro} we present a typical image of a $2kR$ pattern.
Although $k_1$ and $k_2$ are the linearly unstable wavenumbers
\cite{Kumar} excited by $\omega_1$ and $\omega_2$, their
orientation is determined by the 3--wave non-linear interaction
that yields the resonant triad: $\vec k_2' - \vec k_2 = \vec k_1$,
where $|\vec k_2'|=|\vec k_2|=k_2$. The higher harmonics in the
figure may either be real or could occur as an artifact of the
imaging. An additive three--wave resonance occurs for 1/2 forcing
where the resonance $\vec k_1' + \vec k_1 = \vec k_2$ governs the
selected pattern. In this case, the resulting pattern (see section
\ref{Results4}) is a superposition of hexagonal lattices composed
of the two scales.

The observation of $2kR$ states solely for driving ratios 2/3, 4/5
and 1/2 is entirely consistent with Silber and Skeldon's
\cite{SilberSkeldon} predictions (see \ref{Nonlinear-analysis})
that 3--wave interactions coupling the wavenumbers $k_1$ and $k_2$
are only allowed when two odd--parity waves are coupled to a wave
with even parity. Thus, $\vec k_1' + \vec k_1 = \vec k_2$ coupling
is allowed for odd/even frequency ratios such as 1/2, and $\vec
k_2' - \vec k_2 = \vec k_1$ coupling occurs in even/odd forcing,
such as 2/3 and 4/5. It is interesting that we have not observed
these states for other frequency ratios. It is possible that 2MS
states are preferred for all but the simplest frequency ratios
since, for higher ratios, linearly stable tongues corresponding to
wavenumbers close to the value of $k_3$ (defined by
$\omega_3=\omega_2-\omega_1$) are more dense.

A typical phase space in which $2kR$ states are observed is
presented in \Fig{PhaseSpaceBoth} (right). For values of $\chi$
that are far from $\chi_c$, the phase diagram is similar those
described in sections \ref{Results1} and \ref{Results2}. The $2kR$
state exists in the near vicinity of $\chi_c$, and replaces both
the 2MS and unlocked states. This region is bounded for $\chi
> \chi_c$ by squares and for $\chi <
\chi_c$ by $k_1$--dominant DHS states that are mixed with
oscillons \cite{OscillonsNature} (see \cite{Arbell3}). At higher
values of $\nu$ and $h$ (e.g. $\nu=47$ cS, $h=0.3$ cm, $\nu=87$
cS, $h=0.5$ cm) square and hexagonal patterns only exist near
onset. Upon increase of $A$ both types of patterns become rolls.
The $2kR$ state is, however, unaffected by the state preceding it.
They appear for a similar range of $\phi$. The transition to DHS
and oscillon states also occurs as in \Fig{PhaseSpaceBoth}.

\begin{figure}
\vspace{-.5cm} \hspace{0.0cm} \centerline{ \epsfxsize =7.5cm
\epsfysize =2.5cm \epsffile{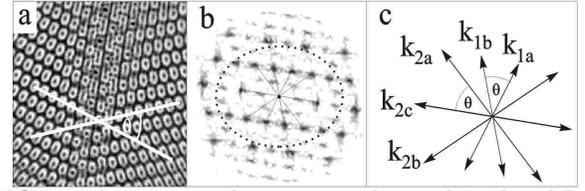} }
 \caption{A common
phenomenon observed in the $2kR$ region of phase space is the
formation of two domains of 2kR states (left) which are oriented
with a relative angle of $\theta$ that is also the angle between
the two larger vectors $\vec{k_2}$, $\vec{k'_2}$ forming the 2kR
state. A similar phenomenon can also be observed in domains formed
by triblock copolymers \protect\cite{Breiner}. } \label{Domain}
\end{figure}

$2kR$ states are not always correlated throughout the entire
system. At relatively low values of $A$, $2kR$ states can
sometimes be found in 2 or 3 domains, as shown in \Fig{Domain}.
The angle separating two such domains is identical to the angle
$\theta$  between $\vec k_2$ and $\vec k_2'$, as defined in
\Fig{2kR_intro}(b). This type of domain separation is also
observed in ``knitting patterns" \cite{Breiner}, formed by
triblock copolymers near a bistable point. (These new materials
have a reciprocal lattice structure similar to $2kR$ states).
\begin{figure}
\vspace{0cm} \hspace{0.0cm} \centerline{ \epsfxsize =8cm
\epsfysize =4.9cm \epsffile{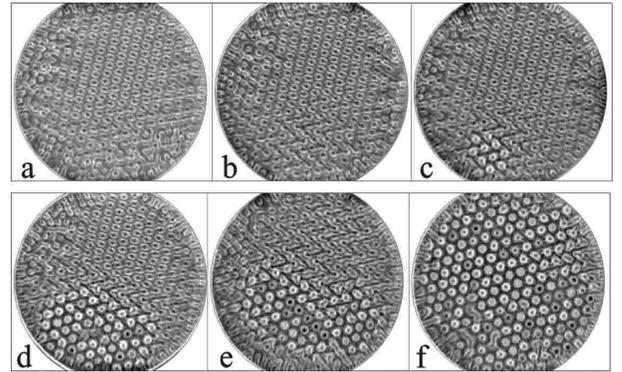} }
 \caption{The
transition from the 2kR state to the DHS state occurs through an
``explosive'' process in which a spatial domain of hexagons forms
and rapidly increases in size. Six images (a-f) of this process
taken at 1.5 second intervals for an experiment performed with
$66/44$ Hz driving for $\nu=23$ cS, and $h=0.2$ cm at mixing angle
$\chi = 70^\circ$. Note areas of mixed 2kR and hexagonal domains
around the perimeter of the hexagonal area (clearly seen in upper
half of (e)). In this process the driving parameters are
\emph{fixed}.
} \label{Explosion}
\end{figure}

 Let us return to the description of the typical phase space. As
increasing the driving amplitude $A$ strengthens the nonlinear
interactions between the waves, $2kR$ domains coalesce at higher
levels of excitation. Further increase of $A$ yields a hysteretic
bifurcation to the ``Double Hexagonal Superlattice'' state (DHS),
where oscillons are formed at the maxima of the pattern (see
\cite{Arbell3}). In \Fig{Explosion} we present a time sequence of
the transition from a global $2kR$ state to a DHS state by means
of rapid front propagation. Only at the final stages of this
process does the pattern bifurcate into the high--amplitude state
consisting of oscillons superimposed on a DHS background (see Sec.
\ref{DHS} and \cite{Arbell3}). The opposite transition from the
DHS to the $2kR$ state has a qualitatively different character.
Small $k_2$--dominated wavelength domains penetrate the DHS from
the perimeter of the cell in a way similar to the transition from
hexagons to SSS. The transition is not reversible and can have a
small hysteresis (under $1\%$).

Both 2MS and $2kR$ states are observed for the driving ratios 2/3
and 4/5. In \cite{Arbell2} the dimensionless parameter $\delta /h
\equiv (\nu / \omega_{ave})^{1/2}/h$ (where
$\omega_{ave}\equiv(\omega_2-\omega_1)/2$) was shown to govern the
selection between the two patterns. For $\phi=0$, 2MS/``Unlocked''
states exist above  $\delta/h \sim 0.12-0.17$, while $2kR$ states
exist below. This critical range of $\delta/h$ was obtained for a
broad range of both $h$ ($0.1\leq h\leq 0.6$cm) and $\nu$
($8.7\leq \nu \leq 186$cS). The parameter $\delta/h$ is the ratio
of two important physical scales of the system: the ratio of the
viscous boundary layer length, where the flow is rotational, to
the fluid layer's height. This parameter, in essence, defines the
region of applicability of the Zhang and Vinals, quasipotential
approximation. $(\delta/h)^2$ is the ratio of the dissipative time
to the driving time scales. In Lioubashevski et al. this number
was critical for determining pattern selection by single frequency
excitations\cite{Oleg1,Oleg2}. Thus, this transition suggests that
high dissipation in the system favors the 2MS over the $2kR$
states. This may result from the (linear) broadening of unstable
tongues \cite{KumarTuckerman} that occurs when dissipation in the
system is increased. This broadening would make the linearly
stable wavenumber observed in 2MS states, $k_3$, more accessible.
\begin{figure}
\vspace{0cm} \hspace{0.0cm} \centerline{ \epsfxsize =8cm
\epsfysize =5.0cm \epsffile{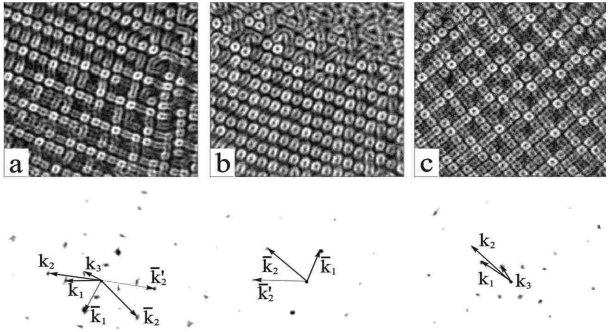} }
\caption{When $\delta/h$ is at the transition between $2kR$ and
2MS stability zones (b) $2kR$ and (c) 2MS  exist in the same
vicinity. ($A=6.59g$, $\chi=60.6^\circ$ for $2kR$ and $A=6.53g$,
$\chi=61.4^\circ$ for 2MS). (a) A transient state combining the
two resonances  was obtained for the parameters of (b). Parameters
here were 2:3 driving, $\omega_0=25$ Hz, and $\delta/h=0.12$.}
\label{2kR_2MS_50_75}
\end{figure}

In the transition regime of $\delta/h$, both 2MS and $2kR$ states
can coexist, as shown in \Fig{2kR_2MS_50_75} for $\delta/h=0.12$.
In this region, at fixed values of $\delta/h$, small changes in
either $A$ or $\chi$ can result in globally stable states of
either type. As shown in \Fig{2kR_2MS_50_75}a, transient states in
which both states are present can also result in this regime. In
these states both resonant mechanisms can operate {\em
concurrently} in different spatial regions of the fluid cell.

\begin{figure}
\vspace{0cm} \hspace{0.0cm} \centerline{ \epsfxsize =8cm
\epsfysize =5.6cm \epsffile{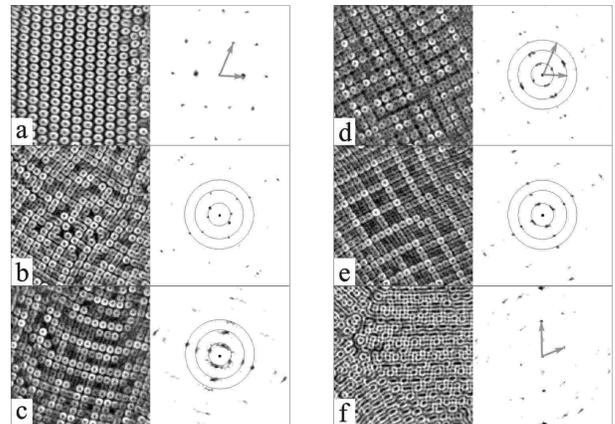} }
 \caption{The effect
of changing $\phi$ in the driving can be seen in the sequence of
images (left) and their power spectra (right) for different values
of $\phi$. While for $\phi=0^\circ$ (a) a 2kR state is stable and
global, increasing $\phi$ from (b) $20^\circ$ and (c) $30^\circ$
transforms the system to a stable 2MS state at (d) $90^\circ$.
Increasing the angle to (e) $120^\circ$, (f) $150^\circ$ reverses
the process and finally for $\phi=180^\circ$ a global $2kR$
re-emerges. The 3 concentric circles indicate the magnitude of the
three 2MS wavevectors ($k_1, k_2, k_3$), while the gray arrows
indicate the wave vectors of the 2kR state $\vec{k_1}, \vec{k_2}$.
These measurements were performed for 2/3 driving with system
parameters: $\omega_0/(2\pi)=25$ Hz, $\nu=23$ cS, and $h=0.2$ cm
at mixing angle $\chi=70.5^\circ$
} \label{50_75_phase}
\end{figure}

As the two driving frequencies are commensurate, the phase
variable, $\phi$ in Eq. \ref{2freqDriving1} is a relevant control
parameter. As was shown in the experimental work of Muller and
Edwards and Fauve (see Sec. \ref{2freq-exp}) changing $\phi$ can
affect pattern selection. Typically, the $2kR$ state exists over
the range $-20^\circ<\phi<+15^\circ$. The phase space presented in
\Fig{PhaseSpaceBoth} (right) is typical for $0.16<h<0.22$ cm and
$\nu=23$ cS. In \Fig{50_75_phase} we show how changing the angle
$\phi$ causes the pattern to change from the $2kR$ to the 2MS
state. The transition is not abrupt and in some regions localized
patches of both states can co--exist. In general, the size of the
region in phase space where a single $2kR$ domain exists decreases
with the distance from $\phi=0^\circ$. Because $\phi=180^\circ$ is
equivalent, for 2/3 driving, to $\phi=0^\circ$, the $2kR$ is also
stable at this angle. The strong effect of changes in $\phi$ is
consistent with the predictions of Zhang et al. (see Sec.
\ref{ModelEq}) who showed how $\phi$ can affect the mode coupling
function $\beta(\theta)$. Silber et al. \cite{SilberTopaz} have
recently demonstrated that changing $\phi$ affects the $2kR$
resonance by varying the non--linear coefficients of the model
equations describing the system.

\subsection{Temporal behavior}

\begin{figure}
\vspace{0cm} \hspace{0.0cm} \centerline{ \epsfxsize =7.5cm
\epsfysize =1.8cm \epsffile{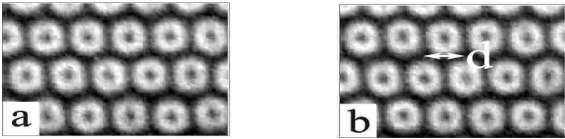} }
\caption{A typical $2kR$ state taken for $\omega_0/(2\pi)=25$ Hz,
$\nu=23$ cS and $h=0.2$ cm, in temporal phases separated by half
of the common excitation time $T/2=\pi/\omega_0$. The time
translation is equivalent to a spatial translation of magnitude
$d=\pi/(|\vec{k_2}+\vec{k'_2}|)$. This feature can be reproduced
by Eq. \protect\ref{height}.
} \label{2kR-Jitter}
\end{figure}

\begin{figure}
\vspace{0cm} \hspace{0.0cm} \centerline{ \epsfxsize =7.5cm
\epsfysize =4.50cm \epsffile{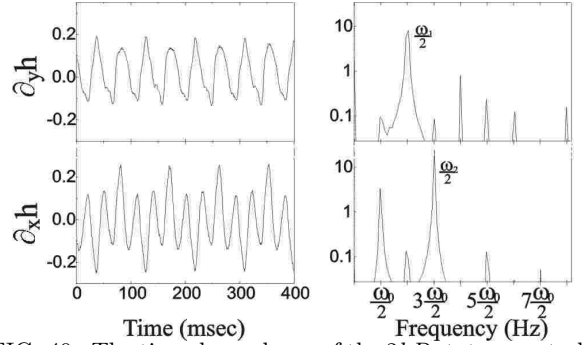} }
 \caption{The time
dependence of the $2kR$ state was studied using the reflection of
a laser by a point on the fluid surface. The directional
derivatives $\partial_{x}h$ and $\partial_{y}h$ as a function of
time (left) and their corresponding power spectra (right) show
that each direction has a different temporal dependence. (The
$x,y$ directions are defined in \protect\Fig{2kR_intro}a).
$\partial_{y}h$ is dominated by the $k_2$ and $k_2'$ components
with dominant frequency $\omega_2/2$ and $\partial_{x}h$
corresponds to the $k_1$ component with dominant frequency
$\omega_1/2$. This figure was taken from \protect\cite{Arbell2}.
The parameters for the above were $\omega_0/(2\pi)=22$ Hz,
$\nu=23$ cS, $\phi=0^\circ$ and $h=0.2$ cm} \label{laser2kR}
\end{figure}

The temporal behavior of the $2kR$ state is similar to that of the
even/odd 2MS states, where the time dependence in different
directions is qualitatively different. Like the even/odd 2MS
states, different directions can have dominant subharmonic or
harmonic components. In \Fig{2kR-Jitter} we see how the harmonic
nature of the $2kR$ is manifested. A time translation of
$\pi/\omega_0$ shifts the observed pattern by a spatial
translation of $\pi/|\vec k_2+\vec k'_2|$ in the x direction (as
defined in \Fig{2kR_intro}a). The overall spatio-temporal behavior
of the $2kR$ state is consistent with \Eq{height}.
\begin{eqnarray}\label{height}
 h\left({t,x,y}\right)
=&&[a_1\cos(\omega_0t)+a_2\cos(2\omega_0t)+...]\cos(\vec k_1\cdot
\vec x) \nonumber \\ 
&&+ \,[b_1\cos(\omega_0t/2)+b_2\cos(3\omega_0t/2)+...]\times \\
 &&\phantom{+\,}[\cos(\vec k_{2}\cdot\vec y)+\cos(\vec
k_2'\cdot \vec y) ]\:.\nonumber
\end{eqnarray}
The fact that the two directions, $\vec x ||\vec k_1$ and $\vec y
\perp x$ (as defined in \Fig{2kR_intro}) each exhibit {\em
different} time dependence is demonstrated in \Fig{laser2kR}. This
typical time series of the $x$ and $y$ components of the surface
gradient of this state at a single point is similar to that
presented in \Fig{Laser30_50_second} for the 2MS state. By our
choice of axes, the $\partial_{x}h$ component contains mainly the
$\omega_1/2$, $2\omega_1/2 \ldots$ peaks, while the dominant
frequencies in the $y$ direction are $(\omega_2-\omega_1)/2$,
$\omega_2/2 \ldots $. Arbitrary $\vec x$ and $\vec y$ directions
will contain both $k_1$ or $k_2$ eigenmodes. This orientational
dependence may prove to be a general characteristic of
superlattice states. When a {\it single} mode is dominant, the
predictions of Zhang and Vinals \cite{Zhang2freq} are in good
quantitative agreement with our measurements of the relative peak
intensities. In the vicinity of $\chi_c$, when 2 modes are
concurrently excited, this analysis does not apply and a new
theoretical framework is needed.

In low viscosity fluids ($\nu=8.7$ cS, $0.1 < h <0.2$ cm, and
$\delta/h< 0.13$) an interesting variant of the pure $2kR$ state
is observed whose symmetry changes with its temporal phase (see
\cite{Arbell2}). At different temporal phases pure hexagonal,
mixed hexagonal and $2kR$  phases can be seen. Interestingly, this
state exists for a significantly broader range of $\phi$
($-70^\circ<\phi <70^\circ$) than the pure $2kR$ state. Together
with the vector triad characteristic of $2kR$ states, coupling
with the difference vector, $\vec{k_2}-\vec{k_1}$, is also
observed in the spatial spectra of these states.

\subsection{Tuning of the resonant angles and quasi--pattern formation}\label{quasi-section}

\begin{figure}
\vspace{0cm} \hspace{0.0cm} \centerline{ \epsfxsize =8cm
\epsfysize=4cm \epsffile{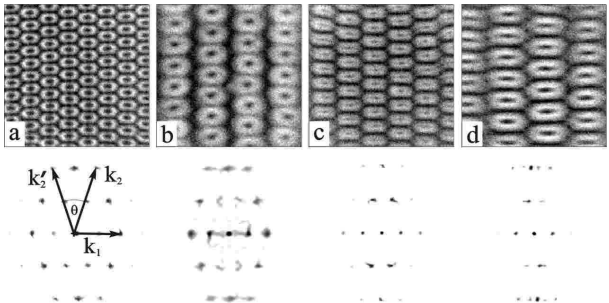} }
 \caption{The resonant
angle between $k'_2$ and $k_2$ varies with the parameters of the
experiment. We obtain angles of (a) $36^\circ$ for system
parameters of $\omega_0/(2\pi)=25$ Hz, $\nu=47$ cS, and $h=0.25$
cm,
(b) $33^\circ$ for system parameters of $\omega_0/(2\pi)=15Hz$,
$\nu=87$ cS, and $h=0.5$ cm,
(c) $32^\circ$ for system parameters of $\omega_0/(2\pi)=25$ Hz,
$\nu=87$ cS, and $h=0.5$ cm and
(d) $29^\circ$ for system parameters of $\omega_0/(2\pi)=20$ Hz,
$\nu=184$ cS, and $h=0.54$ cm.
All images were taken from a 9 x 9 cm$^2$ square in the center of
the circular plate. } \label{2kR_angles_g}
\end{figure}

The angle $\theta$ between the two wavevectors $\vec{k_2}$ and
$\vec{k'_2}$, can be tuned by changing the different system
parameters. Since the values of $k_1$ and $k_2$ are roughly
determined by the dispersion relation $\omega(k)$, the angle
$\theta$ can be varied by changing $\omega_0$, $\nu$, or $h$ while
leaving $m/n$ constant. This is demonstrated in
\Fig{2kR_angles_g}.
 The value of $\theta$ can be calculated
for a specific mixing angle $\chi$ using the numerical method
developed by Kumar and Tuckerman \cite{Besson}.

\begin{figure}
\vspace{0cm} \hspace{0.0cm} \centerline{ \epsfxsize =8.5cm
\epsfysize =1.cm \epsffile{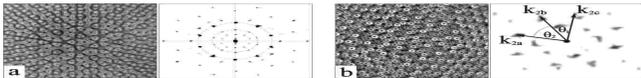} }
 \caption{(a) A
10--fold quasiperiodic pattern and its power spectrum is observed
for $\omega_0/(2\pi)=30$ Hz, $\nu=47$ cS and $h=0.33$ cm. Circles
are drawn with radii $k_1$ (inner) and $k_2$ (outer). For these
parameters $\theta=360^\circ /10$ and five $2kR$ regions combine
to form the quasipattern. For $\omega_0/(2\pi)=30$ Hz, $\nu=23$ cS
and $h=0.2$ cm (b) an nearly 8--fold symmetric pattern is seen.
This pattern, however, is actually a deformed quasipattern since
the $2kR$ value here of $\theta=41^\circ$ does not evenly divide
$360^\circ$. Instead of a single angle, two different angles
$\theta_1=41^\circ$ and $\theta_2=49^\circ$ are observed in the
power spectrum.} \label{quasi}
\end{figure}

As suggested by \cite{MullerModel,Frisch}, when rhomboidal states
exist the tuning of $\theta$ can provide a well--defined mechanism
in which quasipatterns of any desired symmetry may be formed. When
$360/\theta=p$, an integer number of adjacent triads can be
formed. As conjugate pairs of triads are always formed, the
integer $p$ must always be even (as observed in
\cite{Edwards92,Lifshitz}). This is demonstrated in \Fig{quasi}
where the formation of perfect 10--fold quasicrystalline patterns
and approximate 8--fold quasi--patterns occurs for values of
$\theta$ that are tuned to these resonant ($360/\theta=n$) angles.
As the power spectra of these states indicate, each of the inner
circle of peaks of magnitude $k_1$ is coupled by a triad resonance
with two peaks of magnitude $k_2$ along the outer circle. When
$\theta\approx 41^\circ$, a symmetric quasi-pattern is not
possible and a distorted 8--fold quasi-pattern, as shown in
\Fig{quasi}b, occurs.

\subsection{Three frequency driving}

Having observed the distorted 8--fold quasipatterns described in
\Fig{quasi}b, we attempted to stabilize these ``{\em asymmetric}"
quasi--patterns by modifying the driving. Muller in
\cite{Muller2freq} added a third frequency perturbation to break
the spatial phase symmetry in the subharmonic regime and thereby
control the transition between triangles and hexagons. This
motivated us to add a third frequency in order to enable the
excited wavevectors $k_1$ and $k_2$ to spatially lock to the value
of $\theta=45^\circ$ for which 8--fold quasi--patterns can
naturally form. We used the following driving function:

\begin{equation}
A [ a_1 cos(p_1\omega_0t)+a_2 cos(p_2\omega_0t+\phi_1)+a_3
cos(p_3\omega_0t+\phi_2) ]
\end{equation}

\noindent where the total driving amplitude is given by $A$ and
the normalized amplitude ratios by $a_1:a_2:a_3$ with
$a_1+a_2+a_3=1$. $p_1:p_2:p_3$ are the three frequency ratios
$p_1<p_2<p_3$ and $\phi_1$, $\phi_2$ are the phase differences
with respect to the $p_1$ components.

\begin{figure}
\vspace{0cm} \hspace{0.0cm} \centerline{ \epsfxsize =8.5cm
\epsfysize =1.9cm \epsffile{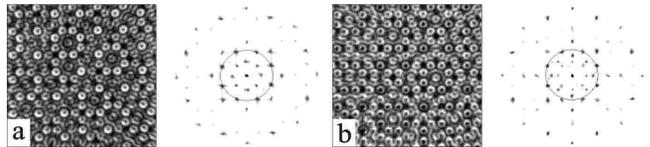} }
\caption{ Two temporal phases of an 8-fold quasi-pattern observed
for three-frequency driving (a,b). This state is temporally
subharmonic and was observed in the region where the distorted
8--fold quasi--patterns were observed (see \protect\Fig{quasi}.
This state was observed in a 50/75/100 Hz experiment with driving
amplitude ratio $a_1:a_2:a_3$ equal to $0.16:0.36:0.48$ and a
phase difference of $180^\circ$ between the 100 Hz component and
the two other components.
} \label{three_Q8_New_first}
\end{figure}

Using 2:3:4 driving we indeed observed a perfect  8--fold
quasi--pattern. In \Fig{three_Q8_New_first} we show images and
power spectra of this state at two temporal phases. This state is
subharmonic in time and can be observed in the region where
10--fold quasi--patterns and 8--fold distorted quasi--patterns
were observed for the 2/3 experiments described above.
Interestingly, this state was exceedingly stable and existed
within a single domain over a wide range of parameters (note the
sharp peaks in \Fig{three_Q8_New_first}). This is in sharp
contrast to the distorted 8--fold state shown in \Fig{quasi} which
existed in both a narrow range of parameters and, as evident in
its diffuse spectrum, had a tendency to break up into domains. It
is possible that the third frequency allows the $2kR$ mechanism to
form the quasi--pattern through the relaxation of the ratio
between $\vec{k_1}$ and $\vec{k_2}$ and thereby the angle between
them. This state was observed in a 50/75/100 Hz experiment with
ratios of driving amplitudes $a_1:a_2:a_3$ equal to
$0.16:0.36:0.48$, where the single--frequency critical
accelerations for these frequencies are, respectively, 2.56g,
4.40g and 6.91g (yielding ratios of a $0.18:0.32:0.49$) and phase
differences $\phi_1=0^\circ$, $\phi_1=180^\circ$. Although the
third frequency acceleration is not small here, it is still below
the critical value for single--frequency excitation at 100Hz and
wavevectors corresponding to the 100 Hz component were not
observed.


\subsection{Transition states to rhomboids}

\begin{figure}
\vspace{0cm} \hspace{0.0cm} \centerline{ \epsfxsize =7.5cm
\epsfysize =4.90cm \epsffile{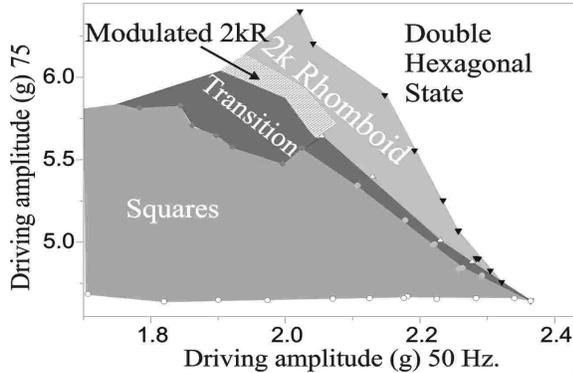} }
\caption{An expanded view of the phase space described in
\protect\Fig{PhaseSpaceBoth} (right). Transition states between
square and $2kR$ states are formed by a superposition of a square
state with one (two) sets of $2kR$ triad vectors. Between the
$2kR$ and transition states, a modulated $2kR$ state exists in
which an additional $\vec{k_2}$ component parallel to the
$\vec{k_1}$ component exists.
} \label{zoom1}
\end{figure}

\Fig{zoom1} shows an expanded view of the $2kR$ phase space around
$\chi_c$ (see \Fig{PhaseSpaceBoth} (right)). The region shown,
lies in the subharmonic region of phase space ($\chi>\chi_c$)
where square patterns form at threshold. For $\chi$ close to
$\chi_c$ the transition region and the $2kR$ region is relatively
narrow compared to the transition region for larger values of
$\chi$.

The transition from the square symmetry region to the $2kR$ state
is perhaps more interesting than the transition between squares to
square 2MS discussed in (subsection \ref{Transitions}). While the
square--2MS transition occurs through the formation of an
``additional'' set of wavevectors (of different wavenumber) that
combine with each of the primary wavevectors that initially formed
the squares, the square--$2kR$ transition has a qualitatively
different nature. Here, the basic square symmetry is not only
broken, but is actually {\em replaced} by a pattern of completely
different symmetry. In $2kR$ states, one of the two $k_2$
wavevectors that are initially perpendicular to each other in the
square state is replaced by a $k_2$ wavevector whose orientation
forms the angle $\theta$ that is determined by the magnitude of
the $k_1$ wavevector which defines the $2kR$ rhomboidal pattern.

\begin{figure}
\vspace{0cm} \hspace{0.0cm} \centerline{ \epsfxsize =8.5cm
\epsfysize =3.8cm \epsffile{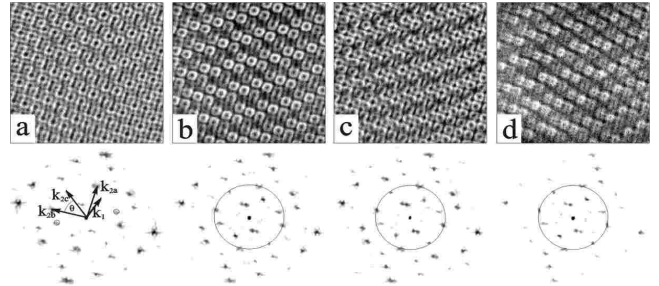} }
 \caption{A typical
time sequence of the transition state is shown with the
corresponding power spectra (bottom). The wavevectors
($\vec{k_{2a}},\vec{k_{2b}}$) generating the squares appear
 together with a third wavevector,
$\vec{k_{2c}}$ that is oriented at an angle of $\theta=41^\circ$
to $\vec{k_{2b}}$. The above experiment was made for system
parameters of $\omega_0/(2\pi)=27Hz$, $\nu=23 cS$, $\phi=0^\circ$,
$\chi=69^\circ$  and $h=0.2cm$.
} \label{SL4_time}
\end{figure}
A typical ``transition'' state at different temporal phases is
shown in \Fig{SL4_time}. The transition occurs through the
formation of one or two additional wavevectors of magnitude $k_2$,
($\vec{k_{2c}}$ in \Fig{SL4_time}a). These new wavevectors are
aligned at the $2kR$ resonant angle $\theta$ with respect to the
$k_2$ wavevectors ($\vec{k_{2a}}$ and $\vec{k_{2b}}$) which form
the squares. Although additional $k_2$ vectors could, in theory,
form $2kR$ triads with {\em all} of the original $k_2$
wavevectors, empirically, we find that only one of the initial
$k_2$ directions is selected. This is possibly due to the fact
that the self-interaction of the {\em harmonic} $k_1$ wavevectors
prefers a hexagonal (rather than a square) arrangement.
\begin{figure}
\vspace{0cm} \hspace{0.0cm} \centerline{ \epsfxsize =8.5cm
\epsfysize =1.8cm \epsffile{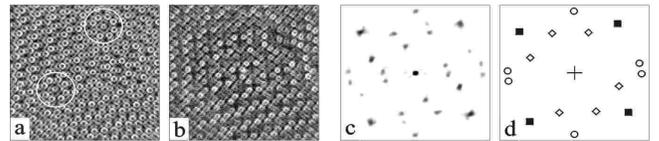} }
\caption{The transition states between square and $2kR$ symmetry
do not usually form a global state. (a) and (b) show different
phases of a typical transition state that is formed by three
domains. In (a) white circles note areas where two $2kR$ domains
intersect. The power spectrum of (a) is shown in (c) and the
wavevectors are noted in (d) by (squares) the primary square
wavevectors, (circles) $2kR$--resonant $k_2$ wavevectors, and
(diamonds) the $k_1$ wavevectors. Since the power spectra reflect
the entire pattern, three separate sets of $2kR$ triads can be
observed in the same spectrum. The system parameters here are
$\omega_0/(2\pi)=30Hz$, $\nu=23 cS$, $h=0.2cm$, $\chi=71.4^\circ$
and $\phi=0^\circ$.
} \label{SL4_domains}
\end{figure}
\noindent Thus the transition pattern that is formed is a
superposition of a $k_2$ {\em square} state with a $2kR$ state.
All other wavevectors seen are formed by secondary interactions of
these wavevectors. Generally, the transition state breaks the
symmetry of the square pattern in a single direction. Since the
transition state is usually not global, two domains can form and
at their common border one can observe structures that retain the
four--fold symmetry, as indicated by the white circles in
\Fig{SL4_domains}a.

In the near vicinity of $\chi_c$, the transition patterns appear
as in \Fig{SL4_time}, and their rhomboidal character is apparent.
Farther away from $\chi_c$, the appearance of the transition state
is more similar to squares, as the orthogonal $k_2$ wavevectors
are more dominant.

\begin{figure}
\vspace{0cm} \hspace{0.0cm} \centerline{ \epsfxsize =7.5cm
\epsfysize =3.5cm \epsffile{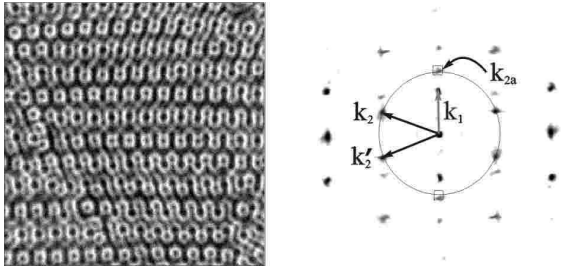} }
 \caption{``Modulated
$2kR$" states (left) are variants of the $2kR$ states formed when
(see the power spectrum (right)) an additional wavevector of
magnitude $k_2$, $\vec{k_{2b}}$, parallel to $\vec{k_1}$ is
generated. The system parameters are $\omega_0/(2\pi)=22$ Hz,
$\nu=23$ cS, $\phi=0^\circ$ and $h=0.2$ cm.
 } \label{SL5}
\end{figure}
For $\chi>70^\circ$ we observe a variant of the $2kR$ state, shown
in \Fig{SL5}, which forms between the ``transition" and the $2kR$
regions (see \Fig{zoom1}). This state, which we call a
``modulated'' $2kR$, consists of a modulation of the regular $2kR$
triad by an additional wavevector of the magnitude and temporal
behavior of the larger frequency component. This state is formed
by the superposition of a $2kR$ state with an additional wave
vector of magnitude $k_2$ that is oriented parallel to the $k_1$
wave vector (see \Fig{SL5} (right)). This wavevector retains the
time dependence of the $k_2$ mode and generates a spatial and
temporal modulation, similar to that induced in the SSS mode, in
the $\vec{k_1}$ direction (x in \Fig{2kR_intro}a). A possible
mechanism that can generate a ``modulated'' $2kR$ state is a 2MS
resonance between the two collinear wavevectors $\vec{k_1}$ and
$\vec{k_2}$ that produces a $\vec{k_3}$ wavevector along the same
direction. This is verified by comparison with the square 2MS
pattern obtained for the same parameters with $\phi=90^\circ$ (see
\Fig{50_75_phase}e). This mechanism is supported by our
observations of the ``modulated'' $2kR$ states for only relatively
low values of $\omega_0$ for which the wavevectors $\vec k_1$,
$\vec k_2$ and $\vec k_3$, forming the 2MS state, are nearly
parallel.

\section{Additional resonant states}\label{Results4}

\subsection{States observed for odd/odd parity driving}

The work of Silber et al. \cite{SilberTopaz} indicates that the
possible three-wave resonant interactions between excited and
damped modes depend on the parity of the driving. For odd/odd
driving no 3-wave resonant interactions between $k_1$ and $k_2$
are expected since both modes are temporally subharmonic (see Sec.
\ref{ModelEq}). When considering four wave interactions, however,
these restrictions are no longer valid. Below, we describe two
experiments conducted with odd/odd driving that indeed show that
4--wave resonant interactions can be found both in the vicinity of
$\chi_c$, where one mode is strongly excited and the other only
weakly damped, and far from $\chi_c$ where the interactions
involve an excited mode and a single strongly damped mode
corresponding to the subharmonic frequency, $\omega_0/2$.
\begin{figure}
\vspace{0cm} \hspace{0.0cm} \centerline{ \epsfxsize =7.5cm
\epsfysize =6.5cm \epsffile{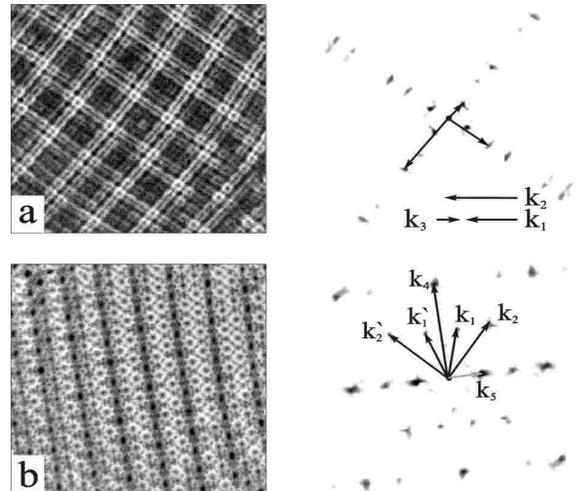} }
 \caption{A 5/7
ratio yields both the $2MS$ resonance (a) and a four--wave
resonance (b). In the $2MS$ power spectrum (a right) the three
vectors are nearly parallel obeying the resonance
$\vec{k_3}=\vec{k_2}-\vec{k_1}$. In (b right) the power spectrum
reveals a qualitatively different resonance:
$\vec{k_1}+\vec{k'_1}=\vec{k_2}+\vec{k'_2}\equiv\vec{k_4}$. This
resonance produces additional vectors such as $\vec{k_5}$ that are
sums and differences of the original wavevectors involved. The
experimental parameters are $55/77$ Hz driving frequencies,
$\nu=23$ cS, and $h=0.2$ cm. In (a) $\phi=90^\circ$,
$\chi=72.8^\circ$ and in (b) $\phi=0^\circ$, $\chi=70.8^\circ$.
} \label{SL_5_7first}
\end{figure}

For $\chi>\chi_c$ a 5/7 frequency ratio can produce a square 2MS
state in the vicinity $\chi_c$. This, however, is not the only
resonantly locked pattern that can be observed at this driving
ratio. Mixing angles typically 3--5 degrees beyond $\chi_c$
results in the formation of the state shown in \Fig{SL_5_7first}b.
The spatial spectrum of this state indicates that it is formed by
a qualitatively different mechanism. The original square pattern,
formed by two orthogonal $k_2$ wavevectors, is broken by an
additional pair of $k_1$ wavevectors, whose orientation is
determined by the four--wave resonance condition:
$\vec{k_1}+\vec{k'_1}=\vec{k_2}+\vec{k'_2}$. This resonance
produces additional vectors that are sums and differences of the
original wavevectors involved. States similar to these have been
previously observed in vertically oscillating convection
experiments \cite{SchatzRogers}, where 3--wave interactions are
forbidden.

\begin{figure}
\vspace{0cm} \hspace{0.0cm} \centerline{ \epsfxsize =7.5cm
\epsfysize =4.5cm \epsffile{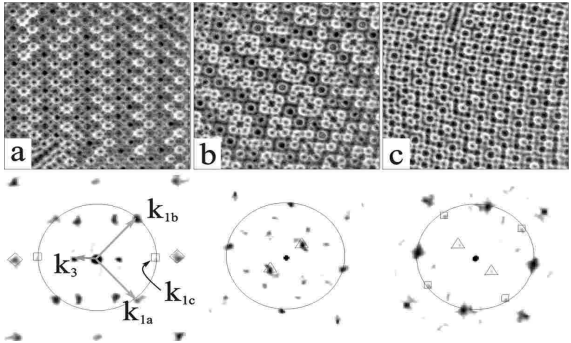} }
\caption{In an experiment with 3/5 ($60/100$ Hz) driving, patterns
with square symmetry (or slightly rhombic) dominate for mixing
angles $\chi<\chi_c$. The primary wavenumber is of magnitude $k_1$
(circles in a,b,c). When the driving is increased, the primary
symmetry is broken by a small wavevector of magnitude $k_3$
(enclosed in triangles). These states are observed for parameters
$\nu=23$ cS, and $h=0.2$ cm both at mixing angle close to $\chi_c$
(a) $\chi= 69.6^\circ$  and far from $\chi_c$ (b) $\chi=
48.1^\circ$ and  (c) $\chi= 58.7^\circ$. A possible mechanism for
the large scale symmetry breaking is a resonant four--wave
interaction. The power spectrum (a bottom) reveals the existence
of additional vectors of magnitude $k_1$ (enclosed by squares).
The interaction of $\vec{k_{1c}}$ with the sum of the original
vectors $\vec{k_{1a}}+\vec{k_{1b}}$ (enclosed by diamonds) results
in a new vector
$\vec{k_3}=\vec{k_{1c}}-(\vec{k_{1a}}+\vec{k_{1b}})$ whose scale
is consistent with $k(\omega_0/2)$.
} \label{60_100_Resonance}
\end{figure}

In odd/odd driving both sides of phase space exhibit subharmonic
temporal response. Unlike the 5/7 driving that produces hexagons
for $\chi<\chi_c$ (despite the subharmonic temporal response),
experiments performed using 3/5 driving with $\omega_0=20$ Hz
yield square symmetric patterns for $\chi>\chi_c$ and nearly
square patterns (rhomboids with an angle of $84^\circ$) for
$\chi<\chi_c$. Here, a state exhibiting a different mode of
4--wave coupling far from $\chi_c$ is presented in
\Fig{60_100_Resonance}. Like the SSS states, these patterns result
from a symmetry breaking bifurcation for $\chi_c +20^\circ> \chi>
\chi_c$. The power spectrum reveals that in addition to the
dominant pair of orthogonal wavevectors, ($\vec{k_{1a}}$ and
$\vec{k_{1b}}$ in \Fig{60_100_Resonance}) an additional wavevector
of the same magnitude, $\vec{k_{1c}}$, is created along the
bisector of $\vec{k_{1a}}$ and $\vec{k_{1b}}$. The vector sum of
these three wavevectors of magnitude $k_1$ produces a smaller
wavevector of magnitude
$k_3=\vec{k_{1c}}-(\vec{k_{1a}}+\vec{k_{1b}})$. The scale of
$\vec{k_3}$ is consistent with $k(\omega_0/2)$, as determined by
the linear dispersion relation. Thus, as in the case of 2MS
states, a symmetry--breaking slaved mode, $\vec{k_3}$ is excited
by a nonlinear resonance.

\subsection{States satisfying more than one resonance condition }
We have observed a number of cases where states which satisfy more
than a single resonance condition were selected by the system.
These states are generally stable in a relatively wide range of
phase space. Here we present a number of examples of such
multiply--resonant nonlinear states.

\begin{figure}
\vspace{0cm} \hspace{0.0cm} \centerline{ \epsfxsize =7.5cm
\epsfysize =3.0cm \epsffile{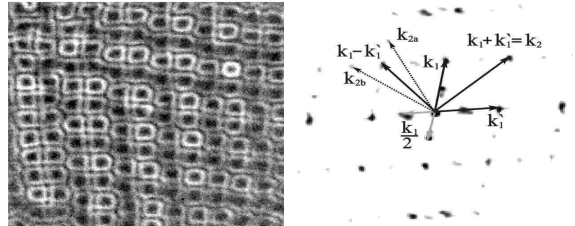} }
 \caption{An image
(left) and power spectrum (right) of a spatially subharmonic state
where a rhomboid is formed by $\vec{k_1}+\vec{k'_1}=\vec{k_2}$
where $k_1$ ($k_2$) are generated by the $5\omega_0$ ($8\omega_0$)
frequencies. The pure rhomboid symmetry is broken by two spatially
subharmonic vectors, $\vec{k_1}/2$ and $\vec{k'_1}/2$.  The
vectors $k_{2a}$ and $k_{2b}$ are formed by additional resonances:
$\vec k_{2a}=\vec{k_1}-\vec k'_1+\vec k_1/2$ and $\vec
k_{2b}=\vec{k_1}-\vec k'_1-\vec k'_1/2$. This state appears for
$50/80$ Hz driving for $\nu=23$ cS and $h=0.2$ cm for
$\chi\sim\chi_c-1^\circ$.
} \label{5_8_Resonance}
\end{figure}

 We have seen that SSS--I states result from a
primary hexagonal symmetry broken by a wavevector of size $k_c/2$.
A similar mechanism can occur for square or rhomboid patterns. In
\Fig{5_8_Resonance} we show a {\em spatially} subharmonic state,
obtained using 5/8 driving, where a rhomboid, is formed by two
wavevectors of magnitude $k_1$ ($\vec{k_1}$ and $\vec{k'_1}$ in
\Fig{5_8_Resonance}) via the additive $2kR$ resonance:
$\vec{k_1}+\vec{k'_1}=\vec{k_2}$. The spatial period of the
rhomboid is doubled by the appearance of two {\em new} wavevectors
of size $k_1/2$, ($\vec{k_1}/2$ and $\vec{k'_1}/2$ in
\Fig{5_8_Resonance}). This state was observed for a mixing angle
$\chi=\chi_c-1^\circ$. An additional spatial resonance is
apparent. The vectors, $\vec{k_1}-\vec{k'_1}+\vec k_1/2$ and
$\vec{k_1}-\vec k'_1-\vec k'_1/2$ both form vectors of length
$k_2$ (see $k_{2a}$ and $k_{2b}$ in \Fig{5_8_Resonance}). This
``extra" 4--wave resonance may be the result of spatial
mode-locking. It may be possible that this extra resonance causes
the selection and resultant stability of this state.

\begin{figure}
\vspace{0cm} \hspace{0.0cm} \centerline{ \epsfxsize =8cm
\epsfysize =5.0cm \epsffile{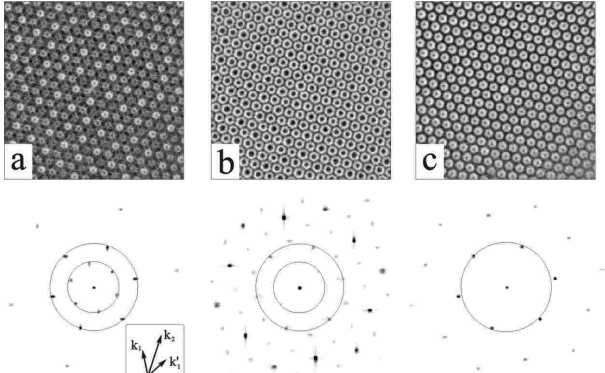} }
\caption{Images (a-c top) and the corresponding power spectra (a-c
bottom) of different temporal phases of a  SSS--II (2MS) state
observed for 1/2 driving in the harmonic region of phase space at
system parameters of $\nu=23$ cS, $40/80$ Hz and $h=0.155$ cm. The
pattern was observed in the vicinity of $\chi_c$  for
$\chi>\chi_c$. This state resembles SSS--II states observed for
2/3 driving although here a typical super--hexagon cell is $\pi/3$
symmetric whereas the SSS-II state generally has $\pi/6$
rotational symmetry.
} \label{SubSupHx_1_f1-2}
\end{figure}

Another example of a state satisfying two resonance conditions is
presented in \Fig{SubSupHx_1_f1-2}. This state, which is only
observed in 1/2 driving experiments, is both a hexagonal 2MS state
as well as an SSS--II state. The driving ratio of 1/2 is unique
since $\omega_3$ (given by $\omega_3=\omega_2-\omega_1$) is equal
to $\omega_1$. Thus, the wavenumber $k_3$, excited by $\omega_3$,
coincides with the wavenumber $k_1$, excited by $\omega_1$.
Therefore \cite{Zhang2freq}, a resonant triad is formed involving
only the two critical wavenumbers $k_1$ and $k_2$. In this case
SSS--II and 2MS states coincide for 20 Hz$<\omega_0<$50 Hz.  The
resonance that is formed is identical to the SSS--II resonance
found for odd/even driving in the harmonic region (see Sec.
\ref{SSS-II}), where the wavevectors of magnitude $q$ and $K$ of
the SSS--II state are replaced by wavevectors of respective
magnitude $k_1$ and $k_2$. The resonance condition for the 1/2
experiment can be written as $\vec{k_1}+ \vec{k'_1}=\vec{k_2}$.
The SSS--II type resonance is possible since, for the system
parameters used, the ratio $k_2/k_1$ is close to $\sqrt{3}$. The
same resonance was also observed by Wagner et al. for a two mode
subharmonic--harmonic interaction in a single--frequency,
lubrication--limit experiment \cite{WagnerNew}. In their
experiment, for special values of $h$ and $\omega$, two modes with
temporal responses of $\omega/2$ and $\omega$ and with spatial
wavenumbers $k_1$ and $k_2$, are excited by single--frequency
forcing.

\begin{figure}
\vspace{0cm} \hspace{0.0cm} \centerline{ \epsfxsize =7.5cm
\epsfysize =2.0cm \epsffile{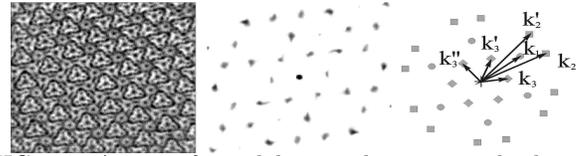} }
 \caption{A
state formed by simultaneous multiple resonances obtained with 2/3
driving with the same parameters used to form a $2kR$ state at
$\phi=0$, $\omega_0/(2\pi)=20$ Hz, $\nu=23$ cS, $\chi=59^\circ$,
and $h=0.2$ cm, but with a phase of $\phi=117^\circ$. The
wavevectors of magnitude $k_3$ are formed both from the resonance
$\vec{k_2}- \vec{k'_2}=\vec{k''_3}$ as well as the resonance
$\vec{k_2}- \vec{k_1}=\vec{k_3}$. In addition, the condition
satisfying the DHS structure is satisfied and the hexagonal
sublattice formed by the $\vec{k_3}$ wavevectors spans the entire
lattice. $k_1$, $k_2$, and $k_3$ are, respectively, noted by
squares, circles, and diamond symbols.
} \label{Transfer_Ph120_first}
\end{figure}

In our last example, we present an example of a state where
multiple resonance conditions are simultaneously satisfied. When
this occurs, we find that the stability of a pattern is
significantly enhanced. This state contains features of most of
the states described above. This single state includes a 2MS
resonance, a DHS resonance (with a $1/\sqrt{7}$ sublattice), an
SSS--II type resonance and an ``additive'' and ``subtractive''
$2kR$ resonance. This state was observed for 2/3 driving
($\omega_0=20$ Hz) with $\phi=117^\circ$ and is shown in
\Fig{Transfer_Ph120_first}.

\begin{figure}
\vspace{0cm} \hspace{0.0cm} \centerline{ \epsfxsize =8cm
\epsfysize =3.2cm \epsffile{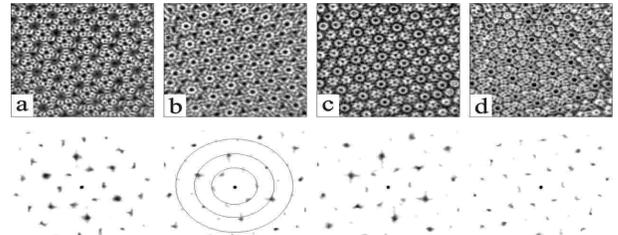} }
\caption{Different temporal phases of the state described in
\protect\Fig{Transfer_Ph120_first} taken for constant values of
the driving parameters. Circles of radii $k_1$ (middle circle),
$k_2$ (outer circle) and $k_3$ (inner circle) are drawn in (b). At
different temporal phases, different scales appear dominant.
 } \label{Transfer_Ph120_time}
\end{figure}

As can be seen in the power spectrum (\Fig{Transfer_Ph120_first}
right) a double lattice with wavenumber $k_2$ at an angle of
$22.2^\circ$ produces the familiar DHS structure described in
Section \ref{Results1}. However, in this case the smallest inner
hexagon has a magnitude of $k_3$, which is associated with
$\omega_3=\omega_2-\omega_1$ frequency, while the second inner
hexagonal set of wavevectors has a magnitude of $k_1$. This
resonance is very stable and exists in a single domain. This
stability is perhaps due its multiply--resonant nature. In
\Fig{Transfer_Ph120_time} a time sequence is shown of the
different phases of this state is presented.

\section{Conclusions}\label{conclusions}
The work described in this paper provides, for the first time, a
partial, but coherent, experimental picture of both the nonlinear
states generated by two--frequency forcing as well as their
domains of existence in phase space and the nonlinear interactions
that generate them. As demonstrated above, the space of nonlinear
patterns formed by two interacting unstable modes is very rich.
Our understanding of the types of structures and their selection
is just beginning. Predicted nonlinear three--wave resonances
\cite{SilberSkeldon}, however, appear to govern nonlinear pattern
selection only for the {\em simplest} ratios (2/3 and 1/2). Recent
theoretical advances made by Silber et al. in the analysis of
two--frequency Faraday system in the vicinity of the co--dimension
2 point \cite{SilberSkeldon,SilberTopaz} suggest that at least
some of the features discovered by our experiments can be
reproduced by amplitude equations derived using the
quasi--potential approximation (e.g. \cite{Zhang2freq}). In
addition, recent work \cite{SilberProctor,SilberRucklidge,RSF}
indicates that many of the superlattice states found far from
$\chi_c$ may be qualitatively understood as representations of
invariant subgroups of broken hexagonal symmetry. A quantitative
theory that describes the important regime of two or more
concurrently unstable modes for arbitrary height and viscosity has
still to be developed.

Some of the major results of this work are summarized below:

\begin{itemize}
\item The temporal symmetry of the driving determines which resonant interactions
{\em can} occur between the primary excited modes. A number of
distinct three and four wave interactions between excited modes
and between slaved and excited modes were experimentally observed.
\item Factors such as the dimensionless dissipation in the system and the driving phase $\phi$
can play an important role in the nonlinear interaction mechanism
selected by the system.  $\phi$ is a convenient parameter for
comparison with theory, as, in contrast to the system's
dissipation, its value does not play a role in the validity of the
theoretical (e.g. \cite{Zhang2freq}) approximation used.
\item Symmetry--breaking can often occur
via modes that nonlinearly couple to the original set of modes.
The symmetry--breaking modes tend to be arranged in invariant
subgroups of the original symmetry group.
\item In many cases, slaved modes can be non--linearly excited by the parametrically amplified
linear modes. The availability of slaved modes (determined by the
ratio of the driving frequencies) is important in the selection of
the final nonlinear states. This provides a nontrivial selection
mechanism for nonlinear states.
\item A theoretically predicted mechanism \cite{MullerModel,Frisch} for producing quasi--patterns
was for the first time experimentally observed. By tuning the
system parameters to satisfy a resonance condition $2n$--fold
quasi patterns can be produced for any desired $n$.
\end{itemize}

We believe that both the states and nonlinear mechanisms described
in this work should be of general importance to a wide class of
parametrically--driven nonlinear systems. Such systems include
parametrically--driven fluid systems, non--linear optical systems
\cite{tlidi,firth,Saffman,Leduc,Logvin}, nonlinear wave
interactions in superfluid helium \cite{ringberg1,ringberg2},
magnetically-driven ferrofluids \cite{Lee}, and possibly
nonlinearly coupled mechanical systems. Although the behavior of
systems driven by 2--frequency forcing is, itself, important.
Understanding the spatio-temporal behavior of such systems is but
a first step in understanding the general behavior in space and
time of nonlinear systems driven by multiple frequencies. The work
presented here is an important building block on the road to
understanding these more complex systems.

We gratefully acknowledge the support of the Israel Academy of
Sciences (Grant no. 203/99).


\begin{thebibliography}{10}

\bibitem{Douday}
S. Douady, Journal of Fluid Mechanics {\bf 221},  383  (1990).

\bibitem{Binks}
D. Binks and W. van~de Water, Physical Review Letters {\bf 78},
4043  (1997).

\bibitem{KudrolliPhysicaD}
A. Kudrolli and J.~P. Gollub, Physica D {\bf 97},  133  (1996).

\bibitem{KumarBajaj}
K. Kumar and K.~M.~S. Bajaj, Physical Review E {\bf 52},  R4606
(1995).

\bibitem{Bosch}
E. Bosch and W. van~de Water, Physical Review Letters {\bf 70},
3420  (1993).

\bibitem{Ciliberto}
S. Ciliberto, S. Douady, and S. Fauve, Europhysics Letters {\bf
15},  23
  (1991).

\bibitem{Ezerskii}
A.~B. Ezerskii, M.~I. Rabinovich, V.~P. Reutov, and I.~M.
Starobinets, Soviet
  Physics JETP {\bf 64},  1228  (1986).

\bibitem{Gluckman}
B.~J. Gluckman, P. Marcq, J. Bridger, and J.~P. Gollub, Physical
Review Letters
  {\bf 71},  2034  (1993).

\bibitem{KudrolliChaos}
A. Kudrolli and J.~P. Gollub, Physical Review E {\bf 54},  R1052
(1996).

\bibitem{Daudet}
L. Daudet, V. Ego, S. Manneville, and J. Bechhoefer, Europhysics
Letters {\bf
  32},  313  (1995).

\bibitem{Zhang2freq}
W. Zhang and J. Vinals, Journal of Fluid Mechanics {\bf 341},  225
(1997).

\bibitem{Muller2freq}
H.~W. Muller, Physical Review Letters {\bf 71},  3287  (1993).

\bibitem{Besson}
T. Besson, W.~S. Edwards, and L.~S. Tuckerman, Physical Review E
{\bf 54},  507
   (1996).

\bibitem{Kumar}
K. Kumar, Proceedings of the Royal Society of London, Series A
(Mathematical,
  Physical and Engineering Sciences) {\bf 452},  1113  (1996).

\bibitem{Arbell1}
H. Arbell and J. Fineberg, Physical Review Letters {\bf 81},  4384
(1998).

\bibitem{Edwards92}
W.~S. Edwards and S. Fauve, Comptes Rendus de l'Academie des
Sciences, Serie II
  (Mecanique, Physique, Chimie Sciences de la Terre et de l'Univers) {\bf 315},
   417  (1992).

\bibitem{Edwards93}
W.~S. Edwards and S. Fauve, Physical Review E {\bf 47},  R788
(1993).

\bibitem{Edwards94}
W.~S. Edwards and S. Fauve, Journal of Fluid Mechanics {\bf 278},
123  (1994).

\bibitem{Arbell3}
H. Arbell and J. Fineberg, Physical Review Letters {\bf 85},  756
(2000).

\bibitem{OscillonsNature}
P.~B. Umbanhowar, F. Melo, and H.~L. Swinney, Nature {\bf 382},
793  (1996).

\bibitem{Kudrolli}
A. Kudrolli, B. Pier, and J.~P. Gollub, Physica D {\bf 123},  99
(1998).

\bibitem{Arbell2}
H. Arbell and J. Fineberg, Physical Review Letters {\bf 84},  654
(2000).

\bibitem{SilberProctor}
M. Silber and M.~R.~E. Proctor, Physical Review Letters {\bf 81},
2450
  (1998).

\bibitem{BinksHeight}
D. Binks, M.~T. Westra, and W. van~de Water, Physical Review
Letters {\bf 79},
  5010  (1997).

\bibitem{MullerModel}
H.~W. Muller, Physical Review E {\bf 49},  1273  (1994).

\bibitem{Frisch}
T. Frisch and G. Sonnino, Physical Review E {\bf 51},  1169
(1995).

\bibitem{Lifshitz}
R. Lifshitz and D.~M. Petrich, Physical Review Letters {\bf 79},
1261  (1997).

\bibitem{Newell}
A.~C. Newell and Y. Pomeau, Journal of Physics A (Mathematical and
General)
  {\bf 26},  L429  (1993).

\bibitem{Zhang1freq}
W. Zhang and J. Vinals, Journal of Fluid Mechanics {\bf 336},  301
(1997).

\bibitem{SilberSkeldon}
M. Silber and A.~C. Skeldon, Physical Review E {\bf 59},  5446
(1999).

\bibitem{Crawford91a}
J.~D. Crawford, Reviews of Modern Physics {\bf 63},  991  (1991).

\bibitem{Crawford91b}
J.~D. Crawford, Physica D {\bf 52},  429  (1991).

\bibitem{SilberTopaz}
M. Silber, C.~M. Topaz, and A.~C. Skeldon, Physica D {\bf 143},
205  (2000).

\bibitem{Tze}
D.~P. Tse, A.~M. Rucklidge, R.~B. Hoyle, and M. Silber, Physica D
{\bf 146},
  367  (2000).

\bibitem{KumarTuckerman}
K. Kumar and L.~S. Tuckerman, Journal of Fluid Mechanics {\bf
279},  49
  (1994).

\bibitem{Oleg2}
O. Lioubashevski, J. Fineberg, and L.~S. Tuckerman, Physical
Review E {\bf 55},
   R3832  (1997).

\bibitem{SimonelliJFM}
F. Simonelli and J.~P. Gollub, Journal of Fluid Mechanics {\bf
199},  471
  (1989).

\bibitem{mudoscillons}
O. Lioubashevski {\it et~al.}, Physical Review Letters {\bf 83},
3190  (1999).

\bibitem{Oleg1}
O. Lioubashevski, H. Arbell, and J. Fineberg, Physical Review
Letters {\bf 76},
   3959  (1996).

\bibitem{SilberRucklidge}
D.~P. Tse, A.~M. Rucklidge, R.~B. Hoyle, and M. Silber, Physica D
{\bf 146},
  367  (2000).

\bibitem{RSF}
S.~M. Rucklidge, A.~M. and F. J., to appear  .

\bibitem{WagnerNew}
C. Wagner, H.~W. Muller, and K. Knorr, Physical Review E {\bf 62},
R33
  (2000).

\bibitem{WagnerVisco}
C. Wagner, H.~W. Muller, and K. Knorr, Physical Review Letters
{\bf 83},  308
  (1999).

\bibitem{Logvin}
A. Logvin Y., T. Ackemann, and W. Lange, Physical Review A
(Atomic, Molecular,
  and Optical Physics) {\bf 55},  4538  (1997).

\bibitem{Lee}
H.~J. Pi, S. Y. Park, J. Lee, and K.~J. Lee, Physical Review
Letters {\bf 84},
  5316  (2000).

\bibitem{Ackemann}
T. Ackemann, A. Logvin~Yu, A. Heuer, and W. Lange, Physical Review
Letters {\bf
  75},  3450  (1995).

\bibitem{Herrero}
R. Herrero {\it et~al.}, Physical Review Letters {\bf 82},  4627
(1999).

\bibitem{Hammack}
J.~L. Hammack and D.~M. Henderson, Annual Review of Fluid
Mechanics {\bf 25},
  55  (1993).

\bibitem{Pampaloni}
E. Pampaloni, S. Residori, S. Soria, and F.~T. Arecchi, Physical
Review Letters
  {\bf 78},  1042  (1997).

\bibitem{JayVictor}
J. Fineberg and V. Steinberg, Physical Review Letters {\bf 58},
1332  (1987).

\bibitem{Saffman}
A.~V. Mamaev and M. Saffman, Physical Review Letters {\bf 80},
3499  (1998).

\bibitem{Zimmermann}
W. Zimmermann {\it et~al.}, Europhysics Letters {\bf 24},  217
(1993).

\bibitem{Scroggie}
A.~J. Scroggie and W.~J. Firth, Physical Review a {\bf 53},  2752
(1996).

\bibitem{Hoyuelos}
M. Hoyuelos, P. Colet, M.~S. Miguel, and D. Walgraef, Physical
Review E {\bf
  58},  2992  (1998).

\bibitem{Dewel}
M. Bachir, S. Metens, P. Borckmans, and G. Dewel, Europhys. Lett.
{\bf 54},
  612  (2001).

\bibitem{Kuznetsov}
E.~A. Kuznetsov, A.~A. Nepomnyashchy, and L.~M. Pismen, Physics
Letters A {\bf
  205},  261  (1995).

\bibitem{Gunaratne}
G. H. Gunaratne, Q. Ouyong, and H. L. Swinney, Physical Review E
{\bf 50},  2802
  (1994).

\bibitem{Matthews}
P.~C. Matthews, Physica D {\bf 116},  81  (1998).

\bibitem{Hall}
P. Hall and R.~E. Kelly, Physical Review E {\bf 52},  3687
(1995).

\bibitem{Yethiraj}
M. Yethiraj, D.~M. Paul, C.~V. Tomy, and E.~M. Forgan, Physical
Review Letters
  {\bf 78},  4849  (1997).

\bibitem{Breiner}
U. Breiner, U. Krappe, E.~L. Thomas, and R. Stadler,
Macromolecules {\bf 31},
  135  (1998).

\bibitem{SchatzRogers}
J.~L. Rogers, M.~F. Schatz, O. Brausch, and W. Pesch, Physical
Review Letters
  {\bf 85},  4281  (2000).

\bibitem{tlidi}
M. Tlidi, P. Mandel, and M. Haelterman, Physical Review E {\bf
56},  6524
  (1997).

\bibitem{firth}
R. Martin, A.~J. Scroggie, G.~L. Oppo, and W.~J. Firth, Physical
Review Letters
  {\bf 77},  4007  (1996).

\bibitem{Leduc}
D. Leduc, M. LeBerre, E. Ressayre, and A. Tailet, Physical Review
a {\bf 53},
  1072  (1996).

\bibitem{ringberg1}
D. Rinberg, V. Cherepanov, and V. Steinberg, Physical Review
Letters {\bf 76},
  2105  (1996).

\bibitem{ringberg2}
D. Rinberg, V. Cherepanov, and V. Steinberg, Physical Review
Letters {\bf 78},
  4383  (1997).

\end{thebibliography}
\end{document}